%% file: template.tex
\documentclass[singlecolumn]{article}

\usepackage{arxiv}

\usepackage[utf8]{inputenc} 
\usepackage[T1]{fontenc}    
\usepackage{hyperref}       
\usepackage{url}            
\usepackage{booktabs}       
\usepackage{amsfonts}       
\usepackage{nicefrac}       
\usepackage{microtype}      
\usepackage{graphicx}
\usepackage[sort]{natbib}
\setcitestyle{comma,numbers,open={[},close={]}}
\usepackage{doi}

\title{Neural networks meet anisotropic hyperelasticity: A framework based on generalized structure tensors and isotropic tensor functions}

\author{
	Karl A. Kalina\\
	Chair of Computational and\\
	Experimental Solid Mechanics\\
	TU Dresden,
	01062 Dresden, Germany \\
	\And
	J\"{o}rg Brummund\\
	Chair of Computational and\\
	Experimental Solid Mechanics\\
	TU Dresden,
	01062 Dresden, Germany \\
	\And
	WaiChing Sun\\
	Department of Civil Engineering\\
	and Engineering Mechanics\\
	Columbia University, 
	NY 10027, New York, United States \\
	\And
	Markus K\"{a}stner\thanks{Corresponding author, email: \texttt{markus.kaestner@tu-dresden.de}.} \\
	Chair of Computational and\\
	Experimental Solid Mechanics\\
	TU Dresden, 
	01062 Dresden, Germany \\
}

\renewcommand{\shorttitle}{Neural networks meet anisotropic hyperelasticity}

\hypersetup{
pdftitle={Preprint_KalinaEtAl_2024},
pdfsubject={},
pdfauthor={Kalina},
pdfkeywords={First keyword, Second keyword, More},
}

\input{StyleSetup}

\theoremstyle{definition} 
\newtheorem{rmk}{Remark}
\usepackage[font=small]{caption}

\begin{document}

\maketitle

\begin{abstract}
We present a data-driven framework for the multiscale modeling of anisotropic finite strain elasticity based on physics-augmented neural networks (PANNs). Our approach allows the efficient simulation of materials with complex underlying microstructures which reveal an overall anisotropic and nonlinear behavior on the macroscale.
By using a set of invariants as input, an energy-type output and by adding several correction terms to the overall energy density functional, the model fulfills multiple physical principles by construction. The invariants are formed from the right Cauchy-Green deformation tensor and fully symmetric 2nd, 4th or 6th order structure tensors which enables to describe a wide range of symmetry groups. Besides the network parameters, the structure tensors are simultaneously calibrated during training so that the underlying anisotropy of the material is reproduced most accurately. In addition, sparsity of the model with respect to the number of invariants is enforced by adding a trainable gate layer and using $\ell_p$ regularization.
Our approach works for data containing tuples of deformation, stress and material tangent, but also for data consisting only of tuples of deformation and stress, as is the case in real experiments.
The developed approach is exemplarily applied to several representative examples, where necessary data for the training of the PANN surrogate model are collected via computational homogenization.
We show that the proposed model achieves excellent interpolation and extrapolation behaviors. In addition, the approach is benchmarked against an NN model based on the components of the right Cauchy-Green deformation tensor.
\end{abstract}

\section*{Graphical abstract}
\includegraphics{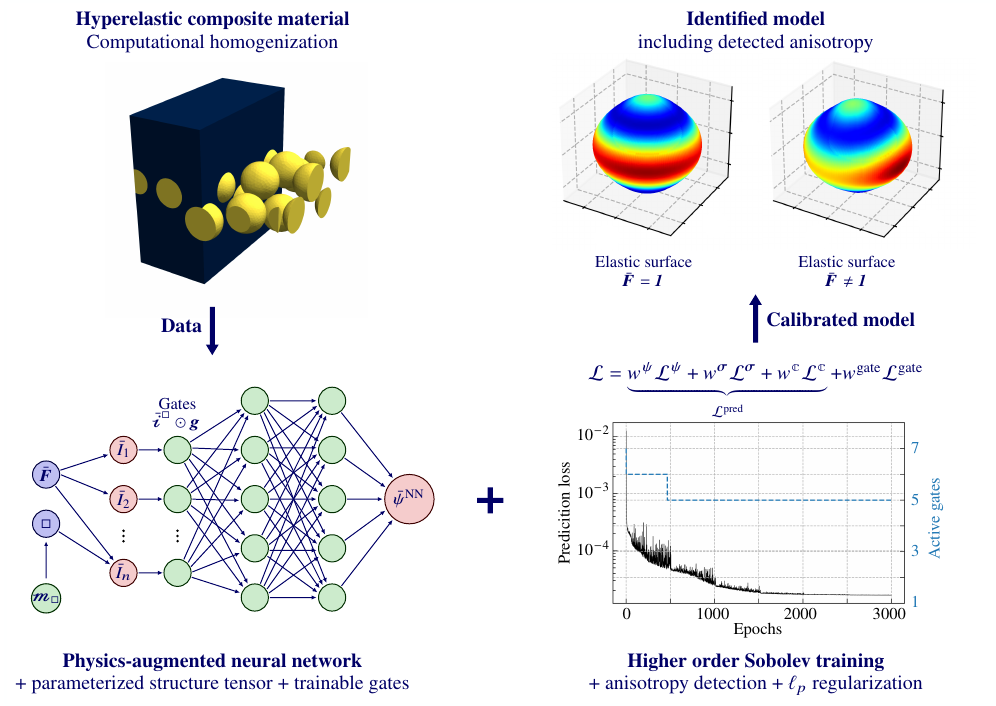}

	
	\section{Introduction}
	\label{sec:int}
	
	One of the cornerstones of continuum solid mechanics are constitutive models which allow to mathematically describe the behavior of a variety of materials such as metals, elastomers or even active materials and even non-Newtonian fluids. During the last century, considerable efforts have been made to understand the mathematical and physical requirements that a constitutive model should fulfill \cite{Haupt2000,Holzapfel2000}. Based on this knowledge, numerous models, referred to as classical constitutive models in the following, have been formulated and parameterized using data from experiments or simulations at lower scales. 
	However, especially for composite materials, which often show an extremely complex anisotropic and nonlinear behavior, these classical models are often not flexible and sufficiently accurate enough.
	For this reason, alternatives based on \emph{machine learning} and in particular the use of \emph{neural networks (NNs)} have recently become increasingly popular in constitutive modeling \cite{Dornheim2023,Fuhg2024}. Approaches of this kind indicate the potential of data-driven constitutive modeling, i.e., without having to decide on a specific model, it is possible to learn complex material behavior. 
	
	\subsection{Application of neural networks in constitutive modeling}
	In the pioneering work of Ghaboussi~et~al.~\cite{Ghaboussi1991} from the early 1990s, NNs, in particular \emph{feedforward neural networks (FNNs)}, were used for the first time to predict hysteresis in uniaxial and multiaxial stress states. To allow the FNN to learn history-dependent behavior, information from several prior time steps are used as input.
	After a brief period in the 1990s, the approach of using NNs for constitutive modeling has not been significantly pursued for the time being. With the increasing popularity of machine learning and the associated rapid advances in efficiency and accessibility, many different data-driven methods\footnote{It should be noted that in addition to NNs, there are several other machine learning/data-driven techniques for constitutive modeling. E.g., Gaussian process regression as applied in \cite{Frankel2020,Fuhg2022a,Ellmer2024} also enables to model elasticity. As shown in \cite{Wiesheier2024}, it is also useful to use splines to formulate the elastic energy density defined on a discretized invariant space. Other approaches based on sparse or symbolic regression allow an automated discovery of constitutive models \cite{Flaschel2021,Flaschel2023,Meyer2023a,Abdusalamov2023,Kissas2024}, i.e., instead of only identifying the parameters of a predefined model, the algorithm selects a model from a large set of candidates. This offers the advantage of interpretability.} have emerged in mechanics in a very short time in recent years, see the review articles \cite{Bock2019,Liu2021,Dornheim2023,Fuhg2024,Zheng2023}.
	
	A rather important development in NN-based constitutive modeling and generally in scientific machine learning is to include fundamental physical knowledge, which is denoted as \emph{physics-informed} \cite{Raissi2019,Henkes2022,Bastek2023,Harandi2024}, \emph{mechanics-informed} \cite{Asad2022}, \emph{physics-augmented} \cite{Klein2022,Linden2023}, \emph{physics-constrained} \cite{Kalina2023}, or \emph{thermodynamics-based} \cite{Masi2021}. 
	This may be accomplished in two ways: either strongly, as in the case of adapted network architectures \cite{Kalina2022a,Linka2021}, or weakly, as in the case of problem-specific loss terms for training, see \cite{Rosenkranz2023,Weber2021,Weber2023}. As shown in \cite{Linden2023,Masi2021,Fuhg2023,Masi2024}, these models enable the usage of sparse training data and a significant improvement of the model's extrapolation capability. 
	In the following, we will give a brief overview on NN-based constitutive modeling which is mainly limited on \emph{elasticity}, i.e., perfectly path-independent behavior. 
	
	There are numerous works that model elasticity with NNs in this sense, e.g., in the initial works \cite{Shen2004,Liang2008} from the 2000s, the elastic potential of isotropic materials is approximated by using an FNN with three deformation-type invariants as input, which leads to the fulfillment of several requirements by construction, e.g., \emph{thermodynamic consistency}, \emph{objectivity}, or \emph{material symmetry}.   
	However, instead of calibrating the model with stress data, the strain energy density data were used directly to train these models.    
	Meanwhile, approaches based on architectures that use the \emph{hyperelastic potential} as output and \emph{invariants} as input are very common, e.g. \cite{Linden2023,Klein2021,Kalina2022a,Thakolkaran2022,Linka2021,Fuhg2022b,Tac2024a,Bahmani2024,Benady2024,Peirlinck2024}.   
	Thereby, a special training method allows direct calibration of the NN using stress and strain tuples. In particular, the loss function involves the derivation of the energy with respect to the deformation, a technique known as \emph{Sobolev training} \cite{Czarnecki2017,Vlassis2020}. 
	If also a loss term for the elasticity tensor is added, which requires to calculate the second derivative of the potential with respect to the deformation, it is called higher-order Sobolev training \cite{Vlassis2021,Kalina2024}. 
	Other models formulate NN-based potentials directly in terms of the components of the strain or deformation tensor \cite{Asad2022,Vlassis2020,Vlassis2022a}, which offers more flexibility in cases of anisotropy. However, this has the disadvantage that the network design no longer enforces the material symmetry by construction. For finite symmetry groups, this problem can be overcome by group symmetrization, see \cite{Fernandez2020a,Klein2021}. As shown in \cite{Garanger2024}, it is also possible to use tensor feature equivariant neural networks to enforce material symmetry.
	
	In addition, \emph{polyconvex} NNs are used in several works \cite{Klein2021,Tac2022a,Chen2022,Linka2023,Tac2024a,Bahmani2024}, which can be favorable in finite element (FE) simulations, improves the extrapolation capability \cite{Linden2023,Kalina2024} and guarantees \emph{rank-one convexity} and thus \emph{ellipticity} \cite{Ebbing2010,Schroder2010}.    
	Various techniques are used for incorporating this condition, with the most widely spread being the application of \emph{fully input convex neural networks (FICNNs)} introduced by Amos~et~al.~\cite{Amos2017}.   
	Recently, Linden~et~al.~\cite{Linden2023} presented an approach based on FICNNs that fulfills all usual conditions of (an)isotropic compressible hyperelasticity by construction, i.e., thermodynamic consistency, symmetry of stress, objectivity, material symmetry, polyconvexity and thus ellipticity, volumetric growth condition, as well a stress- and energy-free undeformed state.
	
	In the context of multiscale problems, NN-based approaches can be used as surrogate models which replace the computationally expensive simulation of \emph{representative volume elements (RVEs)} and thus enable a significant speed up \cite{Fritzen2019,Kalina2023}. Due to the high flexibility and simultaneously excellent prediction quality, NNs enriched with physical knowledge are excellently suited as surrogate models \cite{Kalina2023,Klein2021,Klein2022,Kalina2024,Le2015,Malik2021,Vlassis2021,Maia2023,Jiang2023,Masi2022,Liu2021,Eivazi2023}. In principle, the rule can be established that NNs with as much physics incorporated by design as possible are to be preferred, as these have a significantly better extrapolation behavior and thus less time-consuming RVE simulations have to be carried out \cite{Kalina2023}. 
	
	\subsection{Objectives and contributions of this work}
	
	As discussed in the literature overview given above, numerous very sophisticated approaches for constitutive modeling exist that combine modern machine learning methods with a reasonable physical basis. Restricting to elasticity, a description of the elastic potential by an NN with invariants as input is favorable if the underlying anisotropy can be described. However, most models formulated in this way are either limited to isotropy or assume that the symmetry group and the associated preferred directions are known in advance.  
	
	To the author's knowledge, there are only a few works that deal with the problem of \emph{anisotropy detection} combined with NN-based constitutive modeling in the finite strain regime.\footnote{For linear elasticity, the approaches developed by Cowin~and~Mehrabadi~\cite{Cowin1987} or Moahker~and~Norris~\cite{Moakher2006} can be used to identify the anisotropy class and orientation in the linear-elastic range. However, to apply these techniques, prior knowledge of the elasticity tensor is necessary. Furthermore, it is not possible to find out from the linear elastic regime alone to which anisotropy class the material belongs, since some classes coincide in the linear elastic limiting case, e.g. transverse isotropy and hexagonal anisotropy.} In Linka~et~al.~\cite{Linka2021}, a structure learning block based on a set of 2nd order \emph{structure tensors} built as vector dyads $\ve l_i \otimes \ve l_i$ with $\ve l_i$ being a preferred direction is presented. Besides the NN weights and biases, the preferred directions are determined during training to determine the material's anisotropy. In the same line, Thakolkaran~et~al.~\cite{Thakolkaran2022} detect the fiber orientation for transverse isotropy within the framework NN-EUCLID.
	Finally, Fuhg~et~al.~\cite{Fuhg2022b} have proposed so called tensor-basis NNs which allow to discover both the type and orientation of the anisotropy. To do so, structure tensors and  invariants for isotropy, transverse isotropy and orthotropy are used. The three papers mentioned above already allow the detection of anisotropy for several classes of materials without the need for prior knowledge. However, in all works a restriction to one 2nd order structure tensor or a combination of several 2nd order structure tensors has been made. For the generation of complete invariant sets, however, structure tensors of a higher order are necessary for numerous symmetry groups \cite{Xiao1996,Apel2004,Ebbing2010}. For example, a 4th order structure tensor is required for the cubic group and even a 6th order structure tensor for the hexagonal anisotropy class.
	
	Thus, within this contribution, we present an invariant-based NN approach, where \emph{generalized structure tensors} \cite{Gasser2006} up to 6th order are used for the generation of the invariant set. Simultaneous training of structure tensors and NN weights allows the identification of anisotropy. To do so, parameterized versions of generalized structure tensors fulfilling important properties are used. In addition, we introduce \emph{trainable gates} in combination with a $p$\emph{-norm} type penalty loss which allows us to remove unnecessary invariants from the model, which is also denoted as $\ell_p$ regularization \cite{Flaschel2021,McCulloch2024}. To allow for maximum accuracy in energy, stress and elasticity tensor, a higher-order Sobolev training is applied.    
	Following the idea of \emph{physics-augmented neural networks (PANNs)} \cite{Linden2023,Rosenkranz2024,Klein2022,Kalina2024}, the proposed model is formulated such that as many conditions as possible are fulfilled by construction. These are thermodynamic consistency, compatibility with the balance of angular momentum, objectivity, material symmetry, volumetric growth condition, as well as energy- and stress-free undeformed state.
	The performance of our approach is demonstrated for five different RVEs, where \emph{interpolation} as well as \emph{extrapolation} are considered. In addition, a comparison to a model based on the coordinates of the right Cauchy-Green deformation tensor is shown. The comprehensive RVE databases are generated by a computational homogenization approach. 
	
	The organization of the paper is as follows: In Sect.~\ref{sec:funda}, the
	fundamentals of finite strain continuum mechanics, basic principles of hyperelasticity, a scale transition scheme and the theory of generalized structure tensors are summarized. After this, PANNs based on generalized structure tensors and a training which enables anisotropy detection are introduced in Sect.~\ref{sec:PANNs}. The developed approach is exemplarily shown Sect.~\ref{sec:examples}. After a discussion of the results, the paper is closed by concluding remarks and an outlook to necessary future work in Sect.~\ref{sec:conc}.   
	
	\paragraph{Notation}
	Within this work, tensors of rank one and two are given by boldface italic letters, i.e., $\ve A, \ve B \in \Ln_1$ or $\te C, \te D \in \Ln_2$, where $\Ln_n$ denotes the space of tensors with rank $n\in \N$ with $\N$ being the set of natural numbers without zero.
	Tensors with rank four and six are marked by blackboard symbols and bold upright sans serif letters, i.e., $\tttte A \in \Ln_4$ and $\tttttte A \in \Ln_6$, respectively.
	Single and double contractions of two tensors are given by $\ve C \cdot \te D = C_{kl} D_{li} \ve e_k \otimes \ve e_i$ and $\te C:\te D=C_{kl}D_{kl}$, respectively. Therein, $\ve e_k\in \Ln_1$ and $\otimes$ denote a Cartesian basis vector and the dyadic product, where the Einstein summation convention is used. The cross product of two rank one tensors is given by $\ve A \times \ve B = e_{ijk} A_j B_k \ve e_i$, with $e_{ijk}$ being the antisymmetric Levi-Civita symbol.
	Transpose and inverse of a second order tensor $\te C$ are given by $\te C^T$ and $\te C^{-1}$, respectively.
	Additionally, $\tr \te C$, $\det \te C$, $\cof \te C := \det(\te C) \te C^{-T}$ are used to indicate trace, determinant as well as cofactor, respectively.
	The Hadamard product of tensors, vectors or matrices, i.e., the element-wise product, is given by the symbol $\odot$.
	The space of unit vectors is given by $\Vn:=\{\ve n \in \Ln_1 \, | \, \ve n\cdot \ve n = 1\}$.
	The sets \mbox{$\Sym:=\left\{\te A \in \Ln_2 \, |\, \te A = \te A^T\right\}$}, \mbox{$\Sym_4:=\left\{\tttte A \in \Ln_4 \, |\, A_{ijkl} = A_{jikl} = A_{ijlk} = A_{klij} = A_{kjil} = \ldots \right\}$} and $\Sym_6$ equivalently defined, denote the spaces of fully symmetric 2nd, 4th and 6th order tensors, i.e., with full symmetry. The space of 4th order tensors which only have major and minor symmetry, e.g., the elasticity tensor, is given by the set $\overline{\Sym}_4:=\left\{\tttte A \in \Ln_4 \, |\, A_{ijkl} = A_{jikl} = A_{ijlk} = A_{klij} \right\}$.
	Furthermore, the orthogonal group and special orthogonal group in the Euclidean vector space $\R^3$ are given by $\Othree:=\left\{\te A \in \Ln_2\,|\,\te A^T \cdot \te A = \one\right\}$ and $\SO:=\left\{\te A \in \Ln_2\,|\,\te A^T \cdot \te A = \one,\,\det \te A = 1\right\}$, respectively, while $\GL:=\left\{\te A \in \Ln_2\,|\,\det \te A > 0\right\}$ is the set of invertible second order tensors with positive determinant. Therein, $\one:=\delta_{ij}\ve e_i \otimes \ve e_j\in \Ln_2$ is the second order identity tensor, where $\delta_{ij}$ denotes the Kronecker delta. Norms of rank one and two tensors or matrices are given by $|\ve A| := \sqrt{A_iA_i}$ and $\|\te C\| := \sqrt{C_{ij}C_{ij}}$, respectively. 

	For reasons of readability, the arguments of functions are usually omitted within this work. However, energy functions are given with their arguments to show the dependencies, except when derivatives are written. Furthermore, in the following the symbol of a function is identical with the symbol of the function value itself.

	\section{Fundamentals}
	\label{sec:funda}
	
	In this section, we introduce kinematics and stress measures common in finite strain continuum mechanics. Based on this, general principles of \emph{anisotropic finite strain hyperelasticity} are summarized. Additionally, the concept of \emph{generalized structure tensors} and \emph{isotropic tensor functions} are introduced. Finally, to link the micro- and macroscale, we introduce an appropriate homogenization scheme.

	\subsection{Kinematics and stress measures}
	\label{sec:kin_stress}
	
	Let us consider the motion of a material body with reference configuration \mbox{$\mathcal{B}_0 \subset \R^3$} at time $t_0 \in \R_{\ge 0}$  and current configuration \mbox{$\mathcal{B} \subset \R^3$} at time \mbox{$t\in \mathcal T:=\{\tau\in \R \,|\,\tau \ge t_0\}$}. To describe the body's motion, we introduce a smooth bijective mapping $\ve \varphi: \mathcal{B}_0 \times \mathcal T \to \mathcal{B}$, mapping material points $\ve X\in\mathcal{B}_0$ to $\ve x=\ve \varphi\left(\ve X, t\right) \in \mathcal{B}$.
	The displacement $\ve u \in \Ln_1$ of each material point is given by $\ve u(\ve X, t) := \ve \varphi(\ve X, t) - \ve X$. As further kinematic quantities, the deformation gradient $\te F := (\nablaX \ve \varphi)^T \in \GL$ and its determinant \mbox{$J:=\det \te F \in \R_{>0}$} are defined.  
	Finally, we introduce the right Cauchy-Green deformation tensor \mbox{$\te C:=\te F^T \cdot \te F \in \Sym$} and the Green-Lagrange strain tensor
	$\te E:= \frac{1}{2} (\te C - \one) \in \Sym$ as kinematic quantities which are invariant to rigid body motions. 

	Within finite strain continuum solid mechanics, various stress measures can be defined. In this work, we make use of the Cauchy stress tensor $\te \sigma \in \Sym$, which is symmetric and is also known as true stress, as well as the 1st and 2nd Piola-Kirchhoff stress tensors $\te P \in \Ln_2$ and $\te T \in \Sym$. The latter two stress tensors are linked to the Cauchy stress by the pullback operations $\te P:= J \te \sigma \cdot \te F^{-T}$ and $\te T:= J \te F^{-1}\cdot \te \sigma \cdot \te F^{-T}$, respectively.
	
	For more details on basic principles in continuum solid mechanics the reader is referred to the textbooks of Haupt~\cite{Haupt2000} or Holzapfel~\cite{Holzapfel2000}.
	
	\subsection{Physical conditions for anisotropic finite strain hyperelasticity}
	\label{subsect:phys_cond}
	An elastic constitutive model relates deformation gradient to stress induced at a material point. In \emph{hyperelasticity}, this mapping is not defined directly, but via an elastic potential, i.e., 
	\begin{equation}
		\psi: \GL  \to \R_{\ge 0}, \; \te F \mapsto \psi(\te F) \; \text{and }
		\te P = \diffp{\psi}{\te F} \; . \label{eq:psi}
	\end{equation}
	Relation \eqref{eq:psi}${}_2$ implies energy conservation and path-independency. The model is thus a priori \emph{thermodynamically consistent}, i.e., in accordance with the second law of thermodynamics \cite{Kalina2022a,Linden2023}.
	
	A mapping between the rates of 1st Piola-Kirchhoff stress $\dot{\te P}$ and deformation gradient $\dot{\te F}$ follows by the material time derivative $\dot{(\cdot)}$ of Eq.~\eqref{eq:psi}${}_2$ and by introducing the material tangent $\tttte A \in \Ln_4$, i.e.,
	\begin{equation}
		\dot{\te P} = \tttte A : \dot{\te F} \text{ with }\tttte A := \diffp{{}^2\psi}{\te F \partial \te F} \in \Ln_4 \; .
		\label{eq:tangent}
	\end{equation}
	For the other stress measures introduced in Sect.~\ref{sec:kin_stress}, the equivalent relationships $\overset{\circ}{\te \sigma} = \ttttes c : \te d$ and $\dot{\te T} = \tttte C : \dot{\te E}$ can be found. Thereby, $\te d:=\sym(\te l) \in \Sym$ is the rate of deformation tensor, which is the symmetric part of the velocity gradient $\te l = \dot{\te F} \cdot \te F^{-1}$, and $\overset{\circ}{\te \sigma} = \dot{\te \sigma} - \te l \cdot \te \sigma - \te \sigma \cdot \te l^T + \te \sigma \tr \te l$ is the Truesdell rate of $\te \sigma$. Relations of the introduced material tangents $\ttttes c\in \overline{\Sym}_4$ and $\tttte C\in \overline{\Sym}_4$ to $\tttte A$ can be determined through some tensor calculations and follow to \mbox{$c_{ijkl} = J^{-1} A_{iJkL} F_{jJ} F_{lL} - \delta_{ik} \sigma_{jl}$} as well as \mbox{$C_{IJKL} = \left(A_{iJkL} - \delta_{ik} T_{JL}\right)  F_{Ii}^{-1} F_{Kk}^{-1}$}, respectively. Note that both $\tttte C$ and $\ttttes c$ have a minor and a major symmetry, but are not completely symmetrical.
	
	There are additional mathematical and physical requirements for the hyperelastic potential \cite{Haupt2000,Holzapfel2000}. The most common ones are briefly summarized below. For more details, please refer to the textbooks given above or the recent work Linden~et~al.~\cite{Linden2023}
	
	In order to guarantee \emph{symmetric Cauchy} and \emph{2nd Piola-Kirchhoff stress tensors}, which follow from the balance of angular momentum, $\psi(\te F)$ has to be formulated such that the following condition holds:
	\begin{equation}
		\diffp{\psi}{\te F} \cdot \te F^T =  \te F \cdot \diffp{\psi}{\te F^T} \; .
	\end{equation}

	The principle of \emph{material objectivity} states that the behavior of the material must not change with any rigid body motion. Furthermore, the potential should also account for the material's anisotropy, which is termed \emph{material symmetry}. These two principles are given as
	\begin{align}
		\psi(\te F) &= \psi(\te Q \cdot \te F) 
		\; \forall \te F \in \GL, \; \te Q \in \SO \; \text{and}
		\label{eq:objectivity} \\
		\psi(\te F) &= \psi(\te F \cdot \te Q^T) 
		\; \forall \te F \in \GL, \; \te Q \in \G \subseteq \Othree \; ,
		\label{eq:symmetry}
	\end{align}
	respectively. In Eq.~\eqref{eq:symmetry}, $\mathcal G$ denotes the symmetry group of the material under consideration. There are 11 crystal symmetries following from 32 point groups with finite order. As the numbering of the groups is not uniform in the literature, we use the Schoenflies notation in addition to the group number. The symmetry groups are named \emph{triclinic} $\mathcal G_1$ ($\mathcal C_i$), \emph{monoclinic} $\mathcal G_2$ ($\mathcal C_{2h}$), \emph{rhombic} or \emph{orthotropic} $\mathcal G_3$ ($\mathcal D_{2h}$), \emph{tetragonal} $\mathcal G_4$ ($\mathcal C_{4h}$) and $\mathcal G_5$ ($\mathcal D_{4h}$), \emph{cubic} $\mathcal G_6$ ($\mathcal T_{h}$) and $\mathcal G_7$ ($\mathcal O_{h}$), \emph{trigonal} $\mathcal G_8$ ($\mathcal S_{6}$) and $\mathcal G_9$ ($\mathcal D_{3d}$), and \emph{hexagonal} $\mathcal G_{10}$ ($\mathcal C_{6h}$) and $\mathcal G_{11}$ ($\mathcal D_{6h}$), see \cite{Ebbing2010,Apel2004} for more details. Three additional symmetry groups follow by also considering the continuous cylindrical and spherical groups, which are not finite. These groups are denoted as \emph{transversely isotropic} $\mathcal G_{12}$ ($\mathcal C_{\infty h}$) and $\mathcal G_{13}$ ($\mathcal D_{\infty h}$) and \emph{isotropic} $\mathcal G_{14}$ ($\mathcal K_{h} = \mathcal O(3)$).
	
	Various coercivity conditions can also be taken into account for hyperelasticity, whereby the \emph{volumetric growth condition} is the most widely used \cite{Holzapfel2000,Linden2023}. It is given by $\psi(\te F) \rightarrow \infty$ as $\big(J \rightarrow 0^+ \;\lor\; J\rightarrow\infty\big)$ and states that a material cannot be compressed to a volume of zero or extended to an infinite volume, requiring an increase in energy toward infinity.
	
	Furthermore, the \emph{undeformed} configuration of the material should be \emph{energy- and stress-free}, i.e.,  $\psi(\te F = \one)=0$ and $\te P(\te F = \one) = \zero$ should hold. It is also expected that the stored energy always increases, i.e., it is \emph{non-negative}, if a deformation $\te F$ is applied, thus $\psi(\te F) \ge 0 \; \forall \, \te F \in \GL$. 
	
	Finally, we will also briefly discuss the concept of \emph{ellipticity} which ensures traveling waves with real-valued and non-negative wave speeds \cite{Ebbing2010,Schroder2010,Marsden1984}. In its local form, i.e.,  for a specific state $\te F\in \GL$, (strict) ellipticity requires
	\begin{align}
		\forall \ve a, \ve N \in \Vn: \;  (\ve a \otimes \ve N) : \tttte A(\te F) : (\ve a \otimes \ve N)
		(>) \ge 0  \; .\label{eq:ellipticityLocal}
	\end{align}
	If the above condition~\eqref{eq:ellipticityLocal} applies to all permissible deformation states, we speak of (strict) global ellipticity, i.e., \mbox{$\forall \te F \in\GL: \; \forall \ve a, \ve N \in \Vn: \;   (\ve a \otimes \ve N) : \tttte A(\te F) : (\ve a \otimes \ve N) (>) \ge 0$}. Note that global ellipticity of twice differentiable and smooth energy functions as typically considered in hyperelasticity is equivalent to \emph{rank-one convexity}. Instead of enforcing global ellipticity directly, 
	the mathematical concept of \emph{polyconvexity}, which was introduced by Ball~\cite{Ball1976,Ball1977}, is often used in constitutive modeling\cite{Linden2023,Klein2021,Ebbing2010,Schroder2010,Schroder2003}.\footnote{%
		A potential $\psi(\te F)$ is polyconvex if and only if there is a function such that $\psi(\te F) = \mathcal P(\te F, \cof \te F, \det \te F)$, with $\mathcal P(\te F, \cof \te F, \det \te F)$ convex with respect to its arguments.
		Polyconvexity is sufficient for quasi-convexity, which ensures rank-one convexity and thus global ellipticity \cite{Schroder2010,Ebbing2010}. Furthermore, sequential weak lower semi-continuity (s.w.l.s.) is guaranteed by polyconvexity. It is sufficient for the existence of minimizers if coercivity is also guaranteed \cite{Ebbing2010,Schroder2010}.  The above implications do not apply in reverse.
	}	
	However, it should be noted that polyconvexity can be too restrictive, especially for multiscale modeling \cite{Abeyaratne1984,Klein2024,Kalina2024}
	
	\subsection{Concept of structure tensors and isotropic tensor functions}
	\label{subsec:structure_tensors}
	In order to describe anisotropic constitutive behavior, the concept of \emph{structure tensors} can be used \cite{Haupt2000,Holzapfel2000,Ebbing2010,Apel2004}. Depending on the considered symmetry group, i.e., $\mathcal G_1$ -- $\mathcal G_{14}$, these tensors are of orders up to six, 
	where also odd tensor ranks or antisymmetric tensors are possible, see Xiao~\cite{Xiao1996}. Within this work, we restrict ourselves to fully symmetric tensors of order two $\te G_1, \te G_2,\ldots,\te G_{n_2}\in\Sym$, four $\tttte G_1, \tttte G_2,\ldots,\tttte G_{n_4}\in\Sym_4$ and six $\tttttte G_1, \tttttte G_2,\ldots,\tttttte G_{n_6}\in\Sym_6$.
	The structure tensors reflect the material’s anisotropy and are thus invariant with respect to the symmetry transformations, i.e., 
	\begin{align}
		\te G_\alpha = \te Q \cdot \te G_\alpha \cdot \te Q^T \; , \;
		\tttte G_\beta = \te Q * \tttte G_\beta \; , \;
		\tttttte G_\gamma = \te Q \star \tttttte G_\gamma \; \forall \te Q \in \mathcal G \; ,
	\end{align}
	where $(\te Q * \tttte G_\beta)_{IJKL}=Q_{IM}Q_{JN}Q_{KP}Q_{LQ} G_{MNPQ}^\beta$ and $(\te Q \star \tttttte G_\gamma)_{IJKLMN}= Q_{IO}Q_{JP}Q_{KQ}Q_{LR}Q_{MS}Q_{NT} G_{OPQRST}^\gamma$. If the structure tensors are appended to the list of arguments of $\psi(\te F)$, the energy is an \emph{isotropic tensor function} \cite{Itskov2015} even if the material is anisotropic which means that
	\begin{align}
		\psi(\te F, \mathcal S_2, \mathcal S_4, \mathcal S_6) = \psi(\te F \cdot \te Q^T, \te Q\cdot \mathcal S_2\cdot \te Q^T, \te Q * \mathcal S_4, \te Q \star\mathcal S_6) \; \forall \te Q \in \Othree \; ,
	\end{align}
	whereby the sets $\mathcal S_2 := \{\te G_1, \te G_2,\ldots,\te G_{n_2}\}$, $\mathcal S_4 := \{\tttte G_1, \tttte G_2,\ldots,\tttte G_{n_4}\}$ and $\mathcal S_6 := \{\tttttte G_1, \tttttte G_2,\ldots,\tttttte G_{n_6}\}$ have been used to abbreviate the notation. In the following, for brevity, we summarize all structure tensors in the set $\mathcal S$.
	
	The introduced concept of structure tensors can be used to fulfill numerous of the principles given in Sect.~\ref{subsect:phys_cond} by construction. To this end, the potential is formulated in terms of \emph{invariants} from the right Cauchy-Green deformation tensor $\te C$ and a set of structure tensors $\mathcal S$, i.e., $\psi(\I)$ with $\I:=(I_1,I_2,\ldots,I_n)\in\R^n$. Thus, the constitutive relations according to Eqs.~\eqref{eq:psi}${}_2$ and \eqref{eq:tangent}${}_2$ follow then to
	\begin{align}
		\te P = \sum_{\alpha=1}^n \diffp{\psi}{I_\alpha}\diffp{I_\alpha}{\te F} \text{ and }
		\tttte A = \sum_{\alpha=1}^n \sum_{\beta=1}^n \diffp{{}^2\psi}{I_\alpha\partial I_\beta}\diffp{I_\alpha}{\te F} \otimes 
		\diffp{I_\beta}{\te F} + \sum_{\alpha=1}^n \diffp{\psi}{I_\alpha} \diffp{{}^2I_\alpha}{\te F \partial \te F} \; ,
		\label{eq:constInv}
	\end{align}
	respectively. The partial derivatives of the invariants with respect to $\te F$ are also referred to as tensor generators \cite{Kalina2022a}.
	With that, \emph{thermodynamic consistency}, \emph{symmetry of the Cauchy stress}, \emph{material objectivity} and \emph{material symmetry} are automatically fulfilled \cite{Linden2023}.
	
	\begin{rmk}
		Note that it is not possible to build invariant sets for all introduced symmetry groups $\mathcal G_1$ -- $\mathcal G_{14}$ with two fully symmetric 2nd order structure tensors or one fully symmetric 4th or 6th order structure tensor each, as done here, see Xiao~\cite{Xiao1996}. However, a wide range of groups can be described with theses structure tensors, cf. \ref{app:sym_groups}.
	\end{rmk}
	
	\subsection{Scale transition scheme}
	\label{subsec:homogenization}
	
	In the following, we distinguish between two different scales, the \emph{microscale} and the \emph{macroscale} with characteristic lengths $\ell \in \R_{\ge 0}$ and $\bar \ell \in \R_{\ge 0}$, respectively. The microscale is represented by a heterogeneous structure consisting of matrix and inhomogeneities, while the macroscale is considered to be homogeneous. For the characteristic lengths of both scales, the relationship $\ell \ll \bar \ell$ known as scale separation should hold \cite{Schroder2014}. To label macroscopic quantities, they are marked by an overline in the following, i.e., $\bar{(\cdot)}$.
	
	In order to connect microscopic and macroscopic quantities, a computational homogenization scheme is applied. Consequently, each macroscopic point $\bve X \in \bar\B_0$ gets assigned properties resulting from the behavior of the microscale which is represented by an RVE with domain $\B_0^\text{RVE}$ in the vicinity of $\bve X$ on the microscale. An effective macroscopic quantity can be determined by the volume average
	\begin{align}
		\langle (\cdot) \rangle := \frac{1}{V^\text{RVE}}\int_{\B_0^\text{RVE}} (\cdot) \, \dx V \; ,
	\end{align}
	where $V^\text{RVE}\in\R_{\ge 0}$ is the RVE's volume. With that, macroscopic deformation gradient and 1st Piola-Kirchhoff stress tensor are defined by $\bte F := \langle \te F \rangle$ and $\bte P:= \langle \te P \rangle$, respectively \cite{Schroder2014,Kalina2023}. Boundary conditions (BCs) for the microscopic BVP, which must be solved before volume averaging, can be derived from the Hill-Mandel condition. For the finite strain setting under consideration, it is given by the well known equation $\langle \te P : \dot{\te F}\rangle = \bte P : \dot{\bte F}$ \cite{Schroder2014}. Herein, we use periodic BCs to guarantee the fulfillment of this relation \cite{Kalina2023}. Within FE simulations, we realize the periodic BCs via the concept of master nodes, see Haasemann~et~al.~\cite{Haasemann2006}. This also allows the \emph{algorithmically consistent tangent modulus of the RVE} to be calculated in a straightforward manner, which corresponds to the macroscopic material tangent, i.e., $\btttte A = \btttte A^\text{algo}$.
	
	For the hyperelastic case considered here, the Hill-Mandel condition represents the equality of the rates of the averaged microscopic and macroscopic energies, allowing to calculate the effective potential by volume averaging, i.e. $\bar \psi = \langle \psi \rangle$. Thus, the application of the computational homogenization approach allows us to determine the mapping
	\begin{align}
		\mathcal H: \GL \to \R \times \Ln_2 \times \Ln_4 , \; \bte F \mapsto (\bar \psi, \bte P, \btttte A) \; .
	\end{align}
	The Cauchy stress and the corresponding elasticity tensor can be calculated by applying push forward operations. This allows us to get $\mathcal {P\!F}: \GL \times \Ln_2 \times \Ln_4 \to \Sym \times \Ln_4 , \; (\bte F,\bte P, \btttte A) \mapsto (\bte \sigma, \bttttes c)$.
	All FE simulations in this work were carried out using an in-house code based on Matlab. To close the brief introduction on computational homogenization we want note that all of the definitions given in Sects.~\ref{sec:kin_stress} and \ref{subsect:phys_cond} are valid on the microscopic as well as the macroscopic scales.
	
	\section{Macroscale modeling with generalized structure tensors}
	\label{sec:gen_struct}
	
	As already mentioned in Sect.~\ref{subsec:structure_tensors}, it is favorable to formulate the elastic potential in terms of invariants from $\bte C$ and a set of structure tensors $\mathcal S$. In the following, we introduce the concept of generalized structure tensors \cite{Gasser2006,Ehret2007} and give related sets of invariants.
	
	For the \emph{isotropic} case, i.e., the group $\mathcal G_{14}$, $\mathcal S=\{\one\}$, so that a complete and irreducible \cite{Pennisi1987a} set of invariants is given by the well known set
	\begin{align}
		\bar I_1 := \tr \bte C \; , \; \bar I_2 := \tr(\cof \bte C) \; \text{and} \; \bar I_3 := \det \bte C \; .
		\label{eq:invariants_iso}
	\end{align}
	
	In order to model complex \emph{anisotropic} behavior on the macroscale, generalized structure tensors can be used. Originally, this concept has been introduced to model materials which are characterized by fibers oriented in different directions on the microscale, e.g., biological tissues or fiber reinforced composites \cite{Gasser2006}.
	As already mentioned, we restrict ourselves to fully symmetric structure tensors of 2nd, 4th and 6th order. These tensors and the related invariant basis are introduced in the following. Since these tensors correspond to the macroscale as introduced in Sect.~\ref{subsec:homogenization}, we mark them as macroscopic quantities in the following.
	
	\subsection{2nd order generalized structure tensor}
	A second order generalized structure tensor is defined by
	\begin{align}
		\bte G := \int_\Omega \rho(\theta,\varphi) \ve N(\theta,\varphi) \otimes \ve N(\theta,\varphi) \,\dx A \in \Sym \text{ with }
		\int_\Omega \rho(\theta,\varphi) \, \dx \Omega = 1 \; ,
		\label{eq:GST}
	\end{align}
	where $\ve N(\theta,\varphi)\in \Vn$ is a fiber orientation vector and $\rho(\theta,\varphi) \in \R_{\ge 0}$ is is a probability
	density function describing the orientation of the fibers on the microscale \cite{Gasser2006}.\footnote{Note that the introduced parameterized forms of the generalized structure tensors as introduced in Sect.~\ref{subsec:parametrized} also allow an application to materials with microstructures that are not characterized by a distribution of fibers. This will be demonstrated in the examples part \ref{sec:examples} of this paper.}
	The tensor $\bte G$ is symmetric, positive semi-definite and has a trace of one, i.e., $\bte G^T = \bte G$, $\ve A \cdot \bte G \cdot \ve A \ge 0 \; \forall \ve A \in \Ln_1$ and $\tr \bte G = 1$.
	Following Boehler~\cite{Boehler1977}, with $\mathcal S:=\{\bte G\}$, four additional mixed invariants follow: 
	\begin{align}
		\bar K_4 := \bte C : \bte G \; , \; \bar K_5:= \bte C^2 : \bte G \; , 
		\bar K_6 := \bte C : \bte G^2 \; , \; \bar K_7:= \bte C^2 : \bte G^2 \; .
		\label{eq:inv_G2nd}
	\end{align}
	The set of invariants is thus given by $\bI^{\bte G} := (\bar I_1, \bar I_2, \bar I_3, \bar K_4, \bar K_5, \bar K_6, \bar K_7) \in \R^7$.
	The structure tensor $\bte G$ can also be given by its eigenvalues $\lambda_\alpha\in\R_{\ge 0}$ and the corresponding projection tensors $\te P_\alpha\in\Sym$:
	\begin{align}
		\bte G = \sum_{\alpha=1}^{n_G} \lambda_\alpha \te P_\alpha \text{ with } \te P_\alpha \cdot \te P_\beta = \delta_{\alpha\beta} \te P_{(\beta)} \; , \; 
		\sum_{\alpha=1}^{n_G} \te P_\alpha = \one \; ,
		\label{eq:spectral_G}
	\end{align}
	where $n_G \in\{1,2,3\}$ is the number of non-equal eigenvalues.
	For $n_G=1$ the tensor $\bte G$ describes the \emph{isotropic} group $\mathcal G_{14}$, whereas for $n_G=2$ and $n_G=3$ the material is \emph{transversely isotropic} ($\mathcal G_{13}$) and \emph{orthotropic} ($\mathcal G_3$), respectively \cite{Wollner2023}. 
	Note that $\bar K_6$ and $\bar K_7$ would be omitted if $\bte G$ has only two different eigenvalues, i.e., for transverse isotropy, see \ref{app:ti}.  
	
	\begin{rmk}
		Note that orthotropy is typically modeled with two 2nd order structure tensors given by $\bte G_1:=\ve a_1 \otimes \ve a_1 \in \Sym$ and $\bte G_2:=\ve a_2 \otimes \ve a_2\in\Sym$ with $\ve a_\alpha \cdot \ve a_\beta = \delta_{\alpha\beta}$ \cite{Holzapfel2000}. However, as shown in \ref{app:orth}, the invariant set following from $\bte C$, $\bte G_1$ and $\bte G_2$ can be expressed by $\bI^{\bte G}$ if $\bte G$ has 3 non-equal eigenvalues and two eigenvectors which are (anti)parallel with $\ve a_1$ and $\ve a_2$, respectively.
	\end{rmk}
	
	\subsection{4th order generalized structure tensor}
	In order to model more complex anisotropic behavior, we introduce an extension of Eq.~\eqref{eq:GST} to a 4th order generalized structure tensor given by
	\begin{align}
		\btttte G := \int_\Omega \rho(\theta,\varphi) \ve N(\theta,\varphi) \otimes \ve N(\theta,\varphi) \otimes \ve N(\theta,\varphi) \otimes \ve N(\theta,\varphi) \,\dx A \in \Sym_4 \; , \label{eq:4thorderstruct}
	\end{align}
	see also \cite{Advani1987,Bauer2023}, where a similar quantity is introduced as fiber orientation tensor.\footnote{Note that Fiber orientation tensors are defined in exactly the same way as generalized structure tensors. Fiber orientation tensors of order $n\in\mathbb N$ can be used for the characterization of fiber reinforced composites, generation of RVEs \cite{Mehta2022} or for mean-field homogenization \cite{Kehrer2020}.}
	The 4th order structure tensor given in Eq.~\eqref{eq:4thorderstruct} is fully symmetric, positive semi-definite with respect to vector dyads as well as symmetric 2nd order tensors 
	and has a generalized trace of one: 
	\begin{align}
		\bar G_{KLMN} = \bar G_{LKMN} = \bar G_{KLNM} = \cdots \; , \;
		\te S : \btttte G :  \te S \ge 0 \; \forall \te S \in \Sym  \text{ and } 
		\bar G_{KKLL}&=  \bar G_{KLKL} =  \bar G_{KLLK} = 1 \; .
	\end{align}
	
	In order to identify a corresponding set of invariants for the case $\mathcal S := \{\btttte G\}$, we follow the procedure of Xiao~\cite{Xiao1996}. Thus, we build a set of 2nd order tensors from $\btttte G$ and $\bte C$ by polynomials. Due to the fact that $\btttte G$ is constant and due the Cayley-Hamilton theorem applied for $\bte C$, we get
	\begin{align}
		\bte H_1 := \btttte G : \bte C \; \text{and } \bte H_2 := \btttte G : \bte C^2 \; .
	\end{align}
	According to Boehler~\cite{Boehler1977} we can build a total of 21 invariants from $\bte C$, $\bte H_1$ and $\bte H_2$. This set is complete but not irreducible in general, i.e., for specific cases some invariants can be expressed by others or are even redundant.
	However, in order to reduce the number of invariants for the generalized structure tensor of order four, only invariants up to an order less than or equal to 3 in $\bte C$ are taken into account, i.e., $\bte C^n$ with $n\le3$. This leads to a set comprising the invariants given in Eq.~\eqref{eq:invariants_iso} and the additional invariants
	\begin{equation}
		\begin{split}
			\bar L_4 &:= \tr \bte H_1 \; , \; \bar L_5 := \tr \bte H_1^2 \; , \; \bar L_6 := \tr \bte H_1^3 \; , \; \bar L_7 := \tr \bte H_2 
			\; , \; 
			\bar L_8 := \tr (\bte H_1 \cdot \bte C) \; , \;   \\
			\bar L_9 &:= \tr (\bte H_2 \cdot \bte C) \; , \; 
			\bar L_{10} := \tr (\bte H_1 \cdot \bte H_2)
			\; , \; \bar L_{11} := \tr (\bte H_1^2 \cdot \bte C) \; .
			\label{eq:inv_G4}
		\end{split}
	\end{equation}
	The invariant set is thus given by $\bI^{\btttte G} := (\bar I_1, \bar I_2, \bar I_3, \bar L_4, \bar L_5, \ldots, \bar L_{11}) \in \R^{11}$.
	
	The formulation based on the generalized structure tensor $\btttte G$ includes the \emph{tetragonal} group $\mathcal G_5$ and the \emph{cubic} group $\mathcal G_7$ up to invariants of order 3 in $\bte C$, cf. \ref{app:tetragonal} and \ref{app:cubic}.
	Note that it is also possible to describe further anisotropy classes with $\btttte G$. However, this is not discussed here. The reader is referred to \cite{Olive2022} for a discussion on fully symmetric and traceless 4th order structure tensors.
	
	\subsection{6th order generalized structure tensor}
	Furthermore, to enable the description of anisotropy classes that require structure tensors of orders higher than 4, we introduce a generalized 6th order structure tensor defined as
	\begin{align}
		\btttttte G := \int_\Omega \rho(\theta,\varphi) \ve N(\theta,\varphi) \otimes \ve N(\theta,\varphi) \otimes \ve N(\theta,\varphi) \otimes \ve N(\theta,\varphi) \otimes \ve N(\theta,\varphi) \otimes \ve N(\theta,\varphi) \,\dx A \in \Sym_6 \; .
	\end{align}
	Again, similar to $\bte G\in\Sym$ and $\btttte G\in\Sym_4$, this tensor is fully symmetric, positive semi-definite with respect to $\te S \otimes \ve A \in \Ln_3$ and has a generalized trace of one:
	\begin{align}
		\bar G_{IJKLMN} &= \bar G_{JIKLMN} = \bar G_{IJLKMN} = \bar G_{IKJLMN} = \cdots \; , \\
		S_{IJ} A_K  \bar G_{IJKLMN} S_{LM} A_N &\ge 0 \; \forall \te S \in \Sym, \; \ve A \in \Ln_1  \text{ and} \\
		\bar G_{KKLLMM} &=  1 \; .
	\end{align}
	To build an invariant basis, we adapt again the technique from Xiao~\cite{Xiao1996}, i.e., we build $\bte H_1:= \one : \btttttte G : \bte C$, $\bte H_2:= \bte C : \btttttte G : \bte C$, $\bte H_3:= \one : \btttttte G : \bte C^2$, $\bte H_4:= \bte C : \btttttte G : \bte C^2$, and $\bte H_5:= \bte C^2 : \btttttte G : \bte C^2$
	and use the rules given in Boehler~\cite{Boehler1977}. With that, one gets a total of 98 invariants, which can be reduced to 13 if only invariants with orders up to $\bte C^3$ are used and redundant invariants are neglected. This leads to the set $\bI^{\btttttte G}:=(\bar I_1, \bar I_2, \bar I_3, \bar M_4, \bar M_5, \ldots, \bar M_{13})\in \R^{13}$ comprising the invariants given in Eq.~\eqref{eq:invariants_iso} and the additional invariants
	\begin{equation}
		\begin{split}
			\bar M_4 &:= \tr \bte H_1 \; , \; \bar M_5 := \tr \bte H_1^2 \; , \; \bar M_6 := \tr \bte H_1^3 \; , \; \bar M_7 := \tr \bte H_2 
			\; , \; \bar M_8 := \tr \bte H_3 \; , \; 
			\bar M_9 := \tr \bte H_4 \; , \; \\
			\bar M_{10} &:= \tr (\bte C \cdot \bte H_1^2)
			\; , \; \bar M_{11} := \tr (\bte C \cdot \bte H_2)
			\; , \; \bar M_{12} := \tr (\bte H_1 \cdot \bte H_2)
			\; , \; \bar M_{13} := \tr (\bte H_1 \cdot \bte H_3) \; .
			\label{eq:inv_G6}
		\end{split}
	\end{equation}
	The formulation based on the generalized structure tensor $\btttttte G$ includes the \emph{hexagonal} group $\mathcal G_{11}$ up to 
	invariants of order 3 in $\bte C$, see \ref{app:hexa}.
	
	\subsection{Two 2nd order generalized structure tensors}
	\label{subsec:2xGST2}
	Finally, several generalized structure tensors can also be used to extend the symmetry groups that can be described using the structure tensors introduced so far. 
	By using a set given by two 2nd order structure tensors, i.e.,  $\mathcal S=\{\bte G_1, \bte G_2\}$, we get 12 invariants according to Boehler~\cite{Boehler1977}: 3 of $\bte C$ as defined in Eq.~\eqref{eq:invariants_iso}, 4 each according to Eq.~\eqref{eq:inv_G2nd} for $\bte C, \bte G_1$ and $\bte C, \bte G_2$ and a further invariant defined by $\bar N_{12} := \tr(\bte C \cdot \bte G_1 \cdot \bte G_2)$. Thus, we end up with the set $\bI^{\bte G_{1},\bte G_2}:=(\bar I_1, \bar I_2, \bar I_3, \bar N_4, \ldots, \bar N_{12})\in \R^{12}$. 
	
	This formulation includes five different anisotropy classes. For the following discussion, we are taking up the argumentation from Olive~et~al.~\cite{Olive2022} in a slightly modified form. To this end, we use the spectral decomposition of the structure tensors according to Eq.~\eqref{eq:spectral_G}:
	\begin{align}
		\bte G_1 = \sum_{\alpha=1}^{n_{G_1}} \lambda_\alpha \te P_\alpha \text{ and } 
		\bte G_2 = \sum_{\alpha=1}^{n_{G_2}} \mu_\alpha \te M_\alpha \; .
		\label{eq:spectral_G1/G2}
	\end{align}
	Note that we only consider positive semi-definite structure tensors.
	For the case that both structure tensors have only equal eigenvalues each, i.e, $n_{G_1} = n_{G_2} = 1$, the tensors describe the \emph{isotropic} group $\mathcal G_{14}$. If $\bte G_1$ has $n_{G_1}=2$ non-equal eigenvalues and it holds $\bte G_2 = c_1 \one + c_2 \bte G_1$ with $c_1,c_2 \ge 0 \, \wedge c_1 + c_2 > 0$, the material is \emph{transversely isotropic} ($\mathcal G_{13}$). The same holds if $\bte G_1$ and $\bte G_2$ change the roles.
	For the case that $\bte G_1 \cdot \bte G_2 - \bte G_2 \cdot \bte G_1 = \zero$ holds and $\exists \bte G_\alpha$ with $n_{G_\alpha}=3$ for $\alpha\in\{1,2\}$ or both structure tensors have $n_{G_1} = n_{G_2}=2$ non-equal eigenvalues and it holds $\bte G_1 \ne c_1 \one + c_2 \bte G_2\,\forall c_1,c_2 \in\R$, the material is \emph{orthotropic} ($\mathcal G_3$).
	Both tensors describe the \emph{monoclinic group} $\mathcal G_2$ if it holds $w_L = e_{LMN} G^1_{MP} G^2_{PN} \ne 0$ and $(\bte G_\alpha \cdot \ve w) \times \ve w = \zero\; \forall \alpha\in\{1,2\}$. Finally, $\bte G_1$ and $\bte G_2$ describe the \emph{triclinic group} $\mathcal G_1$ if and only if  $w_L = e_{LMN} G^1_{MP} G^2_{PN} \ne 0$ and $\exists \bte G_\alpha$ with $(\bte G_\alpha \cdot \ve w) \times \ve w \ne \zero$, $\alpha\in\{1,2\}$.

	\section{Physics-augmented neural networks with anisotropy detection}
	\label{sec:PANNs}
	
	To enable an efficient and accurate description of the complex behavior of composite materials at the macroscale, we use NNs as a replacement for the time-consuming RVE simulation in this work. In doing so, we combine the advantages of NNs with a rigorous physical formulation, which we refer to as \emph{physics-augmented neural networks (PANNs)} \cite{Linden2023}.
	In this section, we formulate a PANN that builds on the concept of generalized structure tensors as introduced in Sect.~\ref{sec:gen_struct} and allows the description of a variety of symmetry groups. The training and model identification procedure tailored to the problem makes it possible to recognize the underlying anisotropy only from homogenized data. Note that an application of our approach to data from real experiments is also possible.

	\subsection{Model formulation}
	\subsubsection{Parameterized generalized structure tensors}\label{subsec:parametrized}
	
	To find a set of generalized structure tensors accounting for the underlying anisotropy of the material, it is necessary to determine the probability density function $\rho(\theta,\varphi)$, or, in the case of two structure tensors, the probability density functions. Usually this is done by identifying the orientation distribution from 2D microscopy images or computer tomography (CT) scans of the microstructure, e.g., \cite{Rolf-Pissarczyk2021}.
	
	However, in the following we leave this idea behind us and do not directly use the integral formulations as given in the previous section. Instead, we consider \emph{parameterized versions of the structure tensors}. This procedure has two main advantages: (i) It allows for the efficient determination of $\bte G\in\Sym$, $\btttte G\in \Sym_4$, $\btttttte G\in\Sym_6$ or $\bte G_1,\bte G_2\in\Sym$ directly from available data during training, i.e., from homogenized quantities $(\bte F,\bar \psi, \bte P, \btttte A)$ or also from real experimental data, and (ii)  it enables to apply the modeling framework to arbitrary materials, e.g., crystals, composites or foams. 
	The parameterized generalized structure tensors are constructed in such a way that the properties discussed in Sect.~\ref{sec:gen_struct}, i.e., symmetry, positive semi-definiteness and (generalized) trace of one, are guaranteed while maintaining the high flexibility of the original approach.
	
	\paragraph{2nd order structure tensor(s)}
	For the 2nd order structure tensor, the ansatz
	\begin{align}
		\bte G &:= \frac{1}{g_1+g_2+g_3}\te Q \cdot \diag(g_1,g_2,g_3) \cdot \te Q^T \in \Sym \; ,
		\te Q \in \SO \; ,\\
		[\te Q] &:=
		\begin{bmatrix}
			\cos \varphi_1 \cos \varphi_2 & \cos \varphi_1 \sin \varphi_2 \sin \varphi_3 - \cos \varphi_3 \sin \varphi_1 & \sin \varphi_1 \sin \varphi_3+\cos \varphi_1 \cos \varphi_3 \sin \varphi_2 \\
			\cos \varphi_2 \sin \varphi_1 & \cos \varphi_1 \cos \varphi_3 + \sin \varphi_1 \sin \varphi_2 \sin \varphi_3 &
			\cos \varphi_3 \sin \varphi_1 \sin \varphi_2 - \cos \varphi_1 \sin \varphi_3 \\
			- \sin \varphi_2 & \cos \varphi_2 \sin \varphi_3 & \cos \varphi_2 \cos \varphi_3
		\end{bmatrix}
	\end{align}
	with $\ve{\mathscr m}_{\bte G} \in \mathcal M_{\bte G}:=\left\{ \ve{\mathscr m}_{\bte G} \in \R^6 \, | \, g_1,g_2,g_3 \in[0,1], g_i g_i \ne 0, \, \varphi_1\in[ 0,\pi], \varphi_2\in[ -\pi/2,\pi/2], \varphi_3\in[ -\pi,\pi]  \right\}$ is used.
	Note that it would be possible to reduce from six to five parameters due to the normalization condition \cite{Wollner2023}, i.e., $\tr \bte G =1$. From a numerical point of view, however, the chosen approach has proven to be advantageous, since the conditions on $\ve{\mathscr m}_{\bte G}$ in the optimization can be set quite easily. As already stated, for $g_1=g_2=g_3$, $\bte G$ belongs to the \emph{isotropic} group $\mathcal G_{14}$, for $g_1=g_2\ne g_3$ to the \emph{transversely isotropic} group $\mathcal G_{13}$ and  for $g_1\ne g_2\ne g_3$ to the \emph{orthotropic} group $\mathcal G_{3}$, see \ref{app:ti} and \ref{app:orth}. If two 2nd order tensors are used, the ansatz given above is made for each of the two tensors, respectively. Thus, one gets 12 trainable parameters in this case: $\ve{\mathscr m}_{\bte G_{1},\bte G_2}\in\R^{12}$.

	\paragraph{4th and 6th order structure tensors}
	For the 4th order structure tensor, we propose an ansatz motivated from the crystal symmetries, see also Ebbing \cite{Ebbing2010}. It is given by a sum of dyadic products of three vectors, i.e., 
	\begin{align}
		\btttte G:= \frac{1}{n} \btttte S \text{ with } \btttte S:= \sum_{\alpha=1}^3 \ve A_\alpha \otimes \ve A_\alpha \otimes \ve A_\alpha \otimes \ve A_\alpha 
		\; ,\;  n:= \bar S_{KKLL}
	\end{align}
	This special choice for $\btttte G$ allows to fulfill for symmetry, positive semi-definiteness and normalization. In contrast to the 2nd order structure tensor, only a subset of the structure tensors described by Eq.~\eqref{eq:4thorderstruct} can be described using the selected approach.
	The vectors, given by
	\begin{align}
		\ve A_\alpha := a_\alpha \ve N_{(\alpha)} \in \Ln_1, \; [\ve N]:=
		\begin{bmatrix}
			\sin \vartheta_\alpha \cos \varphi_{(\alpha)} \\
			\sin \vartheta_\alpha \sin \varphi_{(\alpha)} \\
			\cos \vartheta_{\alpha}
		\end{bmatrix} \; , \;
	\end{align}
	are described by the three parameters $a_\alpha \in [0,1]$, $\vartheta_\alpha \in [0,\pi]$, $\varphi_\alpha \in [0, 2\pi]$ each. The parameters of the three vectors, which describe the 4th order structure tensor, are summarized in the set $\ve{\mathscr m}_{\btttte G} \in \mathcal M_{\btttte G}$. For the special case $\ve A_\alpha \cdot \ve A_\beta = |\ve A_{(\alpha)}| |\ve A_{(\beta)}| \delta_{\alpha\beta}$, the resulting tensor $\btttte G$ describes the \emph{cubic} anisotropy group $\mathcal G_7$ and for $\ve A_\alpha \cdot \ve A_\beta =0, \alpha \ne \beta$, $|\ve A_1| = |\ve A_2| \ne |\ve A_3|$ it represents the \emph{tetragonal} group $\mathcal G_5$, see \ref{app:cubic} and \ref{app:tetragonal}.
	
	Note that the choice of parameterization of the unit vectors may affect the uniqueness of the mininizer, see \cite{Mota2016}. However, we were able to achieve very good results with the parameterization used here.
	
	Similarly, for the 6h order structure tensor, we use the ansatz
	\begin{align}
		\btttttte G:= \frac{1}{n} \btttttte S \text{ with } \btttttte S:= \sum_{\alpha=1}^3 \ve A_\alpha \otimes \ve A_\alpha \otimes \ve A_\alpha \otimes \ve A_\alpha \otimes \ve A_\alpha \otimes \ve A_\alpha
		\; ,\;  n:= \bar S_{KKLLMM} \; .
	\end{align}
	Again, this particular choice is made to ensure symmetry, a generalized trace of one and positive semi-definiteness for $\btttttte G$.
	The parameters describing the 6th order structure tensor are summarized in the set $\ve{\mathscr m}_{\btttttte G} \in \mathcal M_{\btttttte G}$. For the special case $\ve A_\alpha \cdot \ve A_\beta = \pm\frac{1}{2}|\ve A_{\alpha}||\ve A_{\beta}|, \alpha \ne \beta$ and $|\ve A_1| = |\ve A_2| = |\ve A_3|$, the tensor describes the \emph{hexagonal} group $\mathcal G_{11}$, see \ref{app:hexa}.

	\subsubsection{Neural network-based model}
	
	In the following, we will restrict ourselves to models that only use a single  2nd, 4th, or 6th order structure tensor or two 2nd order structure tensors. As stated above, these models already cover a wide range of symmetry groups and have proven to be sufficient for the numerical examples shown later. However, the extension to models with other combinations of structure tensors is straightforward.
	
\begin{figure}
		\centering
		\includegraphics{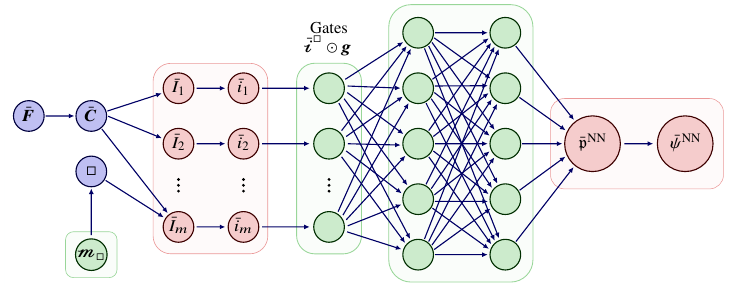}
		\caption{Illustration of the neural network $\bar \psi^\text{NN}(\bI^\square)$ for the representation of the elastic potential described by an invariant set $\bI^\square$ build from $\bte C:=\bte F^T \cdot \bte F$ and one or two structure tensor(s) $\square \in\{\bte G,\btttte G, \btttttte G,(\bte G_{1},\bte G_{2})\}$, i.e., the mapping is $\bar \psi^\text{NN}: \R^n \to \R_{\ge 0}\,,\; \bI^\square \mapsto \bar \psi^\text{NN}(\bI^\square) := (n^\text{out} \circ g^\text{NN} \circ \ve{\mathscr l}^\text{gate} \circ \ve{\mathscr n}^\text{in})(\bI^\square)$. Therein, $\ve{\mathscr n}^\text{in}(\bI^\square)$ and $n^\text{out}(\mathfrak p^\text{NN})$ are non-trainable normalization layers that have to fitted before training, $\ve{\mathscr l}^\text{gate}(\bve{\!\mathscr i}^\square)$ is a trainable gate layer and $g^\text{NN}(\bve{\!\mathscr i}^\square \odot \ve g)$ is a standard PNN guaranteeing positive outputs. The vector $\ve{\mathscr m}_\square$ includes the parameters of the structure tensors and is also trainable.}
		\label{fig:FNN}
	\end{figure}

	\paragraph{Neural network architecture}
	We construct an elastic potential based on a feedforward neural network (FNN) \cite{Kollmannsberger2021} with the invariant set $\bI^\square$ as input, i.e., $\bar \psi^\text{NN}(\bI^\square)$, where $\square\in\{\bte G, \btttte G, \btttttte G, (\bte G_{1},\bte G_2)\}$. This network has a structure tailored to the problem. It consist of a non-trainable input normalization layer, a trainable gate layer, a trainable positive neural network (PNN), and a non-trainable output normalization layer, i.e., $\bar \psi^\text{NN}: \R^n \to \R_{\ge 0}\,,\; \bI^\square \mapsto \bar \psi^\text{NN}(\bI^\square) := (n^\text{out} \circ g^\text{NN} \circ \ve{\mathscr l}^\text{gate} \circ \ve{\mathscr n}^\text{in})(\bI^\square)$. The task of the trainable gate layer is to remove unneeded invariants from the model during training. The architecture is illustrated in Fig.~\ref{fig:FNN}.
	
	To satisfy that the output of the PNN is always greater equal to zero, the activation function in the last hidden layer has to be greater equal to zero for all outputs of the former layer. Here, the softplus activation function is chosen. Furthermore, weights and bias of the output layer have to be non-negative.
	For brevity, the \emph{weights} and \emph{biases} of the  PNN interconnected between the normalization layers and the gate layer are summarized in the parameter set $\w\in\PNN$, where the set $\PNN$ is introduced in \ref{app:NNs}, Eq.~\eqref{eq:setPNN}.
	
	As introduced in \cite{Kalina2024}, \emph{non-trainable normalization layers} for in- and output are used, which allows an efficient optimization without having to normalize the training data. The layers are given by
	\begin{align}
		\ve{\mathscr n}^\text{in}: \R^n \to \R^n, \bI^\square \mapsto \bve{\!\mathscr i}^\square \text{ and }
		\mathscr n^\text{out}: \R_{\ge 0} \to \R_{\ge 0}, \bar{\mathscr p}^\text{NN} \mapsto \bar \psi^\text{NN} \; .
	\end{align}    
	As already mentioned, the weights and biases of these normalization layers are non-trainable. Instead, they have to be fitted to the data before training. 
	
	Furthermore, to allow to automatically remove unneeded invariants from the model during training, we add a \emph{trainable gate layer} between the input normalization layer and the PNN. The gate layer is defined by 
	\begin{align}
		\ve{\mathscr l}^\text{gate}: \R^n \to \R^n, \bve{\!\mathscr i}^\square \mapsto  \bve{\!\mathscr i}^\square \odot \ve g \text{ with }
		g_\alpha := \min(1,\gamma\tanh(\epsilon q_\alpha)) \in[0,1] \; ,
	\end{align}
	where $\gamma,\epsilon\in\R_{>0}$ are hyper parameters and $q_\alpha\in[0,1]$ are trainable variables. Thus, we have the additional set $\ve{\mathscr q} \in \mathscr{G\!a\!t\!e}:=\left\{ \ve{\mathscr q} \in \R^n \, | \, q_\alpha \in[0,1] \right\}$.
	
	The details on the chosen neural network architecture are given in \ref{app:NNs}. 
	
	\paragraph{Additional terms for the construction of the elastic potential}
	In addition, following Linden~et~al.~\cite{Linden2023}, terms which guarantee zero energy and stress in the undeformed state as well as a term for the fulfillment of the volumetric growth condition are added. Thus, we end up with
	\begin{align}
		\bar \psi^\square(\bI^\square,\bar J) := \bar \psi^\text{NN}(\bI^\square) + \bar \psi^\text{en} + \bar \psi^{\text{str,}\square}(\bI^\square,\bar J) + \bar \psi^\text{gr}(\bar J)
		\; .
		\label{eq:NN_model_invariants}
	\end{align}
	Note that the additional terms are chosen in such a way that the material symmetry is not violated.
	
	The term to guarantee zero energy in the undeformed state is independent of the symmetry group and is given by $\bar \psi^\text{en} = - \bar \psi^\text{NN}(\bI^\square)\big|_{\bte F=\one}$.  
	In contrast, the energy expression enforcing $\bte P(\bte F=\one)=\zero$ depends on the chosen set of invariants \cite{Linden2023}, i.e., whether the set is constructed either with $\bte G$, $\btttte G$, $\btttttte G$ or $(\bte G_1,\bte G_2)$.
	The expressions $\bar \psi^{\text{str,}\square}(\bI^\square,\bar J)$ chosen in this work are summarized in \ref{app:stress_norm}. For the growth term, which enforces the fulfillment of the volumetric growth condition introduced in Sect.~\ref{subsect:phys_cond}, the expression $\bar \psi^\text{gr}(\bar J) := \lambda_\text{gr} \left(\bar J + \bar J^{-1} -2 \right)^3$ is chosen. The parameter $\lambda_\text{gr}$ has to be chosen such that the energy grows fast enough during compression. Following \cite{Kalina2023}, a value between \num{1e-2} and \num{1e-3} the material’s initial stiffness has proven to be suitable.
	
	The proposed models fulfill the following conditions by construction: \emph{thermodynamic consistency, symmetric Cauchy and 2nd Piola-Kirchhoff stress, objectivity, material symmetry, energy- and stress-free undeformed state}, as well as \emph{volumetric growth condition}.\footnote{%
		Note that the introduced energy expression is not polyconvex by construction. However, since a polyconvex energy turned out to be too restrictive for the fitting, this does not play a role here. For details on how an NN-based polyconvex energy can be constructed for anisotropic materials, which fulfills the material symmetry and at the same time guarantees a stress-free undeformed configuration, the reader is referred to Linden~et~al.~\cite{Linden2023}.%
	}

	\subsection{Training}
	\label{subsec:train}
	
	To calibrate the NN-based models, the weights and biases of the PNN, collected into $\ve{\mathscr w}\in \PNN$, the parameters of the generalized structure tensors $\ve{\mathscr m}_\square \in \mathcal M_\square$, and the gate variables $\ve{\mathscr q} \in \mathscr{G\!a\!t\!e}$ have to be determined in a suitable training procedure. Thereby, we apply a strategy allowing to optimize all parameters simultaneously, which has the advantage that no further information from the microscopic geometry are necessary.    
	Only data from RVE simulations are required. We collect these data into the set $\mathcal D:=\{{}^1\mathcal T, {}^2\mathcal T,\ldots,{}^k\mathcal T\}$. Thereby, the tuples are given by 
	\begin{equation}
		{}^i \mathcal T:=({}^i \bte F, {}^i\bar \psi,{}^i \bte \sigma, {}^i \bttttes c)^\text{RVE} \in \GL \times \R_{\ge 0} \times \Sym \times \Ln_4 \; .
		\label{eq:T_sig}
	\end{equation}
	Please note that a calibration based on $\bte P$ and $\btttte A$ is equivalent, as the respective quantities can be calculated from each other by pullback or push-forward operations, see Sects.~\ref{sec:kin_stress} and \ref{subsect:phys_cond}.
	
	A model must be assessed against non-calibration data in order to look at its generalizability. Consequently, in accordance with standard machine learning procedure \cite{Linden2023,Klein2021,Vlassis2021,Kollmannsberger2021}, we split the whole dataset $\mathcal D$ into \emph{calibration} and \emph{test} sets, respectively:
	\begin{align}
		\mathcal D = \mathcal D_\text{cal} \cup \mathcal D_\text{test} \text{ and }\varnothing = \mathcal D_\text{cal} \cap \mathcal D_\text{test} \; .
		\label{eq:cal_test}
	\end{align}
	The NN-based model's predictions for $\mathcal D_\text{test}$ can be verified once it has been calibrated with $\mathcal D_\text{cal}$. Thereby, the calibrated model should be able to generate reasonable predictions not only for the calibration but also for the test dataset, which will ensure good generalization of the model and allows to mimic the RVE's effective behavior for arbitrary states $\bte F\in\GL$.
	
	To perform training based on the tuples given in Eq.~\eqref{eq:T_sig}, we define the following loss terms for the \emph{mean squared error (MSE)} for energy, stress and elasticity tensor:
	\begin{align}
		\mathcal L^\psi &:= \frac{1}{n^\psi} \left\| {}^i \bar \psi- \bar \psi({}^i \bte F, \w, \ve{\mathscr m}_\square, \ve{\mathscr q}) \right\|_\text{MSE}
		\text{ with }
		n^\psi := \max\left({}^1\bar \psi^2, {}^2\bar \psi^2, \ldots, {}^{|\mathcal D_\text{cal}|}\bar \psi^2 \right) \; , \\
		\mathcal L^{\te \sigma} &:= \frac{1}{n^{\te \sigma}} \left\| {}^i \bte \sigma- \bte \sigma({}^i \bte F, \w, \ve{\mathscr m}_\square, \ve{\mathscr q}) \right\|_\text{MSE}
		\text{ with }
		n^{\te \sigma} := \frac{1}{3^2}\max\left(\|{}^1\bte \sigma\|^2, \|{}^2\bte \sigma\|^2, \ldots, \|{}^{|\mathcal D_\text{cal}|}\bte \sigma\|^2 \right) \; ,
		\label{eq:loss_sigma}\\
		\mathcal L^{\ttttes c} &:= \frac{1}{n^{\ttttes c}} \left\| {}^i \bcKM- \bcKM({}^i \bte F, \w, \ve{\mathscr m}_\square, \ve{\mathscr q}) \right\|_\text{MSE}
		\text{ with }
		n^{\ttttes c} := \frac{1}{6^2}\max\left(\|{}^1\bcKM \|^2, \|{}^2\bcKM \|^2, \ldots, \|{}^{|\mathcal D_\text{cal}|}\bcKM\|^2 \right) \; . 
		\label{eq:loss_c}
	\end{align}
	Therein, $\bcKM \in \R^{6\times 6}$ denotes a matrix representation of the coordinates of the 4th order material tangent $\bttttes c$ given in Voigt notation and $\left\| (\cdot) -  (\cdot) \right\|_\text{MSE}$ is the standard MSE norm. The normalization factors $n^\psi$, $n^{\te \sigma}$ and $n^{\ttttes c}$ ensure that the loss terms can be compared with each other. Thus, the prediction loss is given by $\mathcal L^\text{pred} = w^\psi \mathcal L^{\psi} + w^{\te \sigma}\mathcal L^{\te \sigma} + w^{\ttttes c}\mathcal L^{\ttttes c}$, where the non-trainable weights $w^\psi, w^{\te \sigma}, w^{\ttttes c} \in \R_{\ge 0}$, with $w^\psi + w^{\te \sigma} + w^{\ttttes c} = 1$, have to be chosen.
	In addition, to enforce sparsity of the model with respect to the number of invariants, a penalty term based on the $p$-norm \cite{Flaschel2021,McCulloch2024} of the gates 
	\begin{align}
		\mathcal L^\text{gate} := \frac{1}{n^\text{gate}}\left[\sum_{\alpha=1}^n(g_\alpha(q_\alpha)+\delta)^p\right]^{\frac{1}{p}} 
		\text{ with } n^\text{gate}:= \left[n(1+\delta)^p\right]^{\frac{1}{p}} 
		\label{eq:gate_loss}
	\end{align}
	is used, where $p \in \R_{\ge 0}$ and $n$ is the number of invariants. The parameter $\delta \ll 1$ prevents division by zero when differentiating.
	By applying the introduced loss terms, the training is defined by the optimization
	\begin{align}
		(\hat{\w}, \hat{\ve{\mathscr m}}, \hat \g) = \underset{\ve{\mathscr w}\in \PNN, \ve{\mathscr m}_\square \in \mathcal M_\square, \ve{\mathscr q} \in \mathscr{G\!a\!t\!e}}{\arg \min} \left( w^\psi \mathcal L^{\psi} + w^{\te \sigma}\mathcal L^{\te \sigma} + w^{\ttttes c}\mathcal L^{\ttttes c} + w^\text{gate} \mathcal L^\text{gate} \right) \; ,
	\end{align}
	where the parameter $w^\text{gate}$ has to be chosen such that the approximation quality of the model does not decrease. Since the second derivatives of the potential with respect to $\bte F$ are necessary to compute $\btttte A$ and from this $\bttttes c$, it is a \emph{higher-order Sobolev training} \cite{Vlassis2021,Kalina2024}. The proposed training method allows to calibrate all four models, $\bar \psi^{\bte G}(\bI^{\bte G},\bar J)$, $\bar \psi^{\btttte G}(\bI^{\btttte G},\bar J)$, $\bar \psi^{\btttttte G}(\bI^{\btttttte G},\bar J)$, as well as $\psi^{\bte G_{1},\bte G_{2}}(\bI^{\bte G_{1},\bte G_{2}},\bar J)$. Note that the internal normalization layers of the PNN have to be fitted to the data before training.
	
	\begin{rmk}
		\label{rmk:train}
		To achieve optimal results after training, a two-step approach has shown to be favorable. Thereby, a pre-training with the \emph{Adam optimizer} and a post-training with the \emph{SLSQP optimizer} (Sequential Least Squares Programming), which is a quasi Newton optimizer, is carried out. As shown in \cite{Kalina2024}, the optimization with SLSQP is clearly superior to Adam for comparatively small networks, since it converges faster and a lower loss is reached. However, it has shown to be difficult to find a suitable structure tensor only with SLSQP, which is due to the fact that a strong dependence on the initialization is typical for this optimizer. However, very good results can be achieved if good starting values are specified by applying the pre-training with Adam.
		The implementation of the models and the calibration workflow was realized using \emph{Python}, \emph{TensorFlow} and \emph{SciPy}.
	\end{rmk}

	\subsection{Data generation and model identification procedure}
	\label{subsec:procedure}
	
	\begin{figure}
		\centering
		\includegraphics{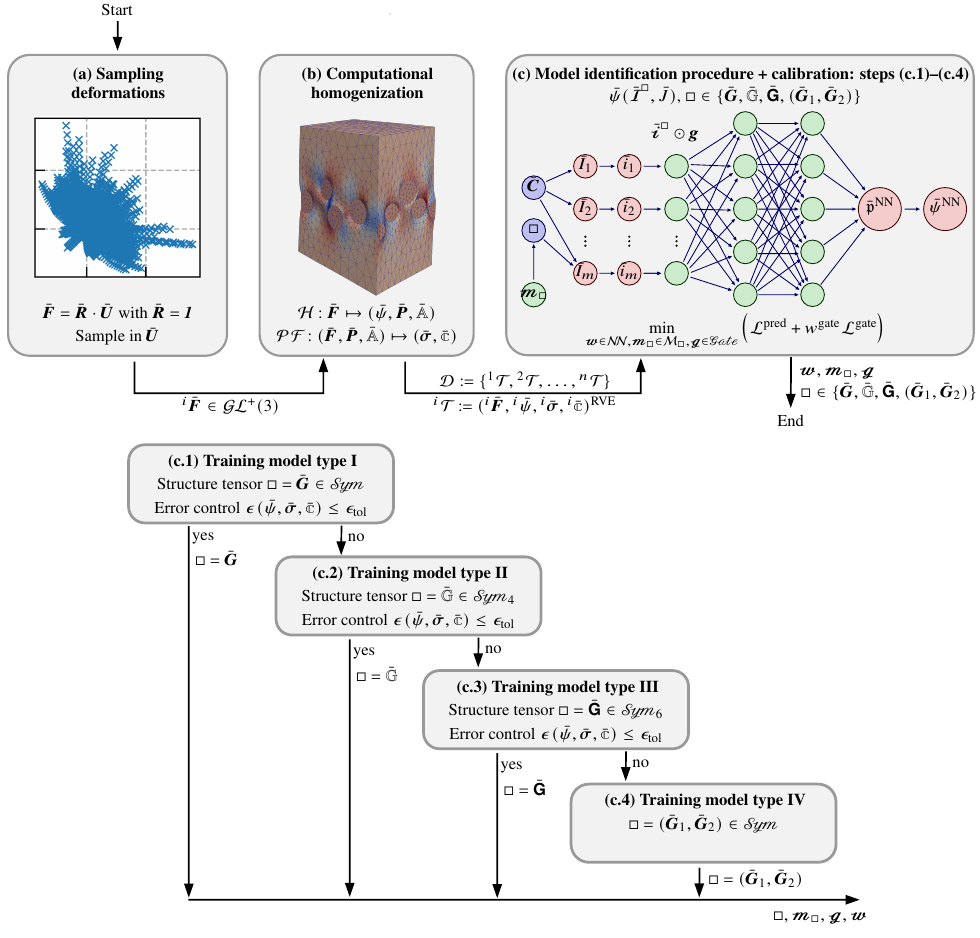}
		\caption{Data generation and model identification procedure: (a) the deformation space is sampled in a prescribed range by Latin Hypercube Sampling (LHS), (b) by prescribing the sampled deformation gradients ${}^i\bte F$ in RVE simulations, corresponding energy ${}^i\bar \psi$, stress ${}^i\bte \sigma$ and elasticity tensor ${}^i\bttttes c$ are calculated, (c) the dataset $\mathcal D$ is used to identify and calibrate the NN-based model $\bar \psi(\bI^\square,\bar J)$ via a higher-order Sobolev training with prediction loss $\mathcal L^\text{pred}= w^\psi \mathcal L^{\psi} + w^{\te \sigma}\mathcal L^{\te \sigma} + w^{\ttttes c}\mathcal L^{\ttttes c}$ and gate loss $\mathcal L^\text{gate}$ for sparsity. The step (c) involves the substeps (c.1)--(c.4): to identify which (set of) structure tensor(s) $\square \in \{\bte G, \btttte G, \btttttte G,(\bte G_{1},\bte G_{2})\}$, i.e., 2nd, 4th, 6th, or two 2nd order structure tensor(s), is needed, the training is performed sequentially from low to high structure tensor order and an error control is performed after each training to decide whether the underlying anisotropy can be described with sufficient accuracy.}
		\label{fig:ident_proc}
	\end{figure}
	
	In order to generate the NN-based surrogate for RVE simulations, the following procedure is applied: \emph{(a) sampling of the deformation space in a prescribed range, (b) generation of a database by computational homogenization}, and \emph{(c) model identification procedure} which consists of a maximum of four substeps. The overall procedure is illustrated in Fig.~\ref{fig:ident_proc}.
	
	The sampling technique, which is similar to \cite{Kalina2024}, is described in \ref{app:samp} and the computational homogenization is done as described in Sect.~\ref{subsec:homogenization}.
	Finally, in step (c), the model which is suitable as a surrogate, i.e.,  $\bar \psi^{\bte G}(\bI^{\bte G},\bar J)$, $\bar \psi^{\btttte G}(\bI^{\btttte G},\bar J)$, $\bar \psi^{\btttttte G}(\bI^{\btttttte G},\bar J)$ or $\bar \psi^{\bte G_{1},\bte G_{2}}(\bI^{\bte G_{1},\bte G_{2}},\bar J)$, is chosen and calibrated.  
	In order to obtain an RVE surrogate model that is as efficient as possible in macroscopic FE simulations \cite{Kalina2023}, we want to find the model with the lowest possible order in the structure tensor that allows to describe the RVE's underlying symmetry group.\footnote{Note that models based on higher order structure tensors contain others, e.g., the model based on the 4th order structure tensor can also be used to describe transverse isotropy and orthotropy.} This is done as follows: 
	First, in (c.1), the model based on the 2nd order structure tensor is calibrated with the RVE data. After training, the relative error measures in energy, stress and material tangent
	\begin{align}
		\epsilon^\psi = \frac{\sum\limits_{i=1}^{|\mathcal D|}|{}^i\bar\psi^\square-{}^i\bar\psi|}{\sum\limits_{j=1}^{|\mathcal D|}|{}^j\bar\psi|} \; , \;
		\epsilon^{\te \sigma} = \frac{\sum\limits_{i=1}^{|\mathcal D|}\|{}^i\bte \sigma^\square-{}^i\bte\sigma\|}{\sum\limits_{j=1}^{|\mathcal D|}\|{}^j\bte \sigma\|} \text{ and }
		\epsilon^{\ttttes c} = \frac{\sum\limits_{i=1}^{|\mathcal D|}\|{}^i\bcKM^\square-{}^i\bcKM \|}{\sum\limits_{j=1}^{|\mathcal D|}\|{}^j\bcKM\|}
		\label{eq:error_measures}
	\end{align}
	are calculated. If all errors are below a given tolerance $\epsilon_\text{tol}$, the model identification is complete. Here, we choose $\epsilon_\text{tol}=\SI{1}{\percent}$. If the tolerance is exceeded, the model based on the 4th order structure tensor is trained and the errors are calculated again, substep (c.2). If this model is also insufficient, the 6th order structure tensor is used, substep (c.3). Finally, if it is not possible to reach errors below $\epsilon_\text{tol}$ with a model based on a single structure tensor, a training based on two 2nd order structure tensors is done (c.4). Note that this final step should lead to an error below the given tolerance, since the formulation then includes the triclinic group, i.e., fully anisotropic behavior.
	In each substep, the training as described in Sect.~\ref{subsec:train} is applied.
	
	\section{Numerical examples}
	\label{sec:examples}
	
	In order to illustrate the ability of the developed NN-based elastic surrogate models to mimic the complex anisotropic behavior of RVEs with high precision,  we will calibrate them with data from computational homogenizations in the following. With these data, the interpolation behavior of the models as well as the extrapolation behavior is investigated. Since we only use data generated by homogenization, we do not consider the case of noisy data in this paper.
	
	\subsection{Data generation via computational homogenization}
	
\begin{figure}[h]
		\centering
		\includegraphics{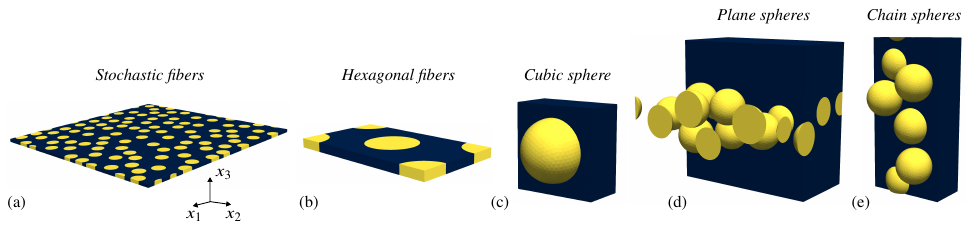}
		\caption{Considered RVEs for data generation: (a) fiber reinforced material (\emph{stochastic fibers}), (b) unit cell with hexagonal fiber arrangement (\emph{hexagonal fibers}), (c) unit cell with one spherical inclusion (\emph{cubic sphere}), (d) particle reinforced plane-like microstructure (\emph{plane spheres}), and (e) particle reinforced chain-like microstructure (\emph{chain spheres}). The volume fractions of the fiber/particle phase are given by $\phi \in \{ 30, 30, 20, 12, 15\}\,\%$ from left to right.}
		\label{fig:RVEs}
	\end{figure}

	In this work, two-phase soft composites consisting of matrix and spherical inclusions or matrix and fibers are considered. In order to provide data with varying overall anisotropic behavior, we consider five different RVEs depicted in Fig.~\ref{fig:RVEs}: An RVE of a fiber reinforced material with stochastic fiber distribution (\emph{stochastic fibers}), a unit cell with a hexagonal fiber arrangement (\emph{hexagonal fibers}), a unit cell with one spherical inclusion (\emph{cubic sphere}), an RVE with a plane-like arrangement of particles (\emph{plane spheres}), and an RVE with an arrangement of particles in a chain-like structure (\emph{chain spheres}). The geometries and the subsequent creation of the periodic mesh has been done with the Python tool \emph{gmshModel}.\footnote{The Python tool gmshModel is freely available under the link \url{https://gmshmodel.readthedocs.io/en/latest/}}
	
	All components, i.e., matrix, particles and fibers are assumed to be compressible and isotropic. For all, we choose the two-parametric neo-Hookean model according to Ciarlet~\cite{Ciarlet1988} given by 
	\begin{equation}
		\psi(I_1,I_3) := \frac{1}{2} \left[\mu (I_1 - \ln I_3 -3) + \frac{\lambda}{2}(I_3 - \ln I_3 - 1) \right]  \text{ with } \mu,\lambda > 0 \; , \; \mu = \frac{E}{2(1+\nu)} \;,\; \lambda = \frac{E \nu}{(1+\nu)(1-2\nu)} \; .
		\label{eq:comp}
	\end{equation}
	In the equation above, $\mu$ and $\lambda$ denote Lam\'{e} parameters and $E$ and $\nu$ are Young's modulus and Poisson's ratio, respectively. Note that the elastic potential \eqref{eq:comp} is also polyconvex \cite{Linden2023} and thus guarantees ellipticity for all possible states $\te F \in \GL$.
	The chosen material parameters for the individual components are given in Tab.~\ref{tab:parameters_micro}.
	
\begin{figure}[t]
		\centering
		\includegraphics{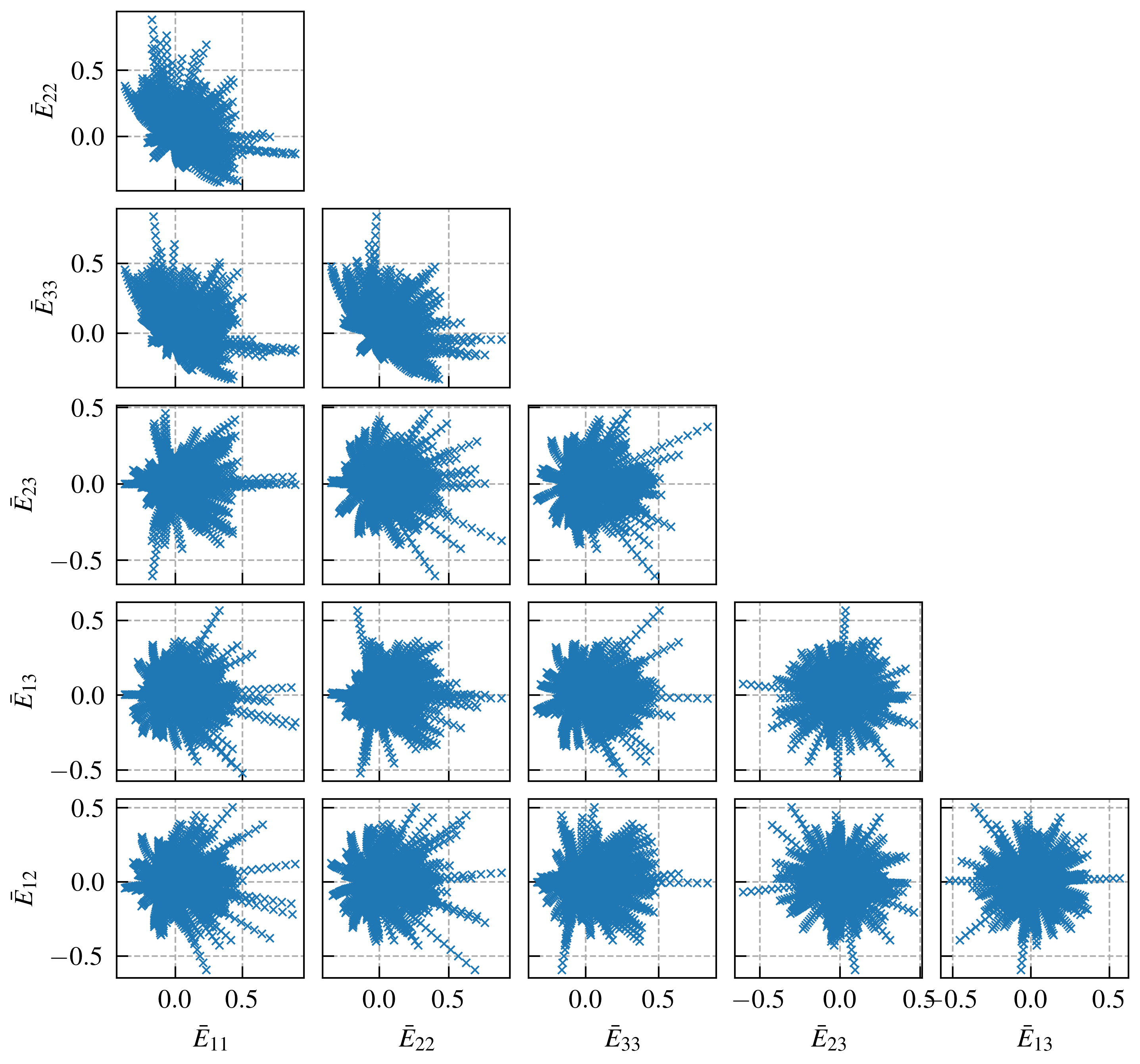}
		\caption{Sampled deformation space comprising 141 loading paths with 20 increments each. Shown are sectional planes of the Green-Lagrange strain tensor $\bte E$.}
		\label{fig:corr_strain}
	\end{figure}

	In order to investigate the RVEs' behavior within a predefined range of effective deformations $\bte F \in \GL$, a sampling technique similar to the approach used in \cite{Kalina2024} is applied. Details on this technique are given in \ref{app:samp}. With that, a total of 154 loading paths comprising 20 increments each are generated. The sampled states are shown within sectional planes of the Green-Lagrange strain tensor $\bte E$ in Fig.~\ref{fig:corr_strain}.
	For all sampled states, the resulting effective material response is calculated by computational homogenization according to Sect.~\ref{subsec:homogenization}, i.e., $\mathcal H: \bte F \mapsto (\bar \psi, \bte P, \btttte A)$ and $\mathcal {P\!F}: (\bte F,\bte P, \btttte A) \mapsto (\bte \sigma, \bttttes c)$.\footnote{%
		Note that some simulations of the RVEs with spherical inclusions diverged before reaching the last increment, which is due to the fact that the large phase contrast can easily lead to extremely high matrix deformations. In that case, the corresponding states have been removed from the dataset of the RVE, respectively.}

	\begin{table}
		\begin{center}
			\caption{Chosen parameters for matrix material, spherical particles and fibers, described by the constitutive model \eqref{eq:comp}.}
			\label{tab:parameters_micro}
			\begin{footnotesize}
				\begin{tabular}{lllll}
					Parameter & Symbol & Matrix material & Spherical particles & Fibers\\
					\hline\hline
					Young's modulus & $E$ & $\SI{1}{\mega\pascal}$ & $\SI{1000}{\mega\pascal}$ & $\SI{10}{\mega\pascal}$ \\
					Poison's ratio & $\nu$ & $0.4$ & $0.3$ & $0.44$ 
				\end{tabular}
			\end{footnotesize}
		\end{center}
	\end{table}
	
	Besides the determination of energy, stress and tangent corresponding to a deformation state,  the \emph{local ellipticity} condition \eqref{eq:ellipticityLocal} has been evaluated for all generated data.
	To this end, the positive semi-definiteness of the acoustic tensor $\bte{\gamma}(\bte F,\ve N)=\bar A_{iJkL}(\bte F)N_J N_L \ve e_i \otimes \ve e_k$ \cite{Ebbing2010,Kalina2024,Schroder2010} has been checked numerically by sampling $\ve N(\vartheta,\varphi)$ parameterized by spherical coordinates $(\vartheta,\varphi)\in[0,\pi] \times [0,2\pi]$ in $\pi/180$ steps for all states $\bte F$ included in the dataset.
	Thereby, a loss of ellipticity was found in some states for all RVEs considered, except for the RVE plane spheres. Thus, the effective material response cannot be represented by a polyconvex model, which would guarantee ellipticity by construction, at least not for the complete data set.   
	
	The generated homogenized data will be made freely available in the final version of the article.
	
	\begin{rmk}
		It is well known that polyconvex microscopic energies in multiscale problems do not necessarily imply macroscopic quasi-convexity and consequently rank-one convexity \cite{Polukhov2020,Abeyaratne1984}. For some regions of the effective deformation $\bte F$, even a loss of ellipticity is possible, cf. \cite{Polukhov2020, Rudykh2013}.
	\end{rmk}

	\subsection{Training and prediction of the neural networks}
	
	The obtained data $\mathcal D$ for the different considered RVEs are now used for the calibration of the proposed NN-based macroscopic models introduced in Sect.~\ref{sec:PANNs}, where the model identification procedure described in Sect.~\ref{subsec:procedure} is applied, i.e., it is automatically determined whether a 2nd, 4th or 6th order structure tensor or a set of two 2nd order structure tensors is required.
	To evaluate the performance of the developed invariant-based NN models, a typical approach based on the coordinates of the right Cauchy-Green deformation tensor $\bte C$ will also be considered as a reference. Similar models are widespread, e.g., \cite{Vlassis2020,Asad2022}. The \emph{coordinate-based reference model} is given by the energy expression
	\begin{align}
		\bar \psi^\text{coord}(\bte C, \bar J) := \bar \psi^\text{PNN}(\bte C) - \bar \psi^\text{PNN}(\bte C)\big|_{\bte F=\one} - \diffp{\bar \psi^\text{PNN}}{\bte C}\Bigg|_{\bte F=\one} : \left(\bte C - \one \right) +  \lambda_\text{gr}\left(\bar J + \bar J^{-1} -2 \right)^2 \; ,
		\label{eq:NN_coord}
	\end{align}
	where a PNN according to \ref{app:NNs}, but without trainable gate, is chosen for the network.
	To enable a fair comparison, it is designed to fulfill the same conditions as the invariant-based models except for the material symmetry, i.e., it ensures thermodynamic consistency, symmetric Cauchy and 2nd Piola-Kirchhoff stress, objectivity,  energy- and stress-free undeformed state, as well as volumetric growth condition.
	
	The parameter $\lambda_\text{gr}$ has to be chosen such that the energy grows fast enough during compression. According to \cite{Kalina2024}, a value of around \num{1e-2} or \num{1e-3} the material's initial stiffness\footnote{In the anisotropic case the maximum initial stiffness has to be considered.} has shown to be reasonable. Here we choose $\lambda_\text{gr}=\SI{0.01}{\mega\pascal}$ for all NN-based models.
	According to the hyperparameter study given in \ref{app:study_hyper}, two hidden layers with 16 neurons each are chosen for the invariant-based NN models and  three hidden layers with 16 neurons each for the coordinate-based model. For training of the invariant-based NN models, the parameter for the penalization of the gates was chosen to $w^\text{gate}=\num{5e-5}$, see \ref{app:study_gate} for a study to determine this value. Following \cite{Flaschel2021}, we have chosen $p=\frac{1}{4}$ for the exponent in the $p$-norm. The parameters in the gate are chosen to $\gamma = 1.025$, $\epsilon = 2.5$ and $\delta = \num{1e-6}$, respectively.
	
	To exclude random effects from initialization, 5 training runs with pre-training and post-training steps as described in Remark~\ref{rmk:train} were carried out each. The best model is then used for the discussion of the results, where this is decided by the measure $\mathcal L^\text{pred}/A$, where $A\in\mathbb N$ is the number of active gates.  
	In the pre-training with Adam, an initial learning rate of 0.01 and an exponential learning rate decay, so that the learning rate is multiplied by $1/3$ every 500 epochs, was selected. In addition, mini-batches were used, which introduces a regularization effect \cite{Kollmannsberger2021}. The mini-batches were recomposed in a random manner after each epoch. For the interpolation studies, the mini-batch size was set to 64, while it was set to 16 for the extrapolation.
	All training runs were carried out with 8 CPUs each, whereby a high performance cluster (HPC) equipped with Intel Xeon Platinum 8470 CPUs was used.
	
	To mark that a structure tensor was calibrated by data from a specific RVE, we use the following superscripts: \emph{stochastic fibers} $\square^\text{stf}$, \emph{hexagonal fibers} $\square^\text{hef}$, \emph{cubic sphere} $\square^\text{cus}$, \emph{plane spheres} $\square^\text{pls}$,  and \emph{chain spheres} $\square^\text{chs}$.

	\subsubsection{Interpolation behavior: Training with stress and tangent}
	\label{subsec:interpolation}
	In this first study we mainly check the interpolation ability of the NN-based models. Thus, as described in Sect.~\ref{subsec:train}, we divide the overall datasets into calibration and test sets with a ratio of $70/30$, respectively, see Eq.~\eqref{eq:cal_test}.
	The prediction loss $\mathcal L^\text{pred}=0.7 \mathcal L^{\te \sigma} + 0.3 \mathcal L^{\ttttes c}$ is chosen, which is to weight the stresses somewhat more than the material tangent. It should be noted that experience has shown that it is not necessary to include the energy $\bar \psi$ itself in the loss, as this is adjusted very well with the Sobolev training used here \cite{Linden2023,Kalina2024,Rosenkranz2023}.
	
	\paragraph{Identified structure tensors}
	For the RVEs \emph{stochastic fibers} and \emph{plane spheres}, a model based on a 2nd order generalized structure tensor $\bte G$ was selected by the algorithm according to Fig.~\ref{fig:ident_proc}, respectively. The coordinates of these structure tensors are given by
	\begin{align}
		[\bte G^\text{stf}] = 
		\begin{bmatrix}
			0.01 & 0.01 & 0.  \\
			0.01 & 0.02 & 0.  \\
			0. & 0. & 0.97
		\end{bmatrix} \; \text{and} \; 
		[\bte G^\text{pls}] = 
		\begin{bmatrix}
			0.01 &  0.  & -0.01 \\
			0.  &  0.  &  0.02 \\
			-0.01 &  0.02 &  0.99
		\end{bmatrix} \; .
		\label{eq:structure2ndorder}
	\end{align}
	The vectors of active/non-active gates for the two RVEs are given by 
	\begin{align}
		G(\ve g^\text{stf}) = (1,0,1,0,1,1,0) \text{ and }  
		G(\ve g^\text{pls}) = (1,1,1,1,0,0,1) \; ,
	\end{align}
	respectively, where $G(x):=[x>0]$ is a function which is zero for all $x\le 0$ and 1 else. The gates correspond to the invariants defined in Eqs.~\eqref{eq:invariants_iso}, \eqref{eq:inv_G2nd}, i.e., the additional invariants needed to describe orthotropy are removed from the model by the gates.
	The identified structure tensor for the RVE \emph{stochastic fibers} corresponds to the expected result, i.e. transverse isotropy $\mathcal G_{13}$ around $\ve e_3$, as the fibers are aligned with the $X_3$ axis, which was identified as a preferred direction during training.\footnote{%
		\label{foot:transverse}
		For the transverse isotropy with preferred direction $X_3$ there is an unlimited number of permissible structure tensors with $\tr \bte G = 1$, since all tensors given by $\bte G = \alpha (\ve e_1 \otimes \ve e_1 + \ve e_2 \otimes \ve e_2) + \beta \ve e_3 \otimes \ve e_3$ with $\alpha,\beta\in \R_{\ge 0}$, $\alpha\ne \beta$ and $2\alpha+\beta = 1$ are equivalent to each other, cf. \ref{app:ti}.
	} The result also makes sense for the RVE \emph{plane spheres}, where the out-of-plane direction is a preferred direction.
	The slight deviation to ideal transverse isotropy with $X_3$ as preferred direction results from the stochastic arrangement of the particles within the plane of particles and from numerical inaccuracies during training.

	In contrast, for the RVE \emph{cubic sphere}, the selection procedure has identified that a 4th order generalized structure tensor $\btttte G$ is needed. 
	This is not surprising, since it is well known that a 4th order structure tensor is needed for cubic anisotropy \cite{Ebbing2010,Xiao1996,Apel2004}. 
	Interestingly, however, a 4th order structure tensor was identified during the optimization, which belongs to the tetragonal group $\mathcal G_5$. This can be explained by the fact that the invariant set belonging to the cubic group $\mathcal G_7$ up to order $\bte C^3$ can be represented by the invariant set of the tetragonal group, cf. \ref{app:cubic}.
	The coordinates of the identified structure tensor in Voigt notation and the vector of active gates are given by
	\begin{align}
		[\btttte G^\text{cus}] = 
		\begin{bmatrix}
			0. &  0. &  0. & -0. & -0. &  0. \\
			0. &  0.5 &  0. & -0. & -0. &  0. \\
			0. &  0. &  0.5 & -0. & -0. &  0. \\
			-0. & -0. & -0. &  0. &  0. & -0. \\
			-0. & -0. & -0. &  0. &  0. & -0. \\
			0. &  0. &  0. & -0. & -0. &  0.    
		\end{bmatrix} \text{ and }
		G(\ve g^\text{cus}) = (1,1,1,1,1,0,1,1,0,1,0) \; ,
		\label{eq:structure4thorder}
	\end{align}
	respectively. The gates correspond to the invariants defined in Eqs.~\eqref{eq:invariants_iso}, \eqref{eq:inv_G4}.
	
	For the RVE \emph{hexagonal fibers}, the algorithm has determined that even a 6th order structure tensor is required. This also corresponds to the expectation for a material that belongs to the hexagonal anisotropy group $\mathcal G_{11}$ \cite{Xiao1996,Apel2004}.
	The three vectors $\ve A_1$, $\ve A_2$ and $\ve A_3$ specifying the 6th order tensor $\btttttte G \in \Sym_6$ have been identified to
	\begin{align}
		[\ve A^\text{hef}_1] = \begin{bmatrix}
			-0.66 \\
			0.38 \\
			0.  
		\end{bmatrix} \; , \;
		[\ve A^\text{hef}_2] = \begin{bmatrix}
			-0.  \\
			0.76 \\
			-0.  
		\end{bmatrix} \text{ and }
		[\ve A^\text{hef}_3] = \begin{bmatrix}
			-0.66 \\
			-0.38 \\ 
			-0.  
		\end{bmatrix} \; .
	\end{align}
	As one can see, the conditions $(|\ve A_\alpha||\ve A_\beta|)^{-1} \ve A_{(\alpha)} \cdot \ve A_{(\beta)} = \pm\frac{1}{2}, \alpha \ne \beta$, $\ve N \cdot \ve A_\alpha = 0$, with $\ve N\in\Vn$ being the fiber direction, and $|\ve A_1| = |\ve A_2| = |\ve A_3|$ apply in good approximation to the three vectors that form $\btttttte G$. The structure tensor determined thus corresponds to the expected structure tensor for the hexagonal group $\mathcal G_{11}$, cf. \ref{app:hexa}. The vector of active/non-active gates is given by $G(\ve g^\text{hef}) = (1,1, 1,1,1,0,1,1,0,0,1,0,1)$, where the gates correspond to the invariants defined in Eqs.~\eqref{eq:invariants_iso}, \eqref{eq:inv_G6}.
	
	Finally, the identification algorithm has determined that two 2nd order structure tensors are needed for the RVE \emph{chain spheres}. These two tensors are given by
	\begin{align}
		[\bte G_1^\text{chs}] = \begin{bmatrix}
			0.08 &  0.02 &  0.04 \\
			0.02 &  0.53 & -0.01 \\
			0.04 & -0.01 &  0.39
		\end{bmatrix} \; \text{and} \;
		[\bte G_2^\text{chs}] =
		\begin{bmatrix}
			0.5 & -0.02 &  0.05 \\
			-0.02 &  0.49 & -0.  \\
			0.05 & -0.  &  0.01
		\end{bmatrix} \; .
	\end{align}
	To determine which symmetry group was determined, we apply the tests given in Sect.~\ref{subsec:2xGST2}. Accordingly, since $w_L = e_{LMN} \bar G^1_{MP} \bar G^2_{PN} \ne 0$ and $\ve v_\alpha = (\bte G_\alpha \cdot \ve w) \times \ve w \ne \zero$ for $\alpha\in\{1,2\}$, the algorithm has detected the \emph{triclinic group} $\mathcal G_1$.
	However, taking a certain numerical tolerance into account, the structural tensors can be assigned to the \emph{monoclinic group} $\mathcal G_2$ to a good approximation. 
	To show this, we consider $\ve w$, $\ve v_1$ and $\ve v_2$ and compute the norms given by $|\ve w| = \num{3.77e-2}$, $|\ve v_1| = \num{6.59e-05}$ and $|\ve v_2| = \num{1.90e-4}$. Thus, since the norms of the latter two vectors are approximately two orders of magnitude smaller compared to the norm of $\ve w$, one can say that the conditions which have to hold for the monoclinic group are fulfilled in good approximation, i.e., $\ve w \ne \zero \wedge \ve v_1 = \zero \wedge \ve v_2 = \zero$. This means, $\bte G_1$ and $\bte G_2$ share one and only one (anti)parallel eigenvector in good approximation.
	The vector of active/non-active gates is given by $G(\ve g^\text{chs}) = (1,0,1,1,0,1,1,1,0,1,1,0)$. The gates correspond to the invariant set as introduced in Sect.~\ref{subsec:2xGST2}.

	\begin{figure}
		\centering
		\includegraphics{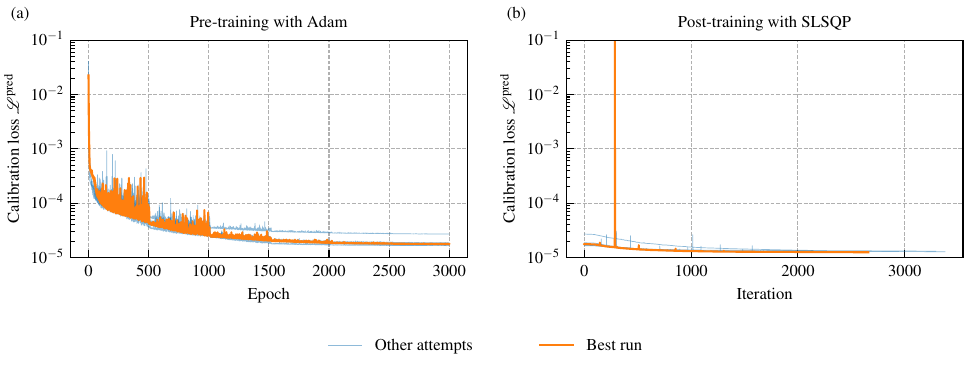}
		\caption{Training process of the invariant-based NN model $\bar \psi^{\bte G}(\bI^{\bte G},\bar J)$ for the RVE plane spheres with the loss $\mathcal L = \mathcal L^\text{pred} + \num{5e-5} \mathcal L^\text{gate}$, $\mathcal L^\text{pred} = 0.7\mathcal L^{\te \sigma} + 0.3 \mathcal L^{\ttttes c}$: (a) pre-training with Adam optimizer and (b) post-training with SLSQP optimizer. Shown is the prediction loss for five training runs.}
		\label{fig:loss_plane}
	\end{figure}
	
	\begin{figure}
		\centering
		\includegraphics{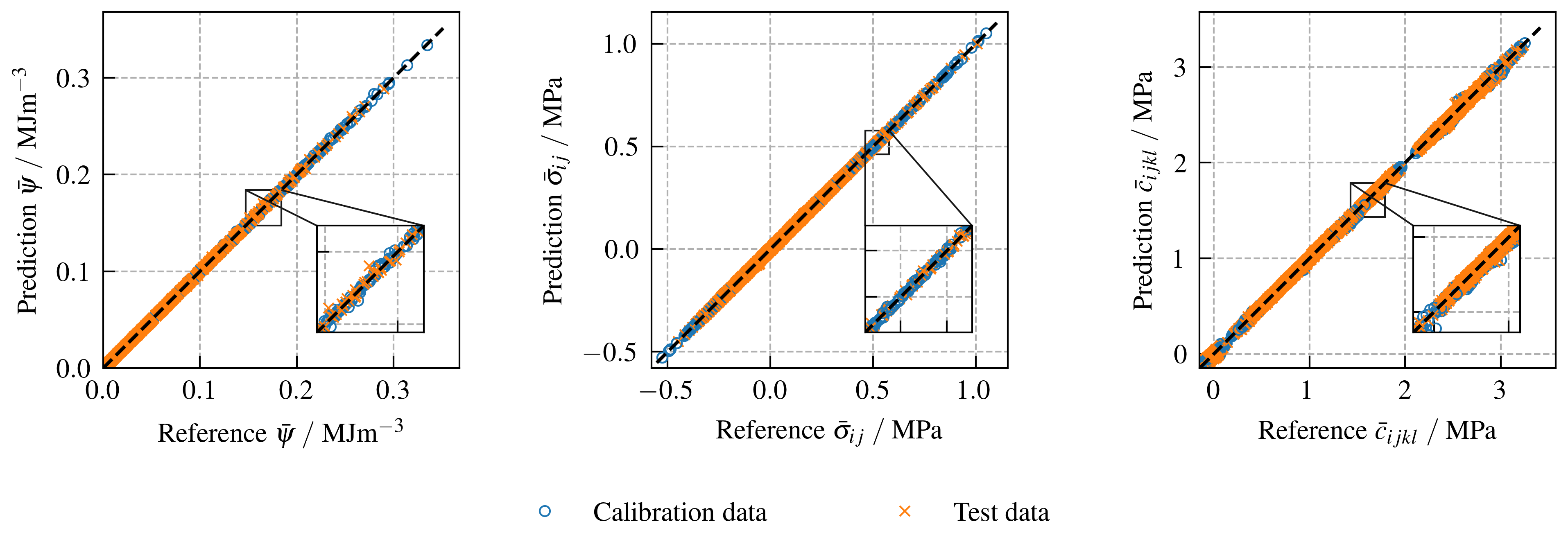}
		\caption{Predictions of the invariant-based NN model $\bar \psi^{\bte G}(\bI^{\bte G},\bar J)$ for the RVE plane spheres compared to reference values: Shown are the energy $\bar \psi\in\R_{\ge 0}$ as well as the coordinates of $\bte \sigma\in\Sym$ and $\bttttes c \in \overline{\Sym}_4$. The ratio of calibration and test data is $70/30$ and the training has been done with the loss $\mathcal L = \mathcal L^\text{pred} + \num{5e-5} \mathcal L^\text{gate}$, where $\mathcal L^\text{pred} = 0.7\mathcal L^{\te \sigma} + 0.3 \mathcal L^{\ttttes c}$.}
		\label{fig:corr_fibers}
	\end{figure}
	
	\paragraph{Training process and model performance}
	The training process is exemplary shown for the RVE \emph{plane spheres} in Fig.~\ref{fig:loss_plane}, where the calibration loss is plotted for all 5 runs with thin lines and the best run is marked with a thick line, respectively.  
	The deviation between the individual runs is relatively small, which demonstrates the robustness of the method.
	As one can be seen in the plots, a very good result is already achieved during pre-training with Adam. Post-training with SLSQP, however, can noticeably reduce the loss again for other RVEs.

	The performance of the calibrated model is depicted in Fig.~\ref{fig:corr_fibers} for energy, stress and material tangent, where the RVE \emph{plane spheres} is exemplary considered here. As one can see, the accuracy of the invariant-based NN model is very good for all three quantities. Thereby, the accurate prediction quality for the material tangent $\bttttes c$ is to be particularly emphasized. 
	As shown in the zoom plot, even this rather difficult to model quantity is predicted with good precision.
	Note that there are only a few studies in the literature in which the NN's predictive quality of this quantity is considered at all, e.g., \cite{Vlassis2021,Eivazi2023,Kalina2024}. Works in which a loss is used for the material tangent and at the same time an energy-based NN model is used are even rarer \cite{Vlassis2021}, which is due to the fact that this requires expensive higher-order Sobolev training. 
	
	To avoid redundancy, the training process and the model predictions are not shown for the other four RVEs here.
	Instead, the final loss terms and the error measures as defined in Eq.~\eqref{eq:error_measures} for all RVEs are given in Tab.~\ref{tab:loss_inter}. There, also the results for the coordinate-based model \eqref{eq:NN_coord} are given for comparison. Although the coordinate-based model achieves slightly higher accuracies for most RVEs, the differences do not play a major role due to the very good precision also achieved with the invariant-based approach. It should also be noted that the coordinate-based model requires 3 instead of two hidden layers in order to achieve acceptable results, cf. the study given in \ref{app:study_hyper}.
	
	\begin{table}
		\begin{center}
			\caption{Interpolation study on the performance of the developed invariant-based NN models $\bar \psi^\square(\bI^\square,\bar J)$ and the coordinate-based reference model $\bar \psi^\text{coord}(\bte C, \bar J)$ for the five considered RVEs and the prediction loss $\mathcal L^\text{pred} = 0.7\mathcal L^{\te \sigma} + 0.3 \mathcal L^{\ttttes c}$. The loss term for training the invariant-based models was $\mathcal L = \mathcal L^\text{pred} + \num{5e-5} \mathcal L^\text{gate}$ and an architecture with two hidden layers with 16 neurons each was chosen. The loss term for the coordinate-based model was $\mathcal L = \mathcal L^\text{pred}$ and an architecture with three hidden layers with 16 neurons each was chosen. The overall datasets were divided into calibration and test sets with a ratio of $70/30$, respectively. All models were trained 5 times, where the best training run was selected. Given are the loss values after training and the error measures for $\bar \psi$, $\bte \sigma$ and $\bttttes c$, cf. Eqs.~\eqref{eq:loss_sigma}, \eqref{eq:loss_c} and \eqref{eq:error_measures}.}
			\label{tab:loss_inter}
			\begin{footnotesize}
				\begin{tabular}{lllllllll}
					RVE & Model & $\square$ & Active gates & $\mathcal L^\text{pred}_\text{cal}$ & $\mathcal L^\text{pred}_\text{test}$ &  $\epsilon^\psi / \%$ & $\epsilon^{\te \sigma}/ \%$ & $\epsilon^{\ttttes c}/ \%$\\
					\hline\hline
					\multirow{2}{*}{Stochastic fibers} & $\bar \psi^\square(\bI^\square,\bar J)$ & $\bte G$ & 4 & \num{2.520e-06} & \num{3.031e-06} & \num{0.19} & \num{0.39} & \num{0.19}\\   
					& $\bar \psi^\text{coord}(\bte C, \bar J)$ & -- & -- & \num{4.145e-06} & \num{6.305e-06} & \num{0.11} & \num{0.31} & \num{0.29}\\ 
					\hline
					\multirow{2}{*}{Hexagonal fibers} & $\bar \psi^\square(\bI^\square,\bar J)$ & $\btttttte G$ & 9 & \num{2.059e-05} & \num{2.352e-05} & \num{0.39} & \num{0.91} & \num{0.47}\\ 
					& $\bar \psi^\text{coord}(\bte C, \bar J)$ & -- & -- & \num{8.319e-06} & \num{1.063e-05} & \num{0.13} & \num{0.42} & \num{0.39}\\
					\hline
					\multirow{2}{*}{Cubic sphere} & $\bar \psi^\square(\bI^\square,\bar J)$ & $\btttte G$ & 8 & \num{2.265e-05} & \num{2.341e-05} & \num{0.27} & \num{0.8} & \num{0.62}\\ 
					& $\bar \psi^\text{coord}(\bte C, \bar J)$ & -- & -- & \num{2.488e-05} & \num{3.363e-05} & \num{0.22} & \num{0.8} & \num{0.85}\\ 
					\hline
					\multirow{2}{*}{Plane spheres} & $\bar \psi^\square(\bI^\square,\bar J)$ & $\bte G$ & 5 & \num{1.349e-05} & \num{1.614e-05} & \num{0.41} & \num{0.92} & \num{0.44}\\ 
					& $\bar \psi^\text{coord}(\bte C, \bar J)$ & -- & -- & \num{4.563e-06} & \num{5.448e-06} & \num{0.1} & \num{0.31} & \num{0.29}\\
					\hline
					\multirow{2}{*}{Chain spheres} & $\bar \psi^\square(\bI^\square,\bar J)$ & $(\bte G_1,\bte G_2)
					$ & 8 & \num{2.037e-05} & \num{2.242e-05} & \num{0.37} & \num{0.92} & \num{0.54}\\ 
					& $\bar \psi^\text{coord}(\bte C, \bar J)$ & -- & -- & \num{9.665e-06} & \num{1.297e-05} & \num{0.13} & \num{0.42} & \num{0.42}\\ 
					\hline
				\end{tabular}
			\end{footnotesize}
		\end{center}
	\end{table}

\begin{figure}
		\centering
		\includegraphics{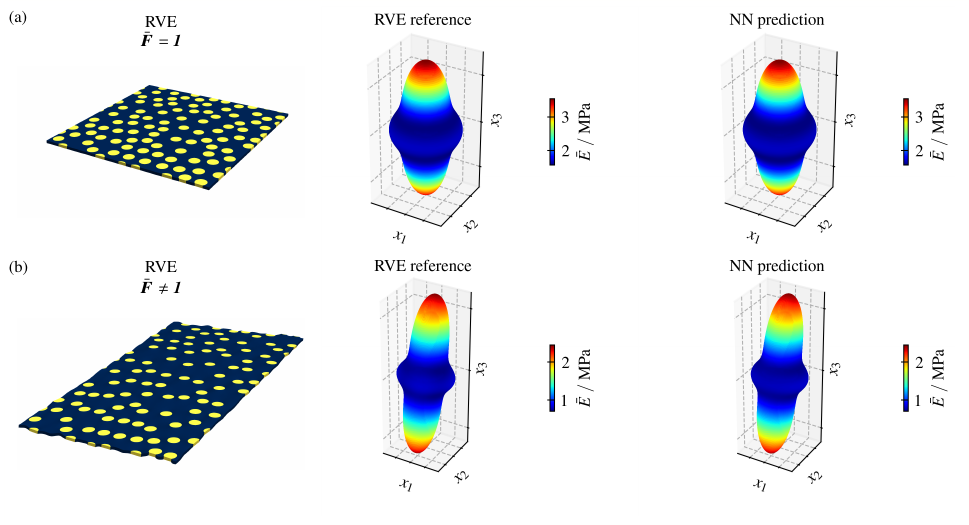}
		\caption{RVE stochastic fibers and corresponding elastic surfaces from the homogenized tangent tensor $\bttttes c$ as well as the NN prediction from the model $\bar \psi^{\bte G}(\bI^{\bte G},\bar J)$: (a) undeformed state with $\bte F = \one$ and (b) deformed state with $\bte F\ne \one$ as given in Eq.~\eqref{eq:F_surface}.}
		\label{fig:surface_fibers_int}
	\end{figure}

\begin{figure}
		\centering
		\includegraphics{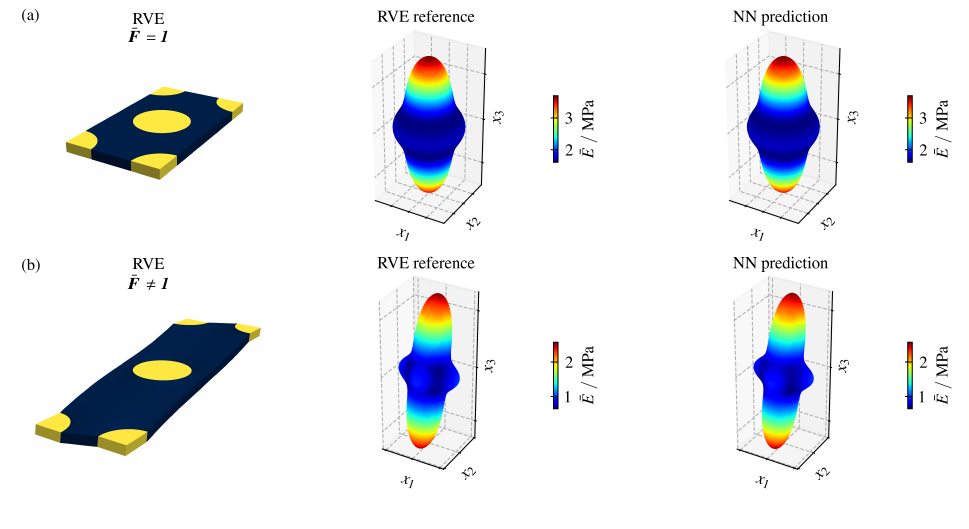}
		\caption{RVE hexagonal fibers and corresponding elastic surfaces from the homogenized tangent tensor $\bttttes c$ as well as the NN prediction from the model $\bar \psi^{\btttttte G}(\bI^{\btttttte G},\bar J)$: (a) undeformed state with $\bte F = \one$ and (b) deformed state with $\bte F\ne \one$ as given in Eq.~\eqref{eq:F_surface}.}
		\label{fig:surface_fibershex_int}
	\end{figure}

\begin{figure}
		\centering
		\includegraphics{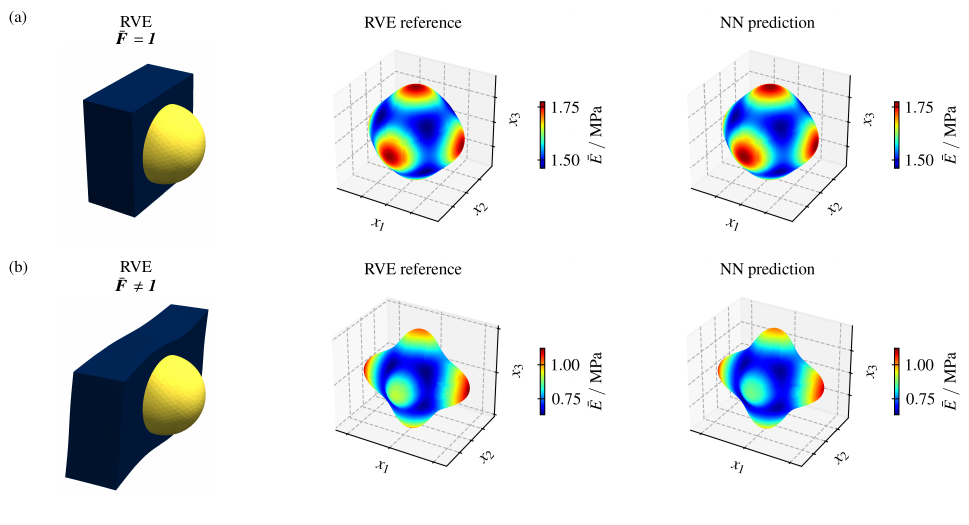}
		\caption{RVE cubic sphere and corresponding elastic surfaces from the homogenized tangent tensor $\bttttes c$ as well as the NN prediction from the model $\bar \psi^{\btttte G}(\bI^{\btttte G},\bar J)$: (a) undeformed state with $\bte F = \one$ and (b) deformed state with $\bte F\ne \one$ as given in Eq.~\eqref{eq:F_surface}.}
		\label{fig:surface_cubic_int}
	\end{figure}

\begin{figure}
		\centering
		\includegraphics{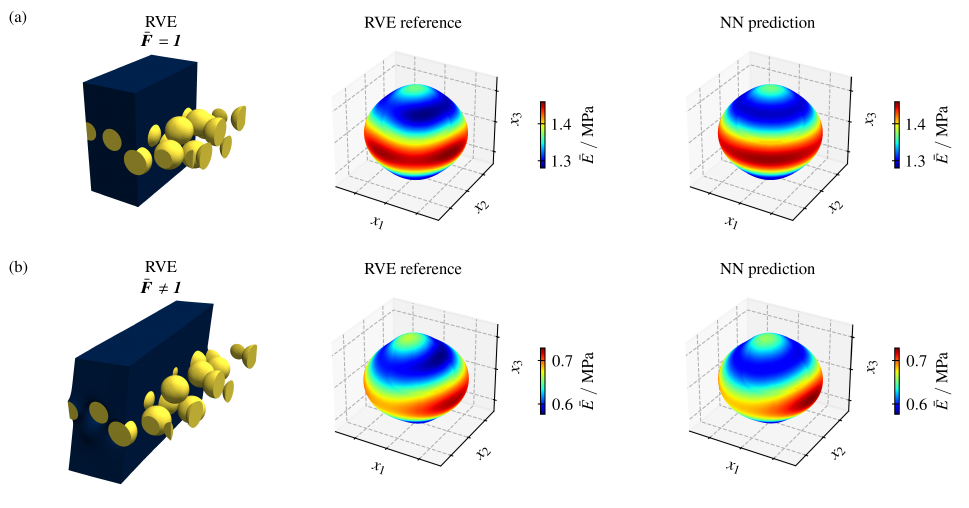}
		\caption{RVE plane spheres and corresponding elastic surfaces from the homogenized tangent tensor $\bttttes c$ as well as the NN prediction from the model $\bar \psi^{\bte G}(\bI^{\bte G},\bar J)$: (a) undeformed state with $\bte F = \one$ and (b) deformed state with $\bte F\ne \one$ as given in Eq.~\eqref{eq:F_surface}.}
		\label{fig:surface_plane_int}
	\end{figure}

\begin{figure}
		\centering
		\includegraphics{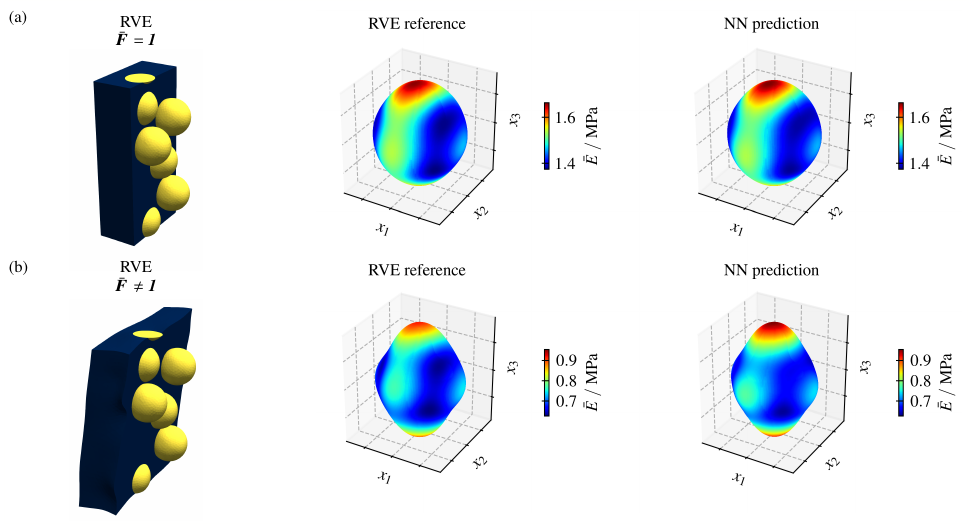}
		\caption{RVE chain spheres and corresponding elastic surfaces from the homogenized tangent tensor $\bttttes c$ as well as the NN prediction from the model $\bar \psi^{\bte G_1,\bte G_2}(\bI^{\bte G_1,\bte G_2},\bar J)$: (a) undeformed state with $\bte F = \one$ and (b) deformed state with $\bte F\ne \one$ as given in Eq.~\eqref{eq:F_surface}.}
		\label{fig:surface_chain_int}
	\end{figure}

	\paragraph{Elastic surface plots}
	Finally, after the discussion of the achieved loss terms and error measures, we want to focus on the material tangents corresponding to the different RVEs.
	To visualize the effective anisotropic behavior of the four considered RVEs and to illustrate the high prediction quality of the calibrated NN models for the effective 4th order tenors $\bttttes c$, the technique described in Nordmann~et~al.~\cite{Nordmann2018} to visualize the material's effective Young's modulus in all spatial directions is applied. We call such a plot \emph{elastic surface} in the following. Since $\bttttes c(\bte F)$ depends on the deformation within the considered nonlinear setting, it is not constant.  
	Thus, the material tangent is not only visualized for the undeformed state, i.e., $\bttttes c(\bte F=\one)$ which is the elasticity tensor, but also for a state with $\bte F\ne \one$. Exemplary, the state 
	\begin{align}
		[\bte F] = 
		\begin{bmatrix}
			0.89 & 0.09 & 0.02 \\
			0.09 & 1.4 & 0.18 \\
			0.02 & 0.18 & 0.94
		\end{bmatrix}
		\label{eq:F_surface}
	\end{align}
	is chosen here. The 3D plots of the Young's modulus are given in Figs.~\ref{fig:surface_fibers_int}--\ref{fig:surface_chain_int} for the five different RVEs, see \cite{Nordmann2018} for details on this visualization technique. As can be clearly seen, each RVE is characterized by a special anisotropy. All RVEs show a significant change in the elastic surface as a result of the imposed deformation. Both the shape of the surface shown and the amount of the maximum modulus $\bar E$ change significantly. This effect is partly due to the already nonlinear behavior of the individual phases given by Eq.~\eqref{eq:comp}, but mainly due to the change in the microstructure as a result of the applied deformation.
	For both, undeformed as well as deformed state, the plots underpin the excellent prediction quality of the invariant-based NN models.
	
	Once again it should be noted that a smaller NN architecture compared to the coordinate-based model is needed to achieve a similar accuracy in the predictions with the invariant-based approach. In addition, the trained model can be interpreted in a certain way, since the anisotropy group can also be inferred beyond the linear elastic case by analyzing the structure tensors. Thus, the advantage of our approach becomes clear even from this first study.
	
	\subsubsection{Interpolation behavior: Training with stress}
	In order to show that the underlying anisotropy and thus a meaningful structural tensor can be identified with the proposed approach by using deformation-stress tuples alone, only $\mathcal L^\text{pred} = \mathcal L^{\te \sigma}$ is now used as loss. The overall datasets are again divided into calibration and test sets with a ratio of $70/30$, respectively, see Eq.~\eqref{eq:cal_test}. In order to decide when the selection procedure shown in Fig.~\ref{fig:ident_proc} terminates, the only the error measures $\epsilon^\psi$ and $\epsilon^{\te \sigma}$, so not $\epsilon^{\ttttes c}$, are chosen since the material tangent is not directly trained within this study, cf. Eq~\eqref{eq:error_measures} for the definition of the errors.
	
	We will discuss the results for the RVE \emph{plane spheres} as an example, cf. Fig.~\ref{fig:RVEs}(d). For this RVE, the following structure tensor was identified after calibration with $\mathcal L^\text{pred} = \mathcal L^{\te \sigma}$:
	\begin{align}
		[\bte G^\text{pls}] = 
		\begin{bmatrix}
			0.08 &  0.  & -0.  \\
			0.  & 0.07 & 0.02 \\
			-0.  &  0.02 &  0.85
		\end{bmatrix} \; .
	\end{align}
	Again, as in the former study, the invariant-based NN model was able to identify the $X_3$ axis as the preferred direction and to find a valid structure tensor. Note that the difference to the identified structure tensor given in Eq.~\eqref{eq:structure2ndorder} is not problematic, since both structure tensors are equivalent, see Footnote~\ref{foot:transverse}. The structure tensor describes \emph{transverse isotropy} $\mathcal G_{13}$ since it is characterized by two approximately equal eigenvalues: $(\lambda_1,\lambda_2,\lambda_3)^\text{pls}=(0.08, 0.07, 0.85)$. In addition, the vector of active/non-active gates is given by $G(\ve g^\text{pls}) =(1,1,1,1,0,0,1)$, i.e., the additional invariants needed to describe orthotropy are removed from the model by the gates.
	
	\begin{figure}
		\centering
		\includegraphics{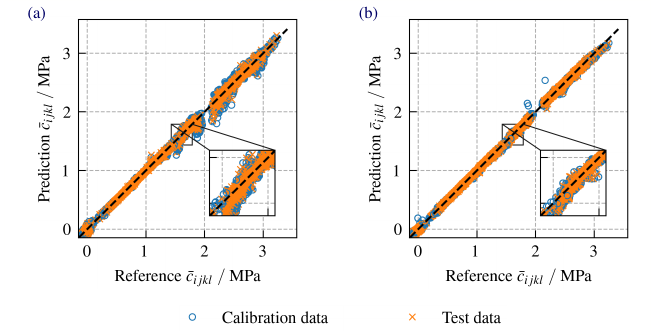}
		\caption{Model predictions of the material tangent $\bttttes c \in \overline{\Sym}_4$ compared to reference values from the RVE plane spheres for NN models trained with the prediction loss $\mathcal L^\text{pred} = \mathcal L^{\te \sigma}$: (a) coordinate-based NN model $\bar \psi^\text{coord}(\bte C,\bar J)$ and (b) invariant-based NN model $\bar \psi^{\bte G}(\bI^{\bte G},\bar J)$. The ratio of calibration and test data is $70/30$.}
		\label{fig:corr_c_fibers}
	\end{figure}
	
	Comparing the predictions of the invariant-based model with the coordinate-based model, it can first be observed that similarly good results for energy and stress are achieved with both models, which is not surprising given the selected loss, see Tab.~\ref{tab:loss_inter_sig} third line. However, as shown in Fig.~\ref{fig:corr_c_fibers}, the accuracy in the material tangent, which was not directly used for training, is noticeably better if the invariant-based NN model is used. 
	This is due to the fact that, if the structure tensor is identified correctly, the model structure ensures the material symmetry condition~\eqref{eq:symmetry} to be fulfilled and is therefore much more suitable for the specific problem of describing the RVE's effective constitutive behavior.
	To be more precise, referring to Eq.~\eqref{eq:constInv}${}_1$, the stress results from the derivative of the energy with respect to the invariants multiplied with derivatives of the invariants with respect to the deformations, the tensor generators \cite{Kalina2022a}. This is similar for the material tangent, which is obtained by further derivation, cf. Eq.~\eqref{eq:constInv}${}_2$.
	
	For the other RVEs considered within this work, similar results are achieved. The final prediction loss values and the error measures defined in Eq.~\eqref{eq:error_measures} are given in Tab.~\ref{tab:loss_inter_sig}. As can be seen there, the invariant-based NN approach is superior regarding the error $\epsilon^{\ttttes c}$ in the material tangent over the coordinate-based model for all RVEs expect for \emph{chain spheres}. 
	
	\begin{table}
		\begin{center}
			\caption{Interpolation study on the performance of the developed invariant-based NN models $\bar \psi^\square(\bI^\square,\bar J)$ and the coordinate-based reference model $\bar \psi^\text{coord}(\bte C, \bar J)$ for the five considered RVEs and the prediction loss $\mathcal L^\text{pred} = \mathcal L^{\te \sigma}$. The loss term for training the invariant-based models was $\mathcal L = \mathcal L^\text{pred} + \num{5e-5} \mathcal L^\text{gate}$ and an architecture with two hidden layers with 16 neurons each was chosen. The loss term for the coordinate-based model was $\mathcal L = \mathcal L^\text{pred}$ and an architecture with three hidden layers with 16 neurons each was chosen. The overall datasets were divided into calibration and test sets with a ratio of $70/30$, respectively. All models were trained 5 times, where the best training run was selected. Given are the loss values after training and the error measures for $\bar \psi$, $\bte \sigma$ and $\bttttes c$, cf. Eqs.~\eqref{eq:loss_sigma} and \eqref{eq:error_measures}.}
			\label{tab:loss_inter_sig}
			\begin{footnotesize}
				\begin{tabular}{lllllllll}
					RVE & Model & $\square$ & Active gates & $\mathcal L^\text{pred}_\text{cal}$ & $\mathcal L^\text{pred}_\text{test}$ &  $\epsilon^\psi / \%$ & $\epsilon^{\te \sigma}/ \%$ & $\epsilon^{\ttttes c}/ \%$\\
					\hline\hline
					\multirow{2}{*}{Stochastic fibers} & $\bar \psi^\square(\bI^\square,\bar J)$ & $\bte G$ & 4 & \num{1.771e-06} & \num{2.593e-06} & \num{0.19} & \num{0.41} & \num{0.28}\\   
					& $\bar \psi^\text{coord}(\bte C, \bar J)$ & -- & -- & \num{6.746e-07} & \num{1.650e-06} & \num{0.08} & \num{0.29} & \num{1.1}\\ 
					\hline
					\multirow{2}{*}{Hexagonal fibers} & $\bar \psi^\square(\bI^\square,\bar J)$ & $\btttttte G$ & 7 & \num{1.194e-05} & \num{1.679e-05} & \num{0.38} & \num{0.93} & \num{0.96}\\ 
					& $\bar \psi^\text{coord}(\bte C, \bar J)$ & -- & -- & \num{1.728e-06} & \num{3.592e-06} & \num{0.1} & \num{0.43} & \num{1.39}\\
					\hline
					\multirow{2}{*}{Cubic sphere} & $\bar \psi^\square(\bI^\square,\bar J)$ & $\btttttte G$ & 8 & \num{1.452e-05} & \num{2.427e-05} & \num{0.22} & \num{0.76} & \num{1.24}\\ 
					& $\bar \psi^\text{coord}(\bte C, \bar J)$ & -- & -- & \num{6.994e-06} & \num{1.187e-05} & \num{0.16} & \num{0.64} & \num{2.03}\\ 
					\hline
					\multirow{2}{*}{Plane spheres} & $\bar \psi^\square(\bI^\square,\bar J)$ & $\bte G$ & 5 & \num{1.000e-05} & \num{1.314e-05} & \num{0.41} & \num{0.92} & \num{0.61}\\ 
					& $\bar \psi^\text{coord}(\bte C, \bar J)$ & -- & -- & \num{1.002e-06} & \num{2.238e-06} & \num{0.06} & \num{0.29} & \num{0.97}\\
					\hline
					\multirow{2}{*}{Chain spheres} & $\bar \psi^\square(\bI^\square,\bar J)$ & $(\bte G_1,\bte G_2)
					$ & 7 & \num{1.274e-05} & \num{1.462e-05} & \num{0.31} & \num{0.89} & \num{1.25}\\ 
					& $\bar \psi^\text{coord}(\bte C, \bar J)$ & -- & -- & \num{1.820e-06} & \num{3.056e-06} & \num{0.08} & \num{0.35} & \num{1.06}\\ 
					\hline
				\end{tabular}
			\end{footnotesize}
		\end{center}
	\end{table}

	\subsubsection{Extrapolation behavior}
	
\begin{figure}
		\begin{center}
			\includegraphics{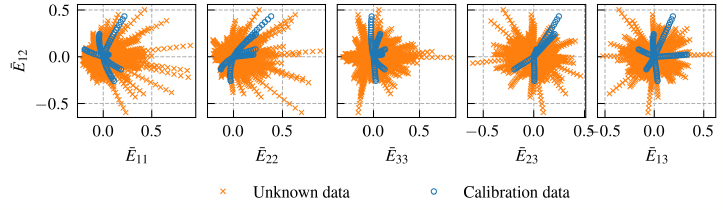}
		\end{center}
		\caption{Dataset $\mathcal D$ for the extrapolation study with: sampled deformation space visualized in five sectional planes of the Green-Lagrange strain tensor. Only 6 load cases out of $\mathcal D$ comprising a total number of 120 tuples are used for calibration.}
		\label{fig:ext_strain}
	\end{figure}

	After the detailed analysis of the model's ability to learn the effective constitutive behavior of different anisotropic RVEs from a large dataset, the \emph{extrapolation behavior} is now studied. 
	To mimic the situation of sparse data for training, only six load cases with 20 increments each are used. For visualization, the chosen load paths are shown in Fig.~\ref{fig:ext_strain} within five sectional planes of the coordinates of the Green-Lagrange strain tensor $\bte E$. Accordingly, large parts of the strain space are not covered by the sparse calibration data for the extrapolation study. In order to decide when the selection procedure shown in Fig.~\ref{fig:ident_proc} terminates, the error measures $\epsilon^\psi$, $\epsilon^{\te \sigma}$ and $\epsilon^{\ttttes c}$ are now calculated using only the calibration data $\mathcal D_\text{cal}$, cf. Eq~\eqref{eq:error_measures}.

	\begin{figure}
		\centering
		\includegraphics{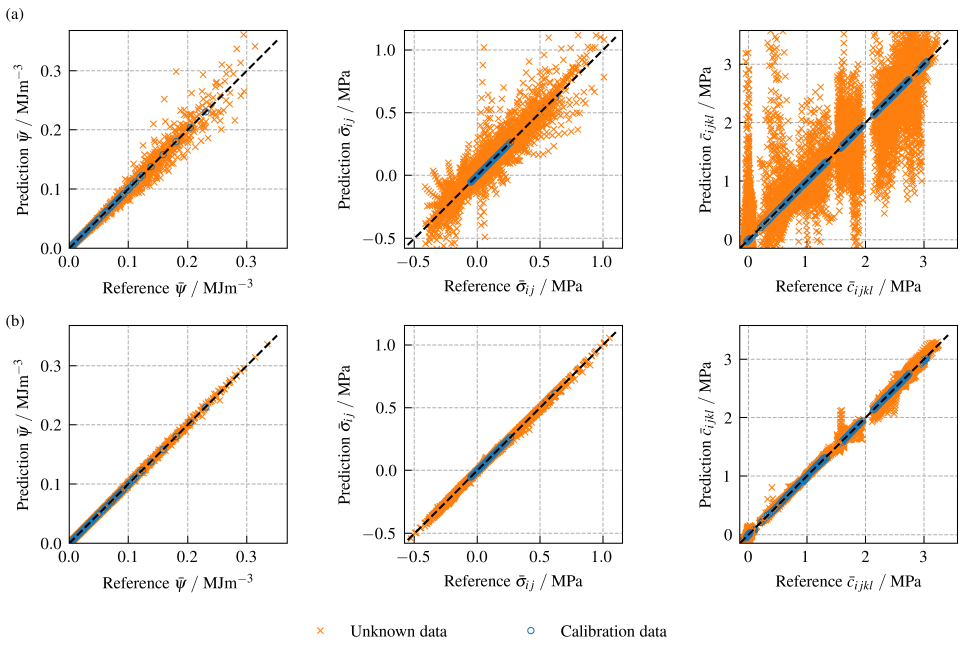}
		\caption{Predictions for the extrapolation behavior of NN-based models for the RVE plane spheres: (a) coordinate-based model $\bar \psi^\text{coord}(\bte C, \bar J)$ with three hidden layers including 16 neurons each and (b) invariant-based model $\bar \psi^{\bte G}(\bI^{\bte G},\bar J)$ with two hidden layers including 16 neurons each.
			The prediction loss was $\mathcal L^\text{pred} = 0.7\mathcal L^{\te \sigma} + 0.3 \mathcal L^{\ttttes c}$ and the loss term for training the invariant-based models was $\mathcal L = \mathcal L^\text{pred} + \num{5e-5} \mathcal L^\text{gate}$. Only six load cases were used for calibration. All models were trained 5 times,  where the best training run was selected.}
		\label{fig:corr_sig_c_ext}
	\end{figure}

	To compare the predictive capability of the invariant-based NN approach and the coordinate-based reference NN model, we again consider the RVE \emph{plane spheres} exemplarily. Even for the sparse data set used for calibration, a reasonable structure tensor was identified during training:
	\begin{align}
		[\bte G^\text{pls}] = 
		\begin{bmatrix}
			0.5 & -0.01 & -0.01 \\
			-0.01 &  0.5 & -0.01 \\
			-0.01 & -0.01 &  0.  
		\end{bmatrix} \; .
	\end{align}
	Thus, the underlying \emph{transverse isotropy} $\mathcal G_{13}$ with the $X_3$-axis as preferred direction has been detected. Also the vector of active/non-active gates given by $G(\ve g^\text{pls})=(1,1,1,0,0,1,1)$ corresponds to this since two reducible invariants have been removed from the model by the gates.
	
	The predictions for energy, stress and material tangent of the coordinate-based model $\bar \psi^\text{coord}(\bte C,\bar J)$ and the invariant-based models $\bar \psi^\square(\bI^\square,\bar J)$ trained with the reduced data and the prediction loss $\mathcal L^\text{pred}:= \mathcal L^{\te \sigma} + \mathcal L^{\ttttes c}$ are given in Fig.~\ref{fig:corr_sig_c_ext}(a),(b).
	As shown there, the calibration data have been learned quite well by both models. However, the predictions for the unknown data, which require the model to extrapolate from the calibration domain, are very poor for the coordinate-based model, cf. Fig.~\ref{fig:corr_sig_c_ext}(a). In contrast, as shown in  Fig.~\ref{fig:corr_sig_c_ext}(b), the extrapolation behavior of the invariant-based approach is significantly improved. 
	
	As shown by the loss terms for the test data given in Tab.~\ref{tab:loss_ext_sigma_c}, the invariant-based approach shows a clearly better extrapolation behavior compared to the coordinated-based approach for all five considered RVEs. The test loss is consistently one or even two orders of magnitude smaller. This advantageous behavior can again be attributed to the fact that the material symmetry is built into the model by construction through the use of structure tensors and invariants. At least if a meaningful structure tensor is detected during training with the sparse data set. 
	It should be noted that for the RVEs \emph{hexagonal fibers} and \emph{cubic sphere}, the algorithm given in Fig.~\ref{fig:ident_proc} did not select 6th and 4th order structure tensors as in the interpolation study, cf. \ref{subsec:interpolation}. This is due to the fact that the tolerance in the error measures $\epsilon^\psi$, $\epsilon^{\te \sigma}$ and $\epsilon^{\ttttes c}$ for the calibration data was not undershot. Nevertheless, although the expected structural tensor set could not be detected for these two cases and a set comprising two 2nd order structure tensors was chosen instead, the extrapolation behavior is still noticeably better than with the coordinate-based approach.

	\begin{table}
		\begin{center}
			\caption{Extrapolation study on the performance of the developed invariant-based NN models $\bar \psi^\square(\bI^\square,\bar J)$ and the coordinate-based reference model $\bar \psi^\text{coord}(\bte C, \bar J)$ for the five considered RVEs and the prediction loss $\mathcal L^\text{pred} = 0.7\mathcal L^{\te \sigma} + 0.3 \mathcal L^{\ttttes c}$. The loss term for training the invariant-based models was $\mathcal L = \mathcal L^\text{pred} + \num{5e-5} \mathcal L^\text{gate}$ and an architecture with two hidden layers with 16 neurons each was chosen. The loss term for the coordinate-based model was $\mathcal L = \mathcal L^\text{pred}$ and an architecture with three hidden layers with 16 neurons each was chosen. Only six load cases were used for calibration. All models were trained 5 times, where the best training run was selected. Given are the prediction loss values after training, cf. Eqs.~\eqref{eq:loss_sigma} and \eqref{eq:loss_c}.}
			\label{tab:loss_ext_sigma_c}
			\begin{footnotesize}
				\begin{tabular}{llllll}
					RVE & Model & $\square$ & Active gates & $\mathcal L^\text{pred}_\text{cal}$ & $\mathcal L^\text{pred}_\text{test}$ \\
					\hline\hline
					\multirow{2}{*}{Stochastic fibers} & $\bar \psi^\square(\bI^\square,\bar J)$ & $\bte G$ & 5 & \num{7.759e-06} & \num{1.141e-04}\\   
					& $\bar \psi^\text{coord}(\bte C, \bar J)$ & -- & -- & \num{2.019e-07} & \num{1.117e-02}\\ 
					\hline
					\multirow{2}{*}{Hexagonal fibers} & $\bar \psi^\square(\bI^\square,\bar J)$ & $(\bte G_1,\bte G_2)
					$ & 8 & \num{2.115e-06} & \num{1.003e-03}\\ 
					& $\bar \psi^\text{coord}(\bte C, \bar J)$ & -- & -- & \num{3.740e-07} & \num{1.522e-02}\\
					\hline
					\multirow{2}{*}{Cubic sphere} & $\bar \psi^\square(\bI^\square,\bar J)$ & $(\bte G_1,\bte G_2)
					$ & 6 & \num{2.214e-05} & \num{3.329e-03}\\ 
					& $\bar \psi^\text{coord}(\bte C, \bar J)$ & -- & -- & \num{5.511e-07} & \num{1.063e-02}\\ 
					\hline
					\multirow{2}{*}{Plane spheres} & $\bar \psi^\square(\bI^\square,\bar J)$ & $\bte G$ & 5 & \num{1.195e-05} & \num{1.248e-04}\\ 
					& $\bar \psi^\text{coord}(\bte C, \bar J)$ & -- & -- & \num{2.904e-07} & \num{1.590e-02}\\
					\hline
					\multirow{2}{*}{Chain spheres} & $\bar \psi^\square(\bI^\square,\bar J)$ & $(\bte G_1,\bte G_2)
					$ & 7 & \num{9.537e-06} & \num{4.285e-04}\\ 
					& $\bar \psi^\text{coord}(\bte C, \bar J)$ & -- & -- & \num{3.246e-07} & \num{1.260e-02}\\ 
					\hline
				\end{tabular}
			\end{footnotesize}
		\end{center}
	\end{table}
	
	\section{Conclusions}
	\label{sec:conc}
	
	In the present work, an NN-based approach for the automated modeling of anisotropic finite strain elasticity including detection of anisotropy type and orientation is proposed. To this end, an invariant-based approach is formulated in such a way that important physical conditions are fulfilled a priori, i.e., by construction, whereby these models are denoted as PANNs \cite{Kalina2024,Klein2022,Linden2023}. The invariants are built from the deformation tensor and generalized structure tensor(s), where 2nd, 4th and 6th order structure tensors are used to cover a wide range of anisotropies. Thus, special attention is paid to the principle of material symmetry.
	The ability of our approach is investigated and compared to an NN model based on the coordinates of the deformation tensor by a calibration to data generated via computational homogenization of five different RVEs.
	
	The article begins with a brief review of the kinematics of finite strains and stress measures in continuum solid mechanics, a summary of common principles in hyperelasticity, the concept of structure tensors and a scale transition scheme. 
	In the following section, the generalized structure tensor approach \cite{Gasser2006} is introduced, where 2nd, 4th and 6th order tensors are given. Based on this theoretical foundation, PANNs are formulated that are based on parameterized versions of the generalized structure tensors, using a network with an additional gate layer and a penalty loss of $p$-norm type \cite{Flaschel2021} to enforce sparsity with respect to the number of invariants involved in the model. Furthermore, a strategy is presented to decide which set of structure tensor(s) is required.
	Finally, the invariant-based NN models are used as macroscopic surrogates for computational homogenizations in an examples section, where a comparison with a model based on the coordinates of the right Cauchy-Green deformation tensor is included. A detailed analysis of the interpolation and extrapolation capability of the models was carried out with the generated data sets.
	It turned out that both approaches, the invariant-based one and the coordinate-based one, deliver very good results for the interpolation case. However, the presented invariant-based approach requires fewer hidden layers for an equivalent prediction quality. This applies to the energy, the stress and the material tangent tensor.	
	In terms of extrapolation behavior, the presented invariant-based approach clearly beats the coordinate-based model. Even with a sparse data set, the prediction quality for states not included in the training remains acceptable, whereas the coordinate-based model exhibits huge errors in the extrapolation regime.
	
	In summary, the presented PANN based on invariants formulated with generalized structural tensors is a very accurate surrogate model for computationally expensive RVE simulations in finite strain elasticity. The a priori inclusion of principles from constitutive modeling in the NN model, in particular the \emph{principle of material symmetry}, ensures that the underlying physics is not violated even during extrapolation, thus guaranteeing good generalization. This also enables comparatively small network architectures. The use of $\ell_p$ regularization enables the elimination of unneeded invariants from the model.
	
	Various applications and extensions of our approach are planned for the future. For example, an additional sparsification of the network as done in \cite{Fuhg2024a,McCulloch2024} is possible. Furthermore, in order to exploit the advantages of the developed invariant-based PANN approach, an integration into multiscale schemes as \emph{FE}${}^\textit{ANN}$ \cite{Kalina2023} and the application to real experimental data are planned. In addition, an extension of our approach to further anisotropy classes is possible, which requires the use of sets of structure tensors.
	Finally, the application of the developed concepts in the NN-based modeling of inelastic material behavior \cite{Rosenkranz2024,Abdolazizi2023a,Tac2024a,Holthusen2024,Vlassis2021,Malik2021,Bahtiri2024,Boes2024}, damage \cite{Tac2024,Zlatic2024a} and coupled problems \cite{Klein2024,Kalina2024,Zlatic2023,Fuhg2024b} is possible.

	\section*{Acknowledgment}
	All presented computations were performed on a PC-Cluster at the Center for Information Services and High Performance Computing (ZIH) at TU Dresden. The authors thus thank the ZIH for generous allocations of computer time. This work was supported by a postdoc
	fellowship of the German Academic Exchange Service (DAAD) to Karl A. Kalina. This support is gratefully acknowledged. Finally, the authors thank the German Research Foundation (DFG) for the support within the Research Training Group GRK 2868 D${}^3$--Project Number 493401063.
	
	\section*{CRediT authorship contribution statement}
	\textbf{Karl A. Kalina:} Conceptualization, Formal analysis, Investigation, Methodology, Visualization, Software, Validation, Visualization, Writing -- original draft, Writing -- review \& editing, Funding acquisition. 
	\textbf{J\"{o}rg Brummund:} Formal analysis, Writing – review \& editing.
	\textbf{WaiChing Sun:} Conceptualization, Resources, Writing -- review \& editing.
	\textbf{Markus Kästner:} Resources, Writing -- review \& editing, Funding acquisition.
	
	\appendix
	
	\section{Structure tensors and invariant sets for specific symmetry groups}
	\label{app:sym_groups}
	Within this appended section, we summarize and discuss the structure tensors and invariants sets for specific symmetry groups. As the numbering of the groups is not uniform in the literature, we use the Schoenflies notation in addition to the group number. In the following, structure tensors and invariants are given for: \emph{isotropy} $\mathcal G_{14}$ ($\mathcal K_{h} = \mathcal O(3)$), \emph{triclinic} anisotropy $\mathcal G_1$ ($\mathcal C_i$), \emph{monoclinic} anisotropy $\mathcal G_2$ ($\mathcal C_{2h}$), \emph{orthotropy} $\mathcal G_{3}$ ($\mathcal D_{2h}$), \emph{tetragonal} anisotropy $\mathcal G_{5}$ ($\mathcal D_{4h}$), \emph{cubic} anisotropy $\mathcal G_{7}$ ($\mathcal O_{h}$), \emph{hexagonal} anisotropy $\mathcal G_{11}$ ($\mathcal D_{6h}$), as well as \emph{transverse isotropy} $\mathcal G_{13}$ ($\mathcal D_{\infty h}$).
	
	For the following discussions, we make use of the spectral decomposition of a symmetric 2nd order tensor 
	\begin{align}
		\te S = \sum_{\alpha=1}^n \lambda_\alpha \te P_\alpha \; \text{with} \; \te P_\alpha \cdot \te P_\beta = \delta_{\alpha(\beta)} \te P_\beta \text{ and } \sum_{\alpha = 1}^n \te P_\alpha = \one  
		\label{eq:projection}
	\end{align}
	with $n\in\{1,2,3\}$ non-equal eigenvalues. The projection tensors $\te P_\alpha \in \Sym$ related to $\te S$ can be represented by Sylvester's formula \cite{Itskov2015}
	\begin{align}
		\te P_\alpha = \delta_{1n} \one + \prod_{\substack{\beta=1\\\beta \ne \alpha}}^n \frac{\te S - \lambda_\beta \one}{\lambda_\alpha - \lambda_\beta} \; .
		\label{eq:sylvester}
	\end{align}
	Furthermore, we use the Cayley-Hamilton theorem \cite{Itskov2015}, which states that a 2nd order tensor fulfills its own eigenvalue equation, i.e.,
	\begin{align}
		\te S^3 - I_1 \te S^2 + I_2 \te S - I_3 \one = \zero \text{ with } I_1 = \tr \te S, \; I_2 = \tr(\cof \te S),\; I_3 = \det \te S \; ,
		\label{eq:CayleyHamilton}
	\end{align}
	where $I_1$, $I_2$, $I_3$ are the principal invariants.
	\subsection{Isotropy $\mathcal G_{14}$}
	By applying the Cayley-Hamilton theorem~\eqref{eq:CayleyHamilton}, one can find the well known relation that the invariant set $\bar J_1 := \tr \bte C$, $\bar J_2 := \frac{1}{2}\tr \bte C^2$ , $\bar J_3 := \frac{1}{3}\tr \bte C^3$ can be expressed by $\bar I_1 := \tr \bte C$, $\bar I_2 := \tr (\cof \bte C)$ , $\bar I_3 := \det \bte C$ as 
	\begin{align}
		\bar J_1 = \bar I_1,\; \bar J_2 = \frac{1}{2} \bar I_1^2 - \bar I_2,\; \bar J_3 = \frac{1}{3} \bar I_1^3 - \bar I_1\bar I_2 + \bar I_3 \; .
	\end{align}
	It is thus equivalent to use the sets $\bI^{\text{iso}} = (\bar I_1,\bar I_2,\bar I_3)\in\R^3$ and $\bI^{\text{iso},*} = (\bar J_1,\bar J_2,\bar J_3)\in\R^3$ for \emph{isotropy} $\mathcal G_{14}$.

	\subsection{Triclinic anisotropy $\mathcal G_1$}
	\label{app:tri}
	The \emph{triclinic anisotropy} $\mathcal G_1$ can be modeled with two symmetric 2nd order structure tensors $\bte A_\text{tri},\bte B_\text{tri}\in\Sym$, cf. Olive~et~al.~\cite{Olive2022}. For these tensors, the following conditions must hold:
	\begin{align}
		w_L = e_{LMN} A^\text{tri}_{MP} B^\text{tri}_{PN} \ne 0 \wedge \left[(\bte A_\text{tri} \cdot \ve w) \times \ve w \ne \zero \vee (\bte B_\text{tri} \cdot \ve w) \times \ve w \ne \zero \right] \; .
		\label{eq:triclinic}
	\end{align}
	The statement of the above condition is that the eigensystems of the tensors $\bte A_\text{tri},\bte B_\text{tri}$ are completely rotated to each other, i.e., there is no eigenvector of $\bte A_\text{tri}$ which is (anti)parallel to an eigenvector of $\bte B_\text{tri}$. Note that it is only possible to fulfill condition~\eqref{eq:triclinic} if both structure tensors have three non-equal eigenvalues each.
	
	According to Boehler~\cite{Boehler1977}, we get 9 additional invariants given by
	\begin{align}
		\bar P_4 &= \tr(\bte C \cdot \bte A_\text{tri}) \; , \; 
		\bar P_5 = \tr(\bte C^2 \cdot \bte A_\text{tri}) \; , \;
		\bar P_6 = \tr(\bte C \cdot \bte A_\text{tri}^2) \; , \; 
		\bar P_7 = \tr(\bte C^2 \cdot \bte A_\text{tri}^2) \; , \;
		\bar P_8 = \tr(\bte C \cdot \bte B_\text{tri}) \; , \; \\
		\bar P_9 &= \tr(\bte C^2 \cdot \bte B_\text{tri}) \; , \;
		\bar P_{10} = \tr(\bte C \cdot \bte B_\text{tri}^2) \; , \; 
		\bar P_{11} = \tr(\bte C^2 \cdot \bte B_\text{tri}^2) \; , \;
		\bar P_{12} = \tr(\bte C \cdot \bte A_\text{tri} \cdot \bte B_\text{tri}) \; .
	\end{align}
	Thus, we end up with the following set $\bI^\text{tri}=(\bar I_1,\bar I_2,\bar I_3,\bar P_4,\bar P_5,\ldots,\bar P_{12})\in\R^{12}$. Note that the given set is complete but might be irreducible.
	It is also possible to model \emph{tricilinic} anisotropy with two skew symmetric 2nd order tensors \cite{Xiao1996,Apel2004}.

	\subsection{Monoclinic anisotropy $\mathcal G_2$}
	\label{app:mono}
	The \emph{monoclinic anisotropy} $\mathcal G_2$ can also be modeled with a set of two symmetric 2nd order structure tensors $\bte A_\text{mon},\bte B_\text{mon}\in\Sym$ \cite{Olive2022}.
	For these tensors, the following conditions must hold:
	\begin{align}
		w_L = e_{LMN} A^\text{mon}_{MP} B^\text{mon}_{PN} \ne 0 \wedge \left[(\bte A_\text{mon} \cdot \ve w) \times \ve w = \zero \wedge (\bte B_\text{mon} \cdot \ve w) \times \ve w = \zero \right] \; .
	\end{align}
	The statement of the above condition is that the tensors $\bte A_\text{mon},\bte B_\text{mon}$ share one and only one eigenvector, i.e., it is parallel or antiparallel.
	
	According to Boehler~\cite{Boehler1977}, we get 9 additional invariants given by
	\begin{align}
		\bar Q_4 &= \tr(\bte C \cdot \bte A_\text{mon}) \; , \; 
		\bar Q_5 = \tr(\bte C^2 \cdot \bte A_\text{mon}) \; , \;
		\bar Q_6 = \tr(\bte C \cdot \bte A_\text{mon}^2) \; , \; 
		\bar Q_7 = \tr(\bte C^2 \cdot \bte A_\text{mon}^2) \; , \;
		\bar Q_8 = \tr(\bte C \cdot \bte B_\text{mon}) \; , \; \\
		\bar Q_9 &= \tr(\bte C^2 \cdot \bte B_\text{mon}) \; , \;
		\bar Q_{10} = \tr(\bte C \cdot \bte B_\text{mon}^2) \; , \; 
		\bar Q_{11} = \tr(\bte C^2 \cdot \bte B_\text{mon}^2) \; , \;
		\bar Q_{12} = \tr(\bte C \cdot \bte A_\text{mon} \cdot \bte B_\text{mon}) \; .
	\end{align}
	Thus, we end up with the following set $\bI^\text{mon}=(\bar I_1,\bar I_2,\bar I_3,\bar Q_4,\bar Q_5,\ldots,\bar Q_{12})\in\R^{12}$. Note that the given set is complete but might be irreducible.
	It is also possible to model \emph{monclinic} anisotropy with one symmetric and one skew symmetric 2nd order tensor \cite{Xiao1996,Apel2004}.

	\subsection{orthotropy $\mathcal G_3$}
	\label{app:orth}
	Typically, \emph{orthotropy} $\mathcal G_3$ is modeled with two structure tensors $\bte G_1:=\ve a_1 \otimes \ve a_1 \in \Sym$ and $\bte G_2:=\ve a_2 \otimes \ve a_2\in\Sym$ with $\ve a_\alpha \cdot \ve a_\beta = \delta_{\alpha\beta}$. Applying the rules from Boehler~\cite{Boehler1977} and accounting for the orthogonality of $\ve a_1$ and $\ve a_2$ as well as the fact that both tensors have two non-equal eigenvalues, respectively, we can build four additional invariants from $\bte C$, $\bte G_1$ and $\bte G_2$:
	\begin{align}
		\bar S_4 = \tr (\bte G_1 \cdot \bte C), \; \bar S_5 = \tr (\bte G_1 \cdot \bte C^2), \; \bar S_6 = \tr (\bte G_2 \cdot \bte C), \; \bar S_7 = \tr (\bte G_1 \cdot \bte C^2) \; .
	\end{align}
	
	As we will show in the following, it is also possible to build an equivalent invariant set by only using a single 2nd order structure tensor $\bte G_\text{orth}\in \Sym$ with three different eigenvalues $\lambda_1 \ne \lambda_2 \ne \lambda_3$.
	According to Boehler~\cite{Boehler1977}, we can build the following four additional invariants form $\bte C$ and $\bte G_\text{orth}$:
	\begin{align}
		\bar T_4 = \tr (\bte G_\text{orth} \cdot \bte C), \; \bar T_5 = \tr (\bte G_\text{orth} \cdot \bte C^2), \; \bar T_6 = \tr (\bte G_\text{orth}^2 \cdot \bte C), \; \bar T_7 = \tr (\bte G_\text{orth}^2 \cdot \bte C^2) \; .
	\end{align}
	By applying Sylvester's formula~\eqref{eq:sylvester}, we get 
	\begin{align}
		\te P_1 &= \ve n_1 \otimes \ve n_1 = \frac{1}{(\lambda_1-\lambda_2)(\lambda_1-\lambda_3)} \left(\bte G_\text{orth}^2 - (\lambda_2+\lambda_3) \bte G_\text{orth} + \lambda_2\lambda_3 \one\right) \text{ and }\\
		\te P_2 &= \ve n_2 \otimes \ve n_2 = \frac{1}{(\lambda_2-\lambda_1)(\lambda_2-\lambda_3)} \left(\bte G_\text{orth}^2 - (\lambda_1+\lambda_3) \bte G_\text{orth} + \lambda_1\lambda_3 \one\right) \; .
	\end{align}
	If $\bte G_\text{orth}$ is chosen in such a way that $\ve n_1=\ve a_1$ and $\ve n_2 = \ve a_2$, we find
	\begin{align}
		\bar S_4 &= \frac{1}{(\lambda_1-\lambda_2)(\lambda_1-\lambda_3)}\left(\bar T_6 - (\lambda_2+\lambda_3) \bar T_4 + \lambda_2\lambda_3\bar I_1\right) ,\\
		\bar S_5 &= \frac{1}{(\lambda_1-\lambda_2)(\lambda_1-\lambda_3)}\left(\bar T_7 - (\lambda_2+\lambda_3) \bar T_5 + \lambda_2\lambda_3(\bar I_1^2 - 2\bar I_2)\right), \\
		\bar S_6 &= \frac{1}{(\lambda_2-\lambda_2)(\lambda_2-\lambda_3)}\left(\bar T_6 - (\lambda_1+\lambda_3) \bar T_4 + \lambda_1\lambda_3\bar I_1\right) ,\\
		\bar S_7 &= \frac{1}{(\lambda_2-\lambda_2)(\lambda_2-\lambda_3)}\left(\bar T_7 - (\lambda_1+\lambda_3) \bar T_5 + \lambda_1\lambda_3(\bar I_1^2 - 2\bar I_2)\right) \; .
	\end{align}
	Thus, it is equivalent to use the set $\bI^\text{orth}=(\bar I_1,\bar I_2,\bar I_3,\bar S_4,\bar S_5,\bar S_6,\bar S_7)\in\R^7$ and $\bI^{\text{orth},*}=(\bar I_1,\bar I_2,\bar I_3,\bar T_4,\bar T_5,\bar T_6,\bar T_7)\in\R^7$.
	
	\subsection{tetragonal anisotropy $\mathcal G_5$}\label{app:tetragonal}
	Following \cite{Apel2004}, \emph{tetragonal anisotropy} $\mathcal G_5$ can be modeled with the structure tensors  
	\begin{align}
		\btttte D_\text{tet}:= \sum_{\alpha=1}^2 \ve A_\alpha \otimes \ve A_\alpha \otimes \ve A_\alpha \otimes \ve A_\alpha \text{ with }
		\ve A_\alpha \cdot \ve A_\beta = \delta_{\alpha\beta} \text{ and } \bte M_\text{tet} := \ve N \otimes \ve N \; , \ve N \in \Vn, \; \ve N \cdot \ve A_\alpha = 0 \; . \label{eq:tetratgonal_GST}
	\end{align}
	A complete invariant set can be formed from $\bte C$, $\bte H_1 := \btttte D_\text{tet} : \bte C$, $\bte H_2 := \btttte D_\text{tet} : \bte C^2$ and $\bte M_\text{tet}$. According to Boehler~\cite{Boehler1977}, one can build a set comprising 30 invariants from these four 2nd order tensors. By exploiting the orthogonality of the three vectors $\ve A_1$, $\ve A_2$ and $\ve N$, we find $\btttte D_\text{tet} : \btttte D_\text{tet} = \btttte D_\text{tet}$. Furthermore, by using $\ve A_\alpha = \te Q \cdot \ve e_\alpha$ and $\ve N = \te Q \cdot \ve e_3$, with $\te Q \in \SO$ and $\ve e_\alpha$ denoting the Cartesian basis vectors, we find that $Q_{I\alpha}Q_{J\beta} \bar C_{IJ} = \bar C^*_{\alpha\beta}$, where $ \bar C_{\alpha\beta}^*$ are the in-plane coordinates of $\bte C$ with respect to $\ve A_1,\, \ve A_2$ and $Q_{I3}Q_{J3} \bar C_{IJ} = \bar C_{33}^*$ is the out-of-plane coordinate in the direction of $\ve N$. From these relations we can find that $\tr(\bte H_1 \cdot \bte M_\text{tet}) = \tr(\bte H_1^2 \cdot \bte M_\text{tet}) = \tr(\bte H_2 \cdot \bte M_\text{tet}) = \tr(\bte H_2^2 \cdot \bte M_\text{tet}) = \tr(\bte C \cdot \bte H_1 \cdot \bte M_\text{tet}) = \tr(\bte C \cdot \bte H_2 \cdot \bte M_\text{tet}) = \tr(\bte H_1 \cdot \bte H_2 \cdot \bte M_\text{tet}) = 0$
	and $\tr(\bte C \cdot \bte M_\text{tet}) = \tr \bte C - \tr \bte H_1 \text{ and } \tr(\bte C^2 \cdot \bte M_\text{tet}) = \tr \bte C^2 - \tr \bte H_2$
	can be expressed by $\bte C$, $\bte H_1$ and $\bte H_2$. Thus, $\bte M_\text{tet}$ can be removed from the list of structure tensors for the tetragonal symmetry group $\mathcal G_5$. By accounting for the relations given above and by neglecting redundant expressions, e.g., $\tr (\bte C \cdot\bte H_1) = \tr \bte H_1^2$, we end up with the following set $\bI^\text{tet}=(\bar I_1,\bar I_2,\bar I_3,\bar U_4,\bar U_5,\ldots,\bar U_{13})\in\R^{13}$, where
	\begin{equation}
		\begin{split}
			\bar U_4 = \tr \bte H_1 \; , \; \bar U_5 = \tr \bte H_1^2 \; , \; \bar U_6 = \tr \bte H_1^3 \; , \; 
			\bar U_7 = \tr \bte H_2 \; , \; \bar U_8 = \tr \bte H_2^2 \; , \; \bar U_9 = \tr \bte H_2^3 \; , \;\\
			\bar U_{10} = \tr (\bte C^2 \cdot \bte H_1) \; , \; \bar U_{11} = \tr (\bte C^2 \cdot \bte H_1^2) \; , \; 
			\bar U_{12} = \tr (\bte C \cdot \bte H_2^2) \; , \; \bar U_{13} = \tr (\bte H_1^2 \cdot \bte H_2^2) \; .
		\end{split}
	\end{equation}
	Note that the given invariant set might be reducible. We do not proof this here. Also note that the 4th order structure tensor 
	\begin{align}
		\btttte D_\text{tet}^*:= \sum_{\alpha=1}^3 \ve A_\alpha \otimes \ve A_\alpha \otimes \ve A_\alpha \otimes \ve A_\alpha \text{ with }
		\ve A_\alpha \cdot \ve A_\beta =0, \alpha \ne \beta, \; |\ve A_1| = |\ve A_2| \ne |\ve A_3|
	\end{align}
	build from three orthogonal vectors $\ve A_\alpha$ is equivalent to \eqref{eq:tetratgonal_GST}.

	\subsection{cubic anisotropy $\mathcal G_7$}\label{app:cubic}
	
	Following \cite{Xiao1996,Apel2004,Ebbing2010}, \emph{cubic anisotropy} $\mathcal G_7$ can be modeled with the structure tensor 
	\begin{align}
		\btttte G_\text{cub}:= \sum_{\alpha=1}^3 \ve A_\alpha \otimes \ve A_\alpha \otimes \ve A_\alpha \otimes \ve A_\alpha \text{ with }
		\ve A_\alpha \cdot \ve A_\beta = \delta_{\alpha\beta} \; .
	\end{align}
	A complete invariant set can be formed from $\bte C$, $\bte H_1 := \btttte G_\text{cub} : \bte C$ and $\bte H_2 := \btttte G_\text{cub} : \bte C^2$, cf Xiao~\cite{Xiao1996}. According to Boehler~\cite{Boehler1977}, one can build a set comprising 21 invariants from the three 2nd order tensors $\bte C$, $\bte H_1$ and $\bte H_2$. We can reduce this set by exploiting the orthogonality of the vectors $\ve A_\alpha$, i.e., it holds $\btttte G_\text{cub} : \btttte G_\text{cub} = \btttte G_\text{cub}$. Furthermore, by using $\ve A_\alpha = \te Q \cdot \ve e_\alpha$, with $\te Q \in \SO$ and $\ve e_\alpha$ denoting the Cartesian basis vectors, we find that $Q_{I\alpha}Q_{J\beta} \bar C_{IJ} = \bar C^*_{\alpha\beta}$, where $\bar C^*_{\alpha\beta}$ are the coordinates of the right Cauchy-Green deformation tensor with respect to the coordinate system formed by $\ve A_\alpha,\, \alpha\in\{1,2,3\}$. By applying these relations and neglecting redundant expressions, we find the set $\bI^\text{cub}=(\bar I_1,\bar I_2,\bar I_3,\bar V_4,\bar V_5,\ldots,\bar V_{11})\in\R^{11}$ with
	\begin{equation}
		\begin{split}
			\bar V_4 &= \tr \bte H_1^2 \; , \; \bar V_5 = \tr \bte H_1^3 \; , \; \bar V_6 = \tr \bte H_2^2 \; , \;   \bar V_7 = \tr \bte H_2^3 \; , \; 
			\bar V_8 = \tr (\bte C^2 \cdot \bte H_1) \; , \; \\
			\bar V_9 &= \tr (\bte C^2 \cdot \bte H_1^2) \; , \; 
			\bar V_{10} = \tr (\bte C \cdot \bte H_2^2) \; , \; \bar V_{11} = \tr (\bte H_1^2 \cdot \bte H_2^2) \; .
		\end{split}
	\end{equation}
	Note that the given invariant set might be reducible. We do not proof this at this point.
	
	\begin{rmk}
		Note that the invariant set belonging to the cubic group can be expressed by invariants from the tetragonal group $\mathcal G_{5}$ up to order $\bte C^3$:
		\begin{align}
			\bar V_4 = \bar U_5 + (\bar I_1 - \bar U_4)^2  \; , \; 
			\bar V_5 = \bar U_6 + (\bar I_1 - \bar U_4)^3  \; , \; 
			\bar V_8 = \bar U_{10} + (\bar I_1 - \bar U_4)(1/2 \bar I_1^2 - \bar I_2 - \bar U_7) \; .
		\end{align} 
	\end{rmk}

	\subsection{hexagonal anisotropy $\mathcal G_{11}$}\label{app:hexa}
	According to \cite{Xiao1996,Apel2004}, the \emph{hexagonal anisotropy} group $\mathcal G_{11}$ can be modeled with the structure tensors 
	\begin{align}
		\btttttte G_\text{hex}:= \sum_{\alpha=1}^3 \ve A_\alpha \otimes \ve A_\alpha \otimes \ve A_\alpha \otimes \ve A_\alpha \otimes \ve A_\alpha \otimes \ve A_\alpha \text{ and } \bte M_\text{hex}:=\ve N \otimes \ve N \; ,
	\end{align}
	where $\ve A_\alpha \cdot \ve A_\beta = \pm\frac{1}{2}, \alpha \ne \beta$, $\ve N \cdot \ve A_\alpha = 0$ and $|\ve A_1| = |\ve A_2| = |\ve A_3| = |\ve N| = 1$. By building the 2nd order tensors $\bte H_1 := \bte C : \btttttte G : \bte C$ and $\bte H_2 := \bte C^2 : \btttttte G : \bte C^2$, cf. \cite{Xiao1996}, one can build a set comprising 30 invariants. However, similar to the tetragonal symmetry discussed in \ref{app:tetragonal}, the following invariants are equal to zero: $\tr(\bte H_1 \cdot \bte M_\text{hex}) = \tr(\bte H_1^2 \cdot \bte M_\text{hex}) = \tr(\bte H_2 \cdot \bte M_\text{hex}) = \tr(\bte H_2^2 \cdot \bte M_\text{hex}) = \tr(\bte C \cdot \bte H_1 \cdot \bte M_\text{hex}) = \tr(\bte C \cdot \bte H_2 \cdot \bte M_\text{hex}) = \tr(\bte H_1 \cdot \bte H_2 \cdot \bte M_\text{hex}) = 0$. By neglecting redundant expressions, we find the set $\bI^\text{hex}=(\bar I_1,\bar I_2,\bar I_3,\bar W_4,\bar W_5,\ldots,\bar W_{24})\in\R^{24}$ with
	\begin{equation}
		\begin{split}
			\bar W_4 &= \tr \bte H_1 \; , \; \bar W_5 = \tr \bte H_1^2 \; , \; \bar W_6 = \tr \bte H_1^3 \; , \; 
			\bar W_7 = \tr \bte H_2 \; , \; \bar W_8 = \tr \bte H_2^2 \; , \; \bar W_9 = \tr \bte H_2^3 \; , \;
			\bar W_{10} = \tr (\bte C \cdot \bte M_\text{hex}) \; , \; \\
			\bar W_{11} &= \tr (\bte C^2 \cdot \bte M_\text{hex}) \; , \;
			\bar W_{12} = \tr (\bte C \cdot \bte H_1) \; , \;
			\bar W_{13} = \tr (\bte C^2 \cdot \bte H_1) \; , \; \bar W_{14} = \tr (\bte C \cdot \bte H_1^2) \; , \; 
			\bar W_{15} = \tr (\bte C^2 \cdot \bte H_1^2) \; , \; \\
			\bar W_{16} &= \tr (\bte C \cdot \bte H_2) \; , \;
			\bar W_{17} = \tr (\bte C^2 \cdot \bte H_2) \; , \; \bar W_{18} = \tr (\bte C \cdot \bte H_2^2) \; , \; 
			\bar W_{19} = \tr (\bte C^2 \cdot \bte H_2^2) \; , \; 
			\bar W_{20} = \tr (\bte H_1 \cdot \bte H_2) \; , \;\\
			\bar W_{21} &= \tr (\bte H_1^2 \cdot \bte H_2) \; , \;
			\bar W_{22} = \tr (\bte H_1 \cdot \bte H_2^2) \; , \; \bar W_{23} = \tr (\bte H_1^2 \cdot \bte H_2^2) \; , \;
			\bar W_{24} = \tr (\bte C \cdot \bte H_1 \cdot \bte H_2) \; .
		\end{split}
	\end{equation}    
	Note again that the given invariant set for the hexagonal anisotropy group $\mathcal G_{11}$ might be reducible. We do not proof this at this point.
	
	Alternatively, if in addition to $\bte H_1$ and $\bte H_2$, the tensors $\bte H_3:= \one : \btttttte G : \bte C$ and $\bte H_4:= \one : \btttttte G : \bte C^2$ are used to build an invariant set for the hexagonal anisotropy group $\mathcal G_{11}$, we can find that the invariants $\tr(\bte C \cdot \bte M_\text{hex}) = \tr \bte C - \frac{2}{3}\tr \bte H_3$ and $\tr(\bte C^2 \cdot \bte M_\text{hex}) = \tr \bte C^2 - \frac{2}{3}\tr \bte H_4$ can be expressed by $\bte C$ and $\btttttte G_\text{hex}$. Thus, $\bte M_\text{hex}$ is not needed anymore in this case.
	
	\subsection{Transverse isotropy $\mathcal G_{13}$}
	\label{app:ti}
	In the case of \emph{transverse isotropy} $\mathcal G_{13}$, the structure tensor is given by $\bte G_\text{ti}:=a_1 \ve a_1 \otimes \ve a_1  + a_2 (\one - \ve a_1 \otimes \ve a_1)\in \Sym$ with $|\ve a_1| = 1$, $a_1,a_2 \in \R_{\ge 0}$ for positive semi-definiteness, $a_1 \ne a_2$, where $a_2$ is chosen to zero typically. 
	Thus, according to Boehler~\cite{Boehler1977},  we get four additional invariants given by
	\begin{align}
		\bar R_4 = \tr(\bte C \cdot \bte G_\text{ti}) , \; \bar R_5 = \tr(\bte C^2 \cdot \bte G_\text{ti})
		, \;  \bar R_6 = \tr(\bte C \cdot \bte G_\text{ti}^2) , \; \bar R_7 = \tr(\bte C^2 \cdot \bte G_\text{ti}^2)
		\label{eq:ti_reduce}
	\end{align}
	However, since $\bte G_\text{ti}$ has only two non-equal eigenvalues $\lambda_1 \ne \lambda_2$, one can find that $\bte G_\text{ti}^2$ can be expressed as
	\begin{align}
		\bte G_\text{ti}^2 = (\lambda_1 + \lambda_2) \bte G_\text{ti} + \lambda_1\lambda_2 \one 
	\end{align}
	by applying Eqs.~\eqref{eq:projection} and \eqref{eq:sylvester}.     
	Thus, the invariants $\bar R_6$, $\bar R_7$ given in Eq.~\eqref{eq:ti_reduce} are redundant since they can be expressed as $\bar R_6 = (\lambda_1 + \lambda_2) \bar R_4 + \lambda_1\lambda_2 \bar I_1$ and $\bar R_7 = (\lambda_1 + \lambda_2) \bar R_5 + \lambda_1\lambda_2 (\bar I_1^2 - 2\bar I_2)$. A complete set for transverse isotropy is thus given by $\bI^\text{ti}=(\bar I_1,\bar I_2,\bar I_3,\bar R_4, \bar R_5)\in\R^5$. Note that it is equivalent to choose the set  $\bI^{\text{ti},*}=(\bar I_1,\bar I_2,\bar I_3,\bar R_6, \bar R_7)\in\R^5$.

	\section{Neural network architecture}
	\label{app:NNs}
	
	In this work, NNs combined with internal in- and output normalization layers are employed to represent the elastic potential \cite{Kalina2024}. This avoids prior normalization of the training data and still limits the weights and biases to a range that is appropriate for an efficient optimization. It also makes the process of integrating calibrated models into FE codes simpler. More specifically, after training, the internal normalization layers are  included into the architecture by just multiplying the normalization weights. In addition, an architecture with an additional gate layer is introduced.
	
	\subsection{Neural network with internal normalization layers}
	A model which have to be trained by the data $\mathcal D$ consisting of tuples ${}^i\mathcal T := ({}^i \ve{\mathscr X}, {}^i Y) \in \R^n \times \R$, with the generalized vector ${}^i \ve{\mathscr X} :=({}^i X_1,{}^i X_2,\ldots,{}^i X_n)$, is given by
	\begin{align}
		f^\text{NN}: \R^n \to \R\,,\; \ve{\mathscr X} \mapsto f^\text{NN}(\ve{\mathscr X}) := (n^\text{out} \circ g^\text{NN} \circ  \ve{\mathscr n}^\text{in})(\ve{\mathscr X}) \; . \label{eq:PNN}
	\end{align}
	Therein, the trainable network is represented by the function $g^\text{NN}(\ve{\mathscr x)}$, where a PNN, i.e., an NN enforcing positive outputs for all possible normalized input vectors $\ve{\mathscr x} \in \R^n$, is used. The internal normalization layers are $\ve{\mathscr n}^\text{in}(\ve{\mathscr X})$ and $\ve{\mathscr n}^\text{out}(y)$.
	In order to not disturb the required positivity, we introduce these normalization layers as
	\begin{align}
		n_\alpha^\text{in}&: \R \to [x_\alpha^\text{min},x_\alpha^\text{max}] \subset \R \,,\; X_\alpha\mapsto n_\alpha^\text{in}(X_\alpha)
		:= X_\alpha \frac{x_\alpha^\text{max}-x_\alpha^\text{min}}{X^\text{max}_\alpha-X^\text{min}_\alpha} + \frac{x_\alpha^\text{max}X_\alpha^\text{min}-x_\alpha^\text{min}X_\alpha^\text{max}}{X_\alpha^\text{min}-X_\alpha^\text{max}} \text{ and}\\
		n^\text{out}&: [y^\text{min},y^\text{max}] \to \R  \,,\; y\mapsto n^\text{out}(y)
		:= \frac{Y^\text{max}-Y^\text{min}}{y^\text{max}-y^\text{min}} y \; ,
	\end{align}
	whereby there is no summation over the index $(\cdot)_\alpha$.
	The values $X_\alpha^\text{min}$, $X_\alpha^\text{max}$, $Y^\text{min}$ and $Y^\text{max}$ have to be determined from the data before training, whereas $x^\text{min}_\alpha$, $x^\text{max}_\alpha$, $y^\text{min}$ and $y^\text{max}$ have to be prescribed. 
	
	The PNN $g^\text{NN}(\ve{\mathscr x)}$ with $H$ hidden layers is given by
	\begin{align}
		\label{eq:PNN}
		o^{[1]}_\alpha &=  
		\mathcal{A}\Big(\sum_{\beta=1}^n w_{\alpha\beta}^{[1]} x_\beta + b_\alpha^{[1]}\Big)\;, \; \alpha \in\{1,2,\ldots,N^\text{nn,1}\}\; , \\
		o^{[h]}_\alpha &=  
		\mathcal{A}\Big(\sum_{\beta=1}^{N^{\text{NN},h-1}}w_{\alpha\beta}^{[h]} o_\beta^{[h-1]}+b_\alpha^{[h]} \Big)\;, \; \alpha \in\{1,2,\ldots,N^\text{nn,h}\}\;, h\in\{2,\dotsc,H\}\;, \\
		g^\text{NN}(\ve{\mathscr x})&= \sum_{\alpha=1}^{N^{\text{NN},H}} W_{\alpha}\, o^{[H]}_\alpha + B \in\R\; ,
	\end{align}
	whereby there are no restrictions for the weights expect for the output layer, i.e., $w_{\alpha\beta}^{[1]}, b_{\alpha}^{[1]}, w_{\alpha\beta}^{[h]}, b_{\alpha}^{[h]} \in \R$. 
	The activation function in the final hidden layer must be greater equal to zero for all possible output values of the previous layer in order to meet the condition that $g^\text{NN}(\ve{\mathscr x}) \ge 0 \; \forall \ve{\mathscr x} \in \R^n$. Thus, the softplus activation function $\softplus(z):= \ln(1+\exp z)$ is chosen and $W_\alpha, B \in \R_{\ge 0}$, the output layer's weights and bias, are restricted to be non-negative.
	All weights and bias values of the PNN are summarized in the $k$-dimensional vector
	\begin{align}
		\w \in \PNN := \left\{
		w_{\alpha\beta}^{[h]}, b_{\alpha}^{[h]} \in \R ; W_\alpha, B \in \R_{\ge 0} \; | \; h\in\{1,\ldots,H\}
		\right\} \; .
		\label{eq:setPNN}
	\end{align}
	For the sake of clarity, a somewhat simplified mathematical notation is used here for the quantity $\PNN$.
	
	\subsection{Neural network with internal normalization layers and gate layer}
	If, to enable sparsity of the model with respect to the number of inputs,  a trainable gate layer is included into the network in addition, the model given by Eq.~\eqref{eq:PNN} is modified according to
	\begin{align}
		f^\text{NN}: \R^n \to \R\,,\; \ve{\mathscr X} \mapsto f^\text{NN}(\ve{\mathscr X}) := (n^\text{out} \circ g^\text{NN} \circ \ve{\mathscr l}^\text{gate} \circ \ve{\mathscr n}^\text{in})(\ve{\mathscr X}) \; . \label{eq:PNN+gate}
	\end{align}
	Thereby, the gate layer is defined by 
	\begin{align}
		\ve{\mathscr l}^\text{gate}: \R^n \to \R^n, \ve x \mapsto \ve x \odot \ve g \text{ with }
		g_\alpha := \min(1,\gamma\tanh(\epsilon q_\alpha)) \in[0,1] \;
	\end{align}
	where $\gamma,\epsilon\in\R$ are hyper parameters and $q_\alpha\in[0,1]$ are trainable variables. Thus, we have the additional set $\ve{\mathscr q} \in \mathscr{G\!a\!t\!e}:=\left\{ \ve{\mathscr q} \in \R^n \, | \, q_\alpha \in[0,1] \right\}$. To enforce sparsity, $\ell_p$ regularization is used within this work.
	
	
	\section{Terms to guarantee zero stress in the undeformed state}
	\label{app:stress_norm}
	
	As discussed in Linden~et~al.~\cite{Linden2023}, the energy expression $\bar \psi^{\text{str},\square}$ enforcing $\bte P(\bte F=\one)=\zero$ depends on the chosen set of invariants, i.e., in our case whether the set is constructed either with $\bte G$, $\btttte G$, $\btttttte G$, or $(\bte G_1,\bte G_2)$.
	
	\paragraph{2nd order structure tensor}
	For the invariant set build with $\bte C$ and the 2nd order structure tensor $\bte G$, i.e., $\bI^{\bte G} \in \R^7$, we use the following expression to enforce a stress-free undeformed state:
	\begin{align}
		\bar \psi^{\text{str,}\bte G}(\bI^{\bte G},\bar J) &= -\mathfrak m^{\bte G} (\bar J -1) - \mathfrak n^{\bte G}(\bar K_4-1)
		- \mathfrak o^{\bte G}(\bar K_6-\bar K_6(\one)) \text{ with } \\
		\mathfrak m^{\bte G} &= 2\left(\diffp{\bar \psi^\text{NN}}{\bar I_1} 
		+ 2 \diffp{\bar \psi^\text{NN}}{\bar I_2} + \diffp{\bar \psi^\text{NN}}{\bar I_3}\right)\Bigg|_{\bte F=\one}
		\; , \\
		\mathfrak n^{\bte G} &= \left(\diffp{\bar \psi^\text{NN}}{\bar K_4} + 2\diffp{\bar \psi^\text{NN}}{\bar K_5} 
		\right)\Bigg|_{\bte F=\one} \; \text{and} \; 
		\mathfrak o^{\bte G} = \left(\diffp{\bar \psi^\text{NN}}{\bar K_6} + 2\diffp{\bar \psi^\text{NN}}{\bar K_7} 
		\right)\Bigg|_{\bte F=\one} \; .
	\end{align}
	
	\paragraph{4th order structure tensor}
	Similarly, for the invariant set $\bI^{\btttte G} \in \R^{11}$ build with $\bte C$ and the 4th order structure tensor $\btttte G$, the expression 
	\begin{align}
		\bar \psi^{\text{str,}\btttte G}(\bI^{\btttte G},\bar J) &= -\mathfrak m^{\btttte G} [\bar J -1] - \mathfrak n^{\btttte G}\left[\bar L_4-\bar L_4(\one)\right] - \mathfrak o^{\btttte G}\left[\bar L_5-\bar L_5(\one)\right] - \mathfrak p^{\btttte G}\left[\bar L_6-\bar L_6(\one)\right] - \mathfrak q^{\btttte G}\left[\bar L_{11}-\bar L_{11}(\one)\right] \text{ with } \\
		\mathfrak m^{\btttte G} &= 2\left(\diffp{\bar \psi^\text{NN}}{\bar I_1} 
		+ 2 \diffp{\bar \psi^\text{NN}}{\bar I_2} + \diffp{\bar \psi^\text{NN}}{\bar I_3} \right)\Bigg|_{\bte F=\one} \; , \; \\
		\mathfrak n^{\btttte G} &= \left(\diffp{\bar \psi^\text{NN}}{\bar L_4} + 2\diffp{\bar \psi^\text{NN}}{\bar L_7} + 2 \diffp{\bar \psi^\text{NN}}{\bar L_8} + 3\diffp{\bar \psi^\text{NN}}{\bar L_9}
		\right)\Bigg|_{\bte F=\one} \; , \\
		\mathfrak o^{\btttte G} &= \left(\diffp{\bar \psi^\text{NN}}{\bar L_5} + \frac{3}{2}\diffp{\bar \psi^\text{NN}}{\bar L_{10}}
		\right)\Bigg|_{\bte F=\one} \; , \\
		\mathfrak p^{\btttte G} &=\diffp{\bar \psi^\text{NN}}{\bar L_6}\Bigg|_{\bte F=\one} \text{ and } 
		\mathfrak q^{\btttte G} =\diffp{\bar \psi^\text{NN}}{\bar L_{11}}\Bigg|_{\bte F=\one}
	\end{align}
	is used. Thereby, only invariants up to the order $\bte C^3$ are used into the set.
	
	\paragraph{6th order structure tensor} Finally, for the model based on the 6th order structure tensor $\btttttte G$, we can use the expression
	\begin{align}
		\bar \psi^{\text{str,}\btttttte G}(\bI^{\btttttte G},\bar J) &= -\mathfrak m^{\btttttte G} [\bar J -1] - \mathfrak n^{\btttttte G}\left[\bar M_4-\bar M_4(\one)\right] - \mathfrak o^{\btttttte G}\left[\bar M_5-\bar M_5(\one)\right] - \mathfrak p^{\btttttte G}\left[\bar M_6-\bar M_6(\one)\right] - \mathfrak q^{\btttttte G}\left[\bar M_{10}-\bar M_{10}(\one)\right] \; , 
	\end{align}
	where the introduced terms are given by
	\begin{align}
		\mathfrak m^{\btttttte G} &= 2\left(\diffp{\bar \psi^\text{NN}}{\bar I_1} 
		+ 2 \diffp{\bar \psi^\text{NN}}{\bar I_2} + \diffp{\bar \psi^\text{NN}}{\bar I_3} \right)\Bigg|_{\bte F=\one} \; , \; \\
		\mathfrak n^{\btttttte G} &= \left(\diffp{\bar \psi^\text{NN}}{\bar M_4} + 2\diffp{\bar \psi^\text{NN}}{\bar M_7} + 2 \diffp{\bar \psi^\text{NN}}{\bar M_8} + 3\diffp{\bar \psi^\text{NN}}{\bar M_9} + 3\diffp{\bar \psi^\text{NN}}{\bar M_{11}} 
		\right)\Bigg|_{\bte F=\one} \; , \\
		\mathfrak o^{\btttttte G} &= \left(\diffp{\bar \psi^\text{NN}}{\bar M_5} + \frac{3}{2}\diffp{\bar \psi^\text{NN}}{\bar M_{12}}
		+ \frac{3}{2}\diffp{\bar \psi^\text{NN}}{\bar M_{13}}
		\right)\Bigg|_{\bte F=\one} \; , \\
		\mathfrak p^{\btttttte G} &=\diffp{\bar \psi^\text{NN}}{\bar M_6}\Bigg|_{\bte F=\one} \text{ and } 
		\mathfrak q^{\btttttte G} =\diffp{\bar \psi^\text{NN}}{\bar M_{10}}\Bigg|_{\bte F=\one}
	\end{align}
	to enforce zero stress in the undeformed state.
	
	\paragraph{Two 2nd order structure tensor}
	For the invariant set build with $\bte C$ and the two 2nd order structure tensors $\bte G_1$ and $\bte G_2$, i.e., $\bI^{\bte G_{1},\bte G_2} \in \R^{12}$, we use the following expression to enforce a stress-free undeformed state:
	\begin{align}
		\begin{split}
			\bar \psi^{\text{str,}\bte G_1,\bte G_2}(\bI^{\bte G_{1},\bte G_2},\bar J) &= -\mathfrak m^{\bte G_{1},\te G_2} (\bar J -1) 
			- \mathfrak n^{\bte G_{1},\bte G_2}(\bar N_4-1) - \mathfrak o^{\bte G_{1},\bte G_2}(\bar N_6-\bar N_6(\one)) \; \ldots \\
			&- \mathfrak p^{\bte G_{1},\bte G_2}(\bar N_8-1) - \mathfrak q^{\bte G_{1},\bte G_2}(\bar N_{10}-\bar N_{10}(\one))
			- \mathfrak r^{\bte G_{1},\bte G_2}(\bar N_{12}-\bar N_{12}(\one)) \text{ with } 
		\end{split}\\
		\mathfrak m^{\bte G_{1},\bte G_2} &= 2\left(\diffp{\bar \psi^\text{NN}}{\bar I_1} 
		+ 2 \diffp{\bar \psi^\text{NN}}{\bar I_2} + \diffp{\bar \psi^\text{NN}}{\bar I_3}\right)\Bigg|_{\bte F=\one}
		\; , \\
		\mathfrak n^{\bte G_{1},\bte G_2} &= \left(\diffp{\bar \psi^\text{NN}}{\bar N_4} + 2\diffp{\bar \psi^\text{NN}}{\bar K_5} 
		\right)\Bigg|_{\bte F=\one} \; , \; 
		\mathfrak o^{\bte G_{1},\bte G_2} = \left(\diffp{\bar \psi^\text{NN}}{\bar N_6} + 2\diffp{\bar \psi^\text{NN}}{\bar K_7} 
		\right)\Bigg|_{\bte F=\one} \; , \\
		\mathfrak p^{\bte G_{1},\bte G_2} &= \left(\diffp{\bar \psi^\text{NN}}{\bar N_8} + 2\diffp{\bar \psi^\text{NN}}{\bar N_9} 
		\right)\Bigg|_{\bte F=\one} \; , \; 
		\mathfrak q^{\bte G_{1},\bte G_2} = \left(\diffp{\bar \psi^\text{NN}}{\bar N_{10}} + 2\diffp{\bar \psi^\text{NN}}{\bar N_{11}} 
		\right)\Bigg|_{\bte F=\one} \; \text{and }\\
		\mathfrak r^{\bte G_{1},\bte G_2} &= \left(\diffp{\bar \psi^\text{NN}}{\bar N_{12}} 
		\right)\Bigg|_{\bte F=\one} \; .
	\end{align}
	
	Note that the proposed stress normalization terms are not polyconvex by construction. However, as the selected NN architecture is also not polyconvex, this is not a disadvantage.
	
	\section{Sampling technique for data generation}
	\label{app:samp}
	
	In this appended section, we describe the details on the sampling technique that has been applied for the data generation. The used technique is similar to the one presented in Kalina~et~al.~\cite{Kalina2024} but with the difference that the invariant space of the underlying anisotropy is not exploited here. 
	
	From the principle of material frame invariance we find that $\bar \psi(\bte F)$ only depends on the right stretch tensor $\bte U$, where $\bte F = \bte R \cdot \bte U$, $\bte R \in \SO$. Thus, for sampling we pragmatically choose $\bte R^\text{samp}:=\one$. Since $\bte U$ has to be positive definite, i.e., $\ve s \cdot \bte U \cdot \ve s > 0 \, \forall \ve s \in \Ln_1, \ve s \ne \zero$, it is not easily possible to directly sample the coordinates $\bar U_{KL}$. Instead, we make use of the relation
	\begin{equation}
		\bte U^\text{samp} = \bte Q^T(\theta_1,\theta_2,\theta_3) \cdot \diag(\bar\lambda_1,\bar\lambda_2,\bar J\,\bar\lambda_1^{-1} \bar\lambda_2^{-1}) \cdot \bte Q(\theta_1,\theta_2,\theta_3) \in \SO
	\end{equation}
	and sample in the six-dimensional space defined by $\boldsymbol{\mathcal S}:=(\bar \lambda_1,\bar \lambda_2,\bar J, \theta_1, \theta_2, \theta_3) \in \R_{>0} \times \R_{>0} \times \R_{>0} \times \R \times \R \times \R$ by Latin Hypercube Sampling (LHS). This guarantees the desired property of all samples ${}^s\bte U^\text{samp}$, $s \in\{1,2,\ldots,n_\text{samp}\}$. For each state ${}^s\bte F^\text{samp}=\bte R \cdot {}^s\bte U^\text{samp} =  {}^s \bte U^\text{samp}$, a loading path from $\bte F=\one$ to the final deformation is generated by linearly interpolating in $\bar \lambda_1, \bar \lambda_2,\bar J$ within $n_\text{inc}$ increments and keeping $\theta_1,\theta_2,\theta_3$ fixed. Thus, we have a set containing $k = n_\text{sampe} \cdot n_\text{inc}$ states: $\mathcal A:=\{{}^1\bte U, {}^2\bte U,\ldots,{}^k\bte U\}$.
	
	Finally, to avoid duplicate states in the data set and to reduce the number of states, filtering is performed using the Green-Lagrange strain $\bte E$ as a measure for comparing the states. To start the process, all states of the first loading path are added to a set $\mathcal U$ which only contains unique states, i.e., $\mathcal U:=\{{}^1\bte E^\text{un},{}^2\bte E^\text{un},\ldots,{}^{n_\text{inc}}\bte E^\text{un}\}$. Then, to decide whether another state ${}^a\bte E$ with $a\in\{n_\text{inc}+1,\ldots,k\}$ is inimitably, the relative distance
	\begin{align}
		{}^{ab}d := \frac{\|{}^a\bte E-{}^b\bte E\|^2}{\relu(\|{}^b\bte E\|^2-\delta)+\delta} \text{ with }
		\delta := \frac{1}{3} \max\left(\|{}^1\bte E\|^2,\|{}^2\bte E\|^2,\ldots,\|{}^k\bte E\|^2\right) \text{ and }
		{}^b\bte E \in \mathcal U
	\end{align}
	of the state to all states currently included in $\mathcal U$ is calculated. Thereby, $\delta$ is a heuristic parameter to prevent small values to be overrepresented \cite{Kalina2024}. 	
	If ${}^{ab}d\ge d_\text{tol}$ only applies to one $b \in \{1,2,\ldots,|\mathcal U|\}$, the state is unique and must be added to $\mathcal U$. To facilitate the subsequent application of a state in the computational homogenization, the complete loading path belonging to a new unique state is added to the set $\mathcal U$. After filtering, the deformation gradients belonging to the identified set of unique states are saved.
	
	To generate the deformation tuples by using the algorithm described above, the parameters given in Tab.~\ref{tab:sampling_params} have been chosen. With that, a total of 154 loading paths have been identified. Note that there exist several alternative sampling technique in the literature, e.g., Kunc~and~Fritzen~\cite{Kunc2019} or Fuhg~and~Bouklas~\cite{Fuhg2022a}.
	
	\begin{table}
		\begin{center}
			\caption{Chosen parameters for the sampling algorithm. With that, a set containing a total of 154 loading paths have been generated.}
			\label{tab:sampling_params}
			\begin{footnotesize}
				\begin{tabular}{lllllllll}
					range $\bar \lambda_1$ & range $\bar \lambda_2$ & range $\bar J$ & range $\theta_1$ & range $\theta_2$ & range $\theta_3$ & $n_\text{samp}$ & $n_\text{inc}$ & $d_\text{tol}$\\
					\hline\hline
					$[0.8,1.4]$ & $[0.8,1.4]$ & $[0.9,1.2]$ & $[0,\pi]$ & $[-\pi/2,\pi/2]$ & $[-\pi,\pi]$ & $\num{5e3}$ & $20$ & $0.15$
				\end{tabular}
			\end{footnotesize}
		\end{center}
	\end{table}

	\section{Hyperparameter study}
	\label{app:hyper}
	
	Within this appended section, we provide a hyperparameter study for the considered network architectures. In addition, a study on the optimal choice of the weight for $\ell_p$ regularization is shown. To exclude random effects from initialization, 5 training runs with pre-training and post-training step as described in Remark~\ref{rmk:train} have carried out each. The best training run is used.
	The overall datasets are divided into calibration and test sets with a ratio of $70/30$, respectively, see Eq.~\eqref{eq:cal_test}.
	To reduce the number of trainings, we set the knowledge of the required order of structure tensor for the respective RVEs here: \emph{stochastic fibers} ($\bte G\in\Sym$), \emph{hexagonal fibers} ($\btttttte G\in\Sym_6$), \emph{cubic sphere} ($\btttte G\in\Sym_4$), \emph{plane spheres} ($\bte G\in\Sym$),  and \emph{chain spheres} ($\bte G_1,\bte G_2\in\Sym$), cf. Tab.~\ref{tab:loss_inter}.
	
	\subsection{Number of hidden layers and neurons}
	\label{app:study_hyper}
	
	\begin{figure}
		\centering
		\includegraphics{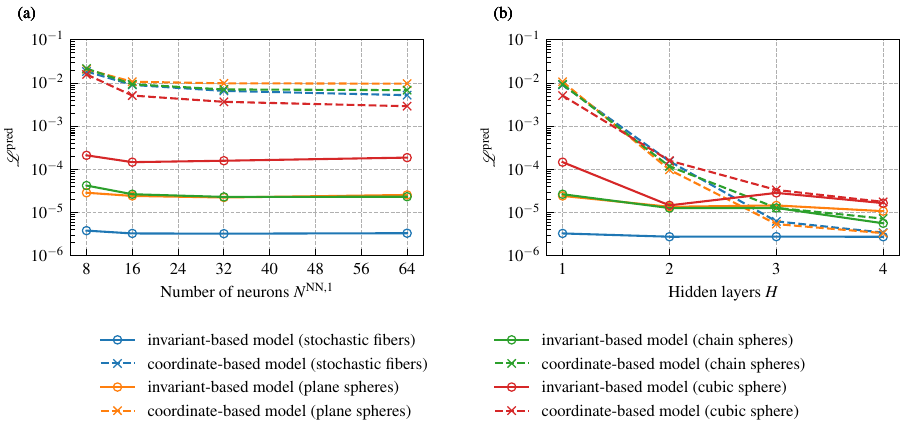}
		\caption{Hyperparameter study for the invariant-based model $\bar \psi^\square(\bI^\square,\bar J)$ and the coordinate-based model $\bar \psi^\text{coord}(\bte C, \bar J)$ with the loss $\mathcal L= 0.7\mathcal L^{\te \sigma} + 0.3\mathcal L^{\ttttes c}$: (a) NNs with one hidden layer and varied number of neurons $N^{\text{NN},1}$ and (b) NNs consisting of $H$ hidden layers with 16 neurons each. The trainable gate layer was deactivated for the invariant-based models. Shown are the results of the best run out of 5 training runs.}
		\label{fig:study_hyperparams}
	\end{figure}
	
	Here, the hyperparameters of the networks used in the invariant-based model $\bar \psi^\square(\bI^\square,\bar J)$ according to Eq.~\eqref{eq:NN_model_invariants} and the coordinate-based model $\bar \psi^\text{coord}(\bte C, \bar J)$ given in Eq.~\eqref{eq:NN_coord} are varied. For the loss, we choose $\mathcal L= 0.7\mathcal L^{\te \sigma} + 0.3\mathcal L^{\ttttes c}$. Within this hyperparameter study, the trainable gate layer is deactivated for the invariant-based models. 
	
	\paragraph{Variation number of neurons in one hidden layer}
	In a first study, we consider NNs with one hidden layer and vary the number of neurons $N^{\text{NN},1}\in\{8,16,32,64\}$. 
	The results are shown in Fig.~\ref{fig:study_hyperparams}(a). As can be seen, the loss achieved with the coordinate-based model does not fall below $\num{5e-3}$ for any of the RVEs, and there is no noticeable improvement when the number of neurons is increased.
	In contrast, the invariant-based approach achieves very good results with just one hidden layer for four of the five RVEs. Only for the RVE cubic sphere does the loss remain above $\num{1e-4}$ even with 64 neurons in the hidden layer.
	
	\paragraph{Variation number of hidden layers}
	Secondly, we consider NNs with $H\in\{1,2,3,4\}$ hidden layers with $N^{\text{NN},h}=16$, $h\in\{1,\ldots,H\}$.
	The results of this study are shown in Fig.~\ref{fig:study_hyperparams}(b). As can be seen, the loss achieved with the coordinate-based model now drops significantly for all RVEs considered when the number of hidden layers is increased. With two hidden layers, the loss values are around $\num{1e-4}$ and with three hidden layers, values clearly below $\num{1e-4}$ and in some cases below $\num{1e-4}$ are achieved for all RVEs. With four hidden layers, there is then no further significant improvement.
	In contrast, the invariant-based approach results in almost no dependence on the number of hidden layers for four of the five RVEs. Even with one hidden layer of 16 neurons, the results here are very good. Only for the RVE cubic sphere there is a noticeable improvement above a number of two hidden layers, so that a loss below $\num{1e-4}$ can be achieved from $H\ge 2$.
	It should be noted that the finding that relatively small networks are required for invariant-based NN approaches is consistent with the literature, cf. \cite{Linden2023,Kalina2024,Linka2021,Rosenkranz2024,Klein2021,Klein2024}.
	
	The study thus results in the following choice for the NNs used in this work: 
	For the invariant-based approach, architectures with 2 hidden layers of 16 neurons each are used. For the coordinate-based model, on the other hand, networks with 3 hidden layers of 16 neurons each are used.

	\subsection{Weighting of the gate loss}
	\label{app:study_gate}
	
	\begin{figure}[t]
		\centering
		\includegraphics{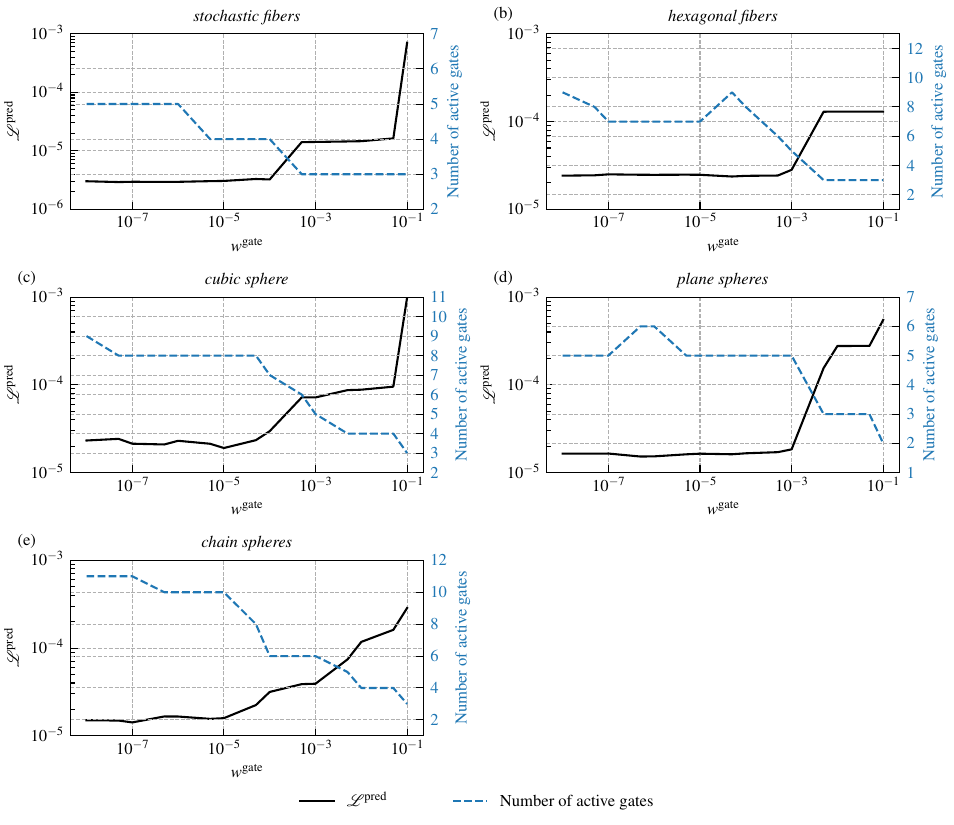}
		\caption{Variation of the weight $w^\text{gate}$ for the gate loss term  $\mathcal L^\text{gate}$: (a)--(e) RVEs stochastic fibers, hexagonal fibers, cubic fibers, plane spheres, and chain spheres, respectively. The prediction loss was chosen to $\mathcal L^\text{pred}= 0.7\mathcal L^{\te \sigma} + 0.3\mathcal L^{\ttttes c}$. Shown are the results of the best run out of 5 training runs.}
		\label{fig:study_sparse}
	\end{figure}
	
	Here, the weight $w^\text{gate}$ for the loss term $\mathcal L^\text{gate}$ defined in Eq.~\eqref{eq:gate_loss} is varied systematically. Thus, only the invariant-based model $\bar \psi^\square(\bI^\square,\bar J)$ is considered. According to the study given in \ref{app:study_hyper}, architectures with 2 hidden layers of 16 neurons each are used.
	The parameters for the gates and the exponent in the $p$-norm are chosen to $\gamma = 1.025$, $\epsilon = 2.5$, $\delta = \num{1e-6}$, and $p=\frac{1}{4}$, respectively. The loss term for the training is given by $\mathcal L = \mathcal L^\text{pred} + w^\text{gate} \mathcal L^\text{gate}$, where the prediction loss is chosen as $\mathcal L^\text{pred}= 0.7\mathcal L^{\te \sigma} + 0.3\mathcal L^{\ttttes c}$. The weight is varied as follows: $w^\text{gate} \in\{\num{1e-8},\num{5e-8},\ldots,\num{1e-1}\}$.
	
	The results of this study are given in Fig.~\ref{fig:study_sparse} for the five considered RVEs. On the left vertical axis of each subplot, the prediction loss is plotted and on the right vertical axis (blue) the number of active gates, i.e., gates for which the condition $g_\alpha > 0$ holds.
	As one can see, the number of active gates after training decreases with an increasing $w^\text{gate}$. However, if the weight is set too high, this leads to an excessive weighting of the penalty term based on the $p$-norm. This initially leads to the elimination of invariants required to describe the anisotropy from the model and, if the value is increased further, to a disproportionate deterioration in the prediction capability.
	Thus, the aim is now to find a value for the weight that leads to a model with as few invariants as possible, but at the same time does not negatively affect the prediction quality. Accordingly, a value of \num{5e-5} has proven to be suitable for all RVEs.

\bibliographystyle{unsrtnat} 
\bibliography{PANN_aniso.bib}

\end{document}

%% file: StyleSetup.tex
\usepackage{amssymb}
\usepackage{amsthm}
\usepackage{amsmath}
\usepackage{siunitx}
\usepackage{mathtools}		
\mathtoolsset{centercolon}	
\usepackage{newtxmath}      
\usepackage{lineno}
\usepackage{pifont}
\usepackage[dvipsnames]{xcolor}
\usepackage{bbm}
\usepackage{multirow}

\newcommand{\R}{\mathbb R} 
\newcommand{\N}{\mathbb N} 
\newcommand{\Sym}{\mathscr{S\! y \! m}} 
\newcommand{\Ln}{\mathcal L} 
\newcommand{\SO}{\mathscr{S\!O}(3)} 
\newcommand{\Othree}{\mathscr{O}(3)} 
\newcommand{\GL}{\mathcal{G\!L}^+(3)} 
\newcommand{\Vn}{\mathcal{N}} 
\newcommand{\G}{\mathcal{G}} 

\newcommand{\B}{\mathcal{B}} 
\newcommand{\I}{\boldsymbol{\mathcal I}}
\newcommand{\bI}{\,\bar{\boldsymbol{\!\I}}}

\newcommand{\bcKM}{\bar{\boldsymbol{\mathscr c}}}

\newcommand{\w}{\boldsymbol{\mathscr w}}
\newcommand{\g}{\boldsymbol{\mathscr g}}

\newcommand{\PNN}{\mathcal{P\!N\!N}} 

\DeclareMathOperator{\relu}{ReLU}
\DeclareMathOperator{\softplus}{SP}
\DeclareMathOperator{\tr}{tr}
\DeclareMathOperator{\cof}{cof}
\DeclareMathOperator{\sym}{sym}

\DeclareMathOperator{\diag}{diag}
\DeclareSIUnit[number-unit-product = \,]{\promille}{\textperthousand}

\newcommand{\dx}{\mathrm d} 
\newcommand{\ve}[1]{\boldsymbol{#1}} 
\newcommand{\te}[1]{\boldsymbol {#1}} 
\newcommand{\tttte}[1]{\mathbb #1} 
\newcommand{\ttttes}[1]{\mathbbm #1} 
\newcommand{\tttttte}[1]{\boldsymbol{\mathsf #1}} 
\newcommand{\one}{\textit{\textbf{1}}}
\newcommand{\zero}{\textit{\textbf{0}}}

\newcommand{\bve}[1]{\bar{\boldsymbol{#1}}} 
\newcommand{\bte}[1]{\bar{\boldsymbol #1}} 
\newcommand{\btttte}[1]{\bar{\mathbb #1}} 
\newcommand{\bttttes}[1]{\bar{\mathbbm #1}} 
\newcommand{\btttttte}[1]{\bar{\boldsymbol{\mathsf #1}}} 

\newcommand{\diffp}[2]{\frac{\partial #1}{\partial #2}} 

\newcommand{\nablaX}{\nabla_{\!\!{\ve X}}}